\documentclass[11pt, oneside, a4paper]{article}

\usepackage[left=2cm,right=2cm,top=2cm,bottom=2cm]{geometry}
\usepackage{amsmath}
\usepackage{amsfonts}

\usepackage{mathtools}
\usepackage{bm}
\usepackage{bbm}
\usepackage{dsfont}

\usepackage{setspace}
\usepackage{float}
\usepackage{rotating}
\usepackage{color}
\usepackage{authblk}   
\usepackage{multirow}

\usepackage{amsthm}   
\newtheorem{theorem}{Theorem}[section]
\newtheorem{corollary}[theorem]{Corollary}

\newcommand\Tstrut{\rule{0pt}{2.6ex}}         

\usepackage{natbib}
\usepackage{tikz}
\usetikzlibrary{calc}

\newcommand{\tikzmark}[1]{\tikz[overlay,remember picture] \node (#1) {};}
\newcommand{\DrawBox}[4][]{%
    \tikz[overlay,remember picture]{%
        \coordinate (TopLeft)     at ($(#2)+(-0.4em,0.9em)$);
        \coordinate (BottomRight) at ($(#3)+(0.2em,-0.4em)$);
        \path (TopLeft); \pgfgetlastxy{\XCoord}{\IgnoreCoord};
        \path (BottomRight); \pgfgetlastxy{\IgnoreCoord}{\YCoord};
        \coordinate (LabelPoint) at ($(\XCoord,\YCoord)!0.5!(BottomRight)$);
        \draw [red,#1] (TopLeft) rectangle (BottomRight);
        \node [below, #1, fill=none, fill opacity=1] at (LabelPoint) {#4};
    }
}

\newcommand\bovermat[2]{%
  \makebox[0pt][l]{$\smash{\overbrace{\phantom{
    \begin{matrix}#2\end{matrix}}}^{\text{#1}}}$}#2}
    
\newcommand{\sss}{\scriptscriptstyle}   
    
\allowdisplaybreaks

\title{\textbf{Familywise error control in multi-armed response-adaptive trials}}

\author[1]{D.\ S.\ Robertson}
\author[1,2]{J.\ M.\ S.\ Wason}
\affil[1]{\small MRC Biostatistics Unit, University of Cambridge, Cambridge, UK}
\affil[2]{\small Institute of Health and Society, Newcastle University, Newcastle, UK}

\date{}

\begin{document}

\maketitle

\begin{abstract}
Response-adaptive designs allow the randomization probabilities to change during the course of a trial based on cumulated response data, so that a greater proportion of patients can be allocated to the better performing treatments. A major concern over the use of response-adaptive designs in practice, particularly from a regulatory viewpoint, is controlling the type I error rate. In particular, we show that the na\"ive $z$-test can have an inflated type I error rate even after applying a Bonferroni correction. Simulation studies have often been used to demonstrate error control, but do not provide a guarantee. In this paper, we present adaptive testing procedures for normally distributed outcomes that ensure strong familywise error control, by iteratively applying the conditional invariance principle. Our approach can be used for fully sequential and block randomized trials, and for a large class of adaptive randomization rules found in the literature. We show there is a high price to pay in terms of power to guarantee familywise error control for randomization schemes with extreme allocation probabilities. However, for proposed Bayesian adaptive randomization schemes in the literature, our adaptive tests maintain or increase the power of the trial compared to the $z$-test. We illustrate our method using a three-armed trial in primary hypercholesterolemia. \\
\end{abstract}


\vspace{3cm}

\noindent Address correspondence to D.\ S.\ Robertson,  MRC Biostatistics Unit, University of Cambridge, IPH Forvie Site, Robinson Way, Cambridge CB2 0SR, UK; E-mail: david.robertson@mrc-bsu.cam.ac.uk

\newpage 

\section{Introduction}
\label{sec:intro}

Clinical trials typically randomize patients using a fixed randomization scheme, where the probabilities of assigning patients to the experimental treatments and control are pre-specified and constant. A common method is to simply use equal randomization to the different arms of the trial. However, such randomization schemes can mean that a substantial proportion of the trial participants will continue to be allocated to treatments that are not the best available, even if interim data indicates that one treatment is likely to be superior.
%
Response-adaptive trials address this concern by adaptively changing the randomization probabilities, so that a greater proportion of patients are allocated to the treatment arm which has a better performance based on the cumulated response data. Hence, as the trial continues and accumulates more data, patients in the trial can benefit from having a higher probability of being assigned to a better treatment. 


Many classes of response-adaptive randomization schemes have been proposed in the literature for binary outcomes. Randomization schemes based on urn models (such as the randomized play-the-winner rule~\citep{Wei1978}), and adaptive biased coin designs (such as the doubly-adaptive biased coin design~\citep{Eisele1994}), have been extensively studied, with a comprehensive presentation given by~\citet{Hu2006}. Many Bayesian adaptive randomization (BAR) schemes have also been proposed~\citep{Thall2007, Trippa2012, Yin2012, Wason2014a}, where the randomization probabilities are recursively updated using a Bayesian model for the patient outcomes.

There is also a growing interest in response-adaptive randomization for continuous responses. For example, there are schemes based on doubly-adaptive biased coin designs~\citep{Hu2006, Zhang2006, Biswas2007}, urn-based drop-the-loser designs~\citep{Ivanova2006} and bandit-based designs~\citep{Smith2017}. A comprehensive recent overview is given by~\citet{Biswas2016}. In this paper, our focus is on normally distributed outcomes, which are encountered in many clinical trials. Indeed, 23 out of the 59 multi-arm clinical trials identified in a review by~\citet{Wason2014} had a continuous outcome.


A comprehensive discussion of the relative advantages and disadvantages of adaptive versus fixed randomization is beyond the scope of this paper. Indeed, the use of adaptive randomization is a widely discussed and somewhat controversial topic in clinical trials. For binary responses, a number of comparisons~\citep{Korn2011, Berry2011, Thall2015, Wathen2017} have focused on the BAR scheme proposed by~\citet{Thall2007}. Particularly in the two-arm setting, fixed randomisation appears to be preferable to this scheme in terms of power and the number of treatment failures, except when the number of patients to be treated beyond the trial is small (as in rare diseases) or where there are large treatment differences~\citep{Lee2012, Du2015}.

However, even in the two-arm setting, optimal response-adaptive schemes (i.e.\ those that target some formal optimality criteria) have been shown to have benefits over fixed randomisation by increasing both power and patient benefit simultaneously~\citep{Rosenberger2001, Rosenberger2004, Tymofyeyev2007, Bello2016}. In the multi-arm setting, which is the focus of this paper, adaptive randomisation can have further advantages over fixed randomization~\citep{Berry2011, Wason2014a, Hey2015, Berry2015}, particularly for more complex trial designs.

Response-adaptive designs also have application outside of the context of clinical trials. For example, multi-arm bandit models are used for market learning in economics~\citep{Bergemann2006} and to improve modern production systems that emphasize `continuous improvement'~\citep{Scott2010}. Some of the ethical concerns surrounding adaptive randomization~\citep{Hey2015} would not apply in these contexts.

Despite the extensive literature on response-adaptive randomization, relatively few clinical trials have actually used such schemes in practice. One of the first examples, which used a randomized play-the-winner rule, was a trial of extracorporeal membrane oxygenation to treat newborns with respiratory failure~\citep{Bartlett1985}. More recent examples include a three-armed trial in untreated patients with adverse karyotype acute myeloid leukemia~\citep{Giles2003}, which used BAR. The ongoing I-SPY~2 trial~\citep{Park2016, Rugo2016}, which screens drugs in neoadjuvant breast cancer, also uses BAR as part of its design. 

A key concern over using response-adaptive randomization, particularly from a regulatory perspective, is ensuring that the type~I error rate is controlled. Indeed, draft regulatory guidance from the U.S.~\citet{FDA2010} includes adaptive randomization under a section entitled ``Adaptive Study Designs Whose Properties Are Less Well Understood''. It then goes on to state that ``particular attention should be paid to avoiding bias and controlling the Type~I error rate"~\citep[pg.~27]{FDA2010} when using adaptive randomization in trials.


In a multi-arm trial, multiple hypotheses are tested simultaneously by design, which leads to a multiple testing problem. To account for this, testing procedures are used that guarantee strong control of the familywise error rate (FWER), which ensures the maximum probability of making at least one type I error is controlled. For confirmatory trials in particular, demonstrating strong control of the FWER is often required by regulators~\citep{FDA2010, European2002}.

For response-adaptive trials, a rigorous proof of FWER control for a particular design is difficult given the complexities of the treatment allocation process. Hence error control has typically either been demonstrated through simulation studies, or by exploiting the asymptotic structure of the adaptive randomization procedure~\citep{Hu2006, Zhu2010}. However, neither method provides a guarantee of FWER control, particularly with small sample sizes. 
%
\citet{Gutjahr2011} showed how to achieve strong control of the FWER for normally distributed outcomes in a two-stage design incorporating response-adaptive randomization. However, our focus is on general response-adaptive trials, without the necessity of restricting to two stages or having a final stage of equal randomization.

In this paper, we show how to guarantee strong control of the FWER for both fully sequential and block randomized response-adaptive trials, for a large class of adaptive randomization rules. Our proposed procedure works by reweighting the usual $z$-statistic through an iterative application of the conditional invariance principle. The resulting adaptive test statistic can then be used to test the elementary null hypothesis that a treatment is superior to the control.


The rest of the paper is organised as follows. In Section~\ref{sec:seq_rar}, we describe the proposed method for fully sequential response-adaptive trials with a fixed allocation to the control. This method is then modified for block randomized response-adaptive trials in Section~\ref{sec:block}, for both a fixed or adaptive control allocation. Simulation studies for the proposed methods are presented in Section~\ref{sec:simul}, and Section~\ref{sec:case_study} gives a case study based on a trial in primary hypercholesterolemia. We conclude with a discussion in Section~\ref{sec:discuss}. All proof details can be found in the Appendices.

\section{Fully sequential response-adaptive trials}
\label{sec:seq_rar}


\subsection{Trial setting}

Suppose a trial is conducted to test $h > 1$ experimental treatments against a common control, using the following design. A total of $n$~patients are allocated to the experimental treatments, and $n_0$~patients are allocated to the control, where~$n_0$ and~$n$ are fixed in advance. Patients are allocated to the different experimental treatments using response-adaptive randomization, where we assume that the randomization rule does not depend on the control information. We also assume the allocation to the control is fixed; that is, the probability of assigning a patient to the control is pre-specified and constant. Maintaining allocation to the control is recommended by the~\citet{FDA2010}, since it best maintains the power of the trial, and helps address the concern about changing patient characteristics over the course of the trial.

The response-adaptive randomization for the experimental treatments starts with a burn-in period~$B$, which uses equal randomization to allocate $r_i > 0$ patients to the $i$th treatment $(i = 1, \ldots, h)$, with the $r_i$ again fixed in advance. Hence a total of~$r = \sum_{i=1}^h r_i$ patients are allocated to the experimental treatments during the burn-in period. 
Let $a_k$ denote the treatment allocation for the $k$th experimental patient ($k = 1, \ldots, n$), where $a_k = i$ if the $k$th patient is allocated to the $i$th treatment. Also, let~$X_k$ denote the efficacy outcome for the $k$th patient. Similarly, let~$X_{0j}$ denote the efficacy outcomes for the $j$th patient on the control ($j = 1, \ldots, n_0$). We assume that \[
X_{0j} \sim N(\mu, \sigma^2), \;\; X_k \! \mid_{a_k = i} \; \sim N(\mu + \delta_i, \sigma^2)
\]
The variance~$\sigma^2$ is assumed known and, without loss of generality, we set~$\sigma^2 = 1$. Here~$\delta_i$ represents the incremental benefit of treatment~$i$ compared to the control, and is the parameter of interest. Finally, let~$n_i$ denote the total number of allocations to the $i$th experimental treatment, including the burn-in period. 

\subsection{Hypothesis testing}

The elementary null hypotheses of interest are $H_i: \delta_i = 0$ against the one-sided alternatives $\bar{H}_i : \delta_i > 0$. We discuss the case when $H_i: \delta_i \leq 0$ at the end of Section~\ref{subsec:cond_inv}. One general method to control for multiple testing is to use the closure principle~\citep{Marcus1976} and consider all intersection hypotheses~$H_I$, where $I \subseteq \{1, \ldots, h\}$. To strongly control the FWER, we reject an elementary null hypothesis~$H_i$ if we also reject every~$H_I$ with~$i \in I$ using a local level-$\alpha$ test. Hence we need to define a valid level-$\alpha$ test for all the intersection hypotheses~$H_I$.
The na\"ive $z$-test for~$H_I$, which does not take into account the response-adaptive randomization used in the trial, rejects $H_I$ if the test statistic \[
T_I = \sum_{k = 1}^{n} \left( \mathds{1}_{\{a_k \in I\}} \frac{X_k}{n_I} \right) - \sum_{j=1}^{n_0}\frac{X_{0j}}{n_0} \] is greater than $z_{\alpha}\left( 1/n_I + 1/n_0\right)^{1/2}$, where $n_I = \sum_{i \in I} n_i$ and~$z_{\alpha}$ is the $(1-\alpha)$ standard normal quantile.

As an alternative to using the closure principle with the test statistic above, we could control the FWER by simply using a Bonferroni correction, or a step-up/step-down procedure such as the Holm procedure. These would only involve calculating test statistics for the~$h$ elementary null hypotheses, i.e.\ calculating $T_I$ for $I = \{i\}$ ($i = 1, \ldots, h$). Hence we present the methodology assuming the closure principle will be used, with the Bonferroni and Holm procedures considered as special cases. We return to this issue in Section~\ref{sec:simul}.

\subsection{Inflation of the familywise error rate}
\label{subsec:inflate_fwer_seq}

Since the $z$-test ignores the adaptive randomization used, it is possible to inflate the FWER. As an example, consider the following adaptive randomization scheme for $h = 2$ treatments: \[
a_{k+1} = \begin{cases}
2 & \text{if } \; \sum_{j = 1}^k (\mathds{1}_{\{a_j = 1\}} \frac{X_j}{n_{1k}}) > 0.5\\
1 & \text{otherwise}
\end{cases}
\] where $n_{1k} =  \sum_{j = 1}^k \mathds{1}_{\{a_j = 1\}}$. This can be viewed as implementing early stopping for efficacy for treatment~1, which is not taken into account using the na\"ive $z$-test.

We ran a simulation study to calculate the type~I error rate using the above randomization scheme. We set~$\alpha = 0.05$, $n_0 = n = 60$, $r_1 = r_2 = 5$ and the true treatment means $\mu = 0, \delta_1 = 0$, $\delta_2 = 1$. The type~I error rate as averaged over $10^5$ simulations is $10.4\%$, more than double the nominal~$5\%$ level. We subsequently refer to allocation rules of this type as `type~I error inflator' rules (which clearly would never be used in practice).

%

\subsection{Auxiliary design}
\label{subsec:aux_seq}

Working with the actual design of the trial is difficult because the response-adaptive randomization affects the distribution of the usual $z$-test statistics. Hence for each~$H_I$ we introduce a simpler design, called the auxiliary design, for which we do know the distribution. The actual trial design can then be viewed as a series of data-dependent modifications of the auxiliary design, where we account for the modifications using the conditional invariance principle. The auxiliary designs are purely hypothetical, and are only used to construct the modified tests for the actual design. As well, the allocations in the auxiliary designs are fixed before the start of the actual trial.

The auxiliary design for hypothesis~$H_I$ is as follows. As in the actual design, a total of $n$~patients are allocated to the experimental treatments, and $n_0$~patients are allocated to the control. The allocations and responses to the control treatment are the same as the actual design. For the patients allocated to the experimental treatments, the auxiliary design starts with a burn-in period~$B$ with $r$ patients that is identical to the actual design. The subsequent~$n-r-1$ allocations are given by a fixed sequence $(b_{r+1},\ldots, b_{n-1})$, which can be chosen arbitrarily. These allocations can be considered as a `guess' of a likely allocation sequence of the actual trial design. One possibility would be to randomize equal numbers of patients for each treatment. The final allocation $b_n$ must be to one of the treatments in~$I$. 

%

We now introduce some notation for the auxiliary design. Let $n_i' =  \sum_{j = 1}^n \mathds{1}_{\{b_j = i\}}$ denote the total number of allocations to the $i$th experimental treatment. Also let  $m_{i,k} = \sum_{j = k}^n \mathds{1}_{\{b_j = i\}} $ denote the total number of allocations to the $i$th treatment for patients $(k, k+1, \ldots, n)$. We define $n_I' = \sum_{i \in I} n_i'$ and $m_{I,k} = \sum_{i \in I} m_{i,k}$. Under the auxiliary design, $n_i'$ is fixed for all~$i$, and hence under $H_I$, the usual $z$-statistic \[
T_I' = \sum_{k = 1}^{n} \left( \mathds{1}_{\{b_k \in I\}} \frac{Y_k}{n_I'} \right) - \sum_{j=1}^{n_0}\frac{X_{0j}}{n_0}\]
is normally distributed with mean zero and variance $(1/n_I' + 1/n_0)$. Hence we reject~$H_I$ if $T_I'$ is greater than $z_{\alpha}\left( 1/n_I' + 1/n_0 \right)^{1/2}$.



\subsection{Adaptive test statistic}
\label{subsec:cond_inv}

Adaptive designs, such as the trial being considered, follow a common conditional invariance principle in order to control the type~I error rate~\citep{Brannath2007}. 
For our response-adaptive trial in question, we apply the conditional invariance principle sequentially, where each step considers the next patient recruited into the trial. Below we give the test statistic for testing hypothesis~$H_I$ under the actual design, given that the allocation is fully sequential. The proof of Theorem~\ref{theorem_seq} can be found in Appendix~A.



\begin{theorem}
\label{theorem_seq}

Under $H_I$, the following test statistic is normally distributed with mean 0 and variance $(1/n_I' + 1/n_0)$: \vspace{-12pt} \[
\tilde{T}_I = \sum_{k = 1}^{n} \left( \mathds{1}_{\{a_k \in I\}} \frac{X_k}{w_k^{(I)}} \right) - \sum_{j=1}^{n_0} \frac{X_{0j}}{w_{n, j}^{(0)}} \vspace{-12pt}
\]  
where \vspace{-12pt} \begin{align*}
& w_k^{(I)} = n_I', \quad w_k^{(0)} = n_0 \qquad (k = 1, \ldots, r)\\
& w_{r+l}^{(I)} = f(\lambda_{r+l}, \eta_{r+l}, \tilde{m}_{I,r+l}), \quad  w_{r+l}^{(0)} = g(\lambda_{r+l}, w_{r+l}^{(I)}, \tilde{m}_{I,r+l}) \qquad (l = 1, \ldots, n-r) \\
& w_{n, j}^{(0)} = F_1(w_{n-1}^{(0)}, m_{0,1}, m_{0,2})  \qquad (j = 1, \ldots, m_{0,1}) \\
&
w_{n, j}^{(0)} = F_2(w_{n-1}^{(0)}, m_{0,1}, m_{0,2}) \qquad (j = m_{0,1}+1, \ldots, n_0) \\[0pt]
& \lambda_{r+l} = \frac{m_{I,r+l}}{w_{r+l-1}^{(I)}} - \frac{n_0}{w_{r+l-1}^{(0)}}, \quad \eta_{r+l} = \frac{m_{I,r+l}}{\left[w_{r+l-1}^{(I)}\right]^2} + \frac{n_0}{\left[w_{r+l-1}^{(0)}\right]^2} \qquad (l = 1, \ldots, n-r) \\[0pt]
& \tilde{m}_{I,r+l} = m_{I,r+l} + \mathds{1}_{\{ a_{r+l} \in I,  b_{r+l} \notin I\}}  - \mathds{1}_{\{ a_{r+l} \notin I,  b_{r+l} \in I\}} \qquad (l = 1, \ldots, n-r) \\[0pt]
& f(\lambda, \eta, m) = \frac{\lambda m - \{m n_0 (n_0 \eta - \lambda^2)\}^{1/2} }{\lambda - n_0 \eta}, \quad g(\lambda, w, m) = \frac{n_0 w}{m - \lambda w } \\
& F_1(w, m_1, m_2) = \mathds{1}_{\{ a_n \in I \}}w - \mathds{1}_{\{ a_n \notin I \}}
\frac{2m_{1} \lambda_{n} + [m_1 m_2\{\eta_{n}(m_1 + m_2) - \lambda_{n}^2\}]^{1/2}}{\lambda_{n}^2 - m_{2} \eta_{n}}  \\
& F_2(w, m_1, m_2) = \mathds{1}_{\{ a_n \in I \}}w -   \mathds{1}_{\{ a_n \notin I \}}
\frac{m_{2} F_1(w, m_1, m_2)}{m_{1} + \lambda_{n} F_1(w, m_1, m_2)} \\
& m_{0,1} + m_{0,2} = n_0, \quad m_{0,1} > 0, \quad m_{0,2}  > 0
\end{align*}

\end{theorem}

\noindent Hence we reject $H_I$ if $\tilde{T}_I$ is greater than $z_{\alpha} \left( 1/n_I' + 1/n_0 \right)^{1/2}$. In Appendix~B, we give some simple numerical examples of how the weights change over the course of a trial. In practice, to keep the weights as close to the natural weight~$n_0$ for as many of the control observations as possible, we recommend setting $m_{0,1} = n_0 - 1$ and $m_{0,2} = 1$, as used for the simulation studies in Section~\ref{subsec:simul_seq}.


In all of the scenarios that we have investigated, the weights~$w_k^{(I)}$ for the experimental treatments have been positive. Hence in these cases, the test procedure also controls the FWER for the composite null hypotheses~$H_i: \delta_i \leq 0$. To see this, suppose the elementary null hypotheses are $H_i^* : \delta_i = \delta_i^* < 0$. Under $H_I^*$, we can rewrite the distribution of the responses $X_k^*$ as $X_k + \delta_i^*$, where $X_k \sim N(0,1)$. Hence under $H_I^*$ \[
\text{pr}(\tilde{T}_I^* > c) = \text{pr}\left(\tilde{T}_I > c - \sum_{k = 1}^n \mathds{1}_{\{a_k \in I\}} \frac{\delta_k^*}{w_k^{(I)}} \right) < \text{pr}(\tilde{T}_I > c )
\] where $\tilde{T}_I^*$ and $\tilde{T}_I$ are the adaptive test statistics for $H_I^*$ and $H_I$ respectively.

\section{Block randomized response-adaptive trials}
\label{sec:block}


\subsection{Trial setting}
\label{subsec:block1}

It may not be feasible or desirable to randomize patients one-by-one in a fully sequential manner. Instead one can use block randomization, where after the burn-in period~$B$, patients are adaptively randomized to the experimental treatments in blocks of size $(d_1, \ldots, d_J)$ over~$J$ stages, with $\sum_{j=1}^J d_j = n$. The randomization of the $j$th block depends on the data up to block~$(j-1)$, as well as any external information available at the time. Defining~$d_0 = 0$, let $D_l = \sum_{j = 0}^l (r + d_j)$ for $l = 0, \ldots, J$, which represents the total number of allocations by the end of $l$th block, with the zeroth block corresponding to the burn-in period. For notational convenience, we let $D_{-1} = 0$. The allocation to the control is again assumed to be fixed throughout the trial.

Due to the block structure of the trial, we can  relax the assumption that the randomization rule used for the experimental treatments does not depend on the control information. This is achieved by splitting up the $n_0$ patients allocated to the control into blocks. More explicitly, suppose that during the burn-in period, $r_0 > 0$ patients are allocated to the control, where~$r_0$ is fixed in advance. Subsequently, in the $j$th block, $d_{0j}$ patients are allocated to the control, where $\sum_{j = 1}^J (r_0 + d_{0j}) = n_0$. We assume that for the final block $d_{0J} > 1$.

The response-adaptive randomization at block~$l$ may now depend on the control information available at the end of block~$(l-1)$; that is, the outcome data available from the first $\sum_{j = 1}^{l-1} (r_0 + d_{0j})$  patients allocated to the control. For notational convenience, define $d_{00} = 0$ and let $D_{0,l} = \sum_{j = 0}^l (r_0 + d_{0j})$ ($l = 0, \ldots, J$), which represents the total number of allocations to the control by the end of $l$th block. For notational convenience, let $D_{0,-1} = 0$.
%
%


To control the FWER, we can modify the approach described in Section~\ref{sec:seq_rar} to account for the block structure. As before, we have an auxiliary design for the patients on the experimental treatments, but now in step~$l$ of the process ($l \in \{1, \ldots, J\}$) the actual design is a data-dependent modification of all the allocations for the patients in block~$l$. Hence the weights for the observations in each block will be the same, and are updated block-by-block. 



\subsection{Auxiliary design and adaptive test statistic}
\label{subsec:aux_block}


The auxiliary design for an intersection hypothesis~$H_I$ is the same as described in Section~\ref{subsec:aux_seq}, except that we now impose a block structure on the auxiliary assignments to the experimental treatments. 
%
%
%
%
%
As before, the auxiliary and actual designs are identical during the burn-in period~$B$, and we require $b_n \in I$. For the auxiliary design, let $n_i'$ denote the total number of allocations to the $i$th treatment ($i = 1, \ldots, h$), including the burn-in period. Also let \[
m_{0,j} = \sum_{k = j}^J d_{0k}, \qquad m_{i,j} = \sum_{k = D_j + 1}^n \mathds{1}_{\{b_k = i\}} \] denote the total number of allocations to the control and $i$th treatment respectively for patients in blocks $(j, j+1, \ldots, J)$. We define $n_I' = \sum_{i \in I} n_i'$ and $m_{I,j} = \sum_{i \in I} m_{i,j}$.

We apply the conditional invariance principle block-by-block,  where each step considers an additional block of patients recruited into the trial. This gives the following test statistic for testing~$H_I$, with a proof and the formulae for the weights given in Appendix~C.


\begin{theorem}
\label{theorem_block}
If $m_{I,J} > 0$ then under $H_I$, the following test statistic is normally distributed with mean 0 and variance $(1/n_I' + 1/n_0)$ :\[
\tilde{T}_{I} = \sum_{j = 0}^{J} \sum_{k = D_{j-1} + 1}^{D_{j}} \mathds{1}_{\{a_k \in I\}} \frac{X_k}{w_j^{(I)}}  - 
\sum_{j = 0}^{J} \sum_{k = D_{0,j-1} + 1}^{D_{0,j}}\frac{X_{0k}}{w_j^{(0)}} \vspace{6pt}\]
\end{theorem}
 
\begin{corollary}

If $m_{I,J} = 0$, then let $n_{0,J,1} + n_{0,J,2} = m_{0,J}$, where $n_{0,J,1}, n_{0,J,2} > 0$.  Under $H_I$, the following test statistic is normally distributed with mean 0 and variance $(1/n_I' + 1/n_0)$:\[
\tilde{T}_{I} = \sum_{j = 0}^{J-1} \sum_{k = D_{j-1} + 1}^{D_{j}} \mathds{1}_{\{a_k \in I\}} \frac{X_k}{w_j^{(I)}}  - \sum_{j = 0}^{J-1} \sum_{k = D_{0,j-1} + 1}^{D_{0,j}} \frac{X_{0k}}{w_j^{(0)}} - \sum_{k = D_{0,J-1}+1}^{D_{0,J,1}} \frac{X_{0k}}{w_{J,1}^{(0)}} - \sum_{k = D_{0,J,1}+1}^{n_0} \frac{X_{0k}}{w_{J,2}^{(0)}} \vspace{6pt}\]

\end{corollary}

\noindent We reject $H_I$ if $\tilde{T}_I$ is greater than $z_{\alpha} \left( 1/n_I' + 1/n_0\right)^{1/2} $.  In order to keep the weights as close to the natural weight $n_0$ for as many of the control observations as possible, we recommend setting $n_{0,J,1} = m_{0,J} - 1$ and $n_{0,J,2} = 1$, as used for the simulation studies in Section~\ref{subsec:simul_block}. In all of the scenarios that we have investigated, the weights~$w_j^{(I)}$ for the experimental treatments have all been positive. Hence in these cases, the test procedure also controls the FWER for the composite null hypotheses~$H_i: \delta_i \leq 0$.


\subsection{Extension for adaptive control allocations}
\label{subsec:block_with_control}

Thus far, we have assumed that the allocations to the control follow some fixed scheme. We now relax this assumption in the block-randomized setting. 
Since the form of the adaptive test statistic $\tilde{T}_I$ is similar to the one presented above, the formula for~$\tilde{T}_I$ can be found in Appendix~D. 
Note that it is possible the procedure will fail to give a valid test statistic in this setting, as shown in Appendix~E.1. 



\section{Simulation studies}
\label{sec:simul}




As we have already seen in Section~\ref{subsec:inflate_fwer_seq}, using the closure principle with the usual $z$-test does not strongly control the FWER. An alternative method of control is to use the Bonferroni correction on the elementary null hypotheses $H_1, \ldots, H_h$.  
%
We also consider the Holm procedure, which is a step-down procedure that is uniformly more powerful than Bonferroni~\citep{Holm1979}.
%
%
An advantage of both these procedures is that only $h$ test statistics are calculated, rather than $(2^h - 1)$ test statistics when using the closure principle. This motivates also applying the Holm procedure to the $p$-values derived from the adaptive test statistics $\tilde{T}_i$ for $i = 1, \ldots, h$. More precisely, we use the adjusted $p$-values $\tilde{p}_i = 1 - \Phi((1/n_i' + 1/n_0')^{-1/2} \, \tilde{T}_i)$, instead of the usual $p$-values $p_i = 1 - \Phi((1/n_i + 1/n_0)^{-1/2} \, T_i)$ derived from the $z$-test.

To distinguish between the different methods, we call our proposed procedure that uses the closure principle the `adaptive closed test'. Similarly, applying the closure principle to the usual $z$-test gives the `closed $z$-test'. Applying the Holm procedure to our adjusted $p$-values gives the `Holm adaptive test', while applying the Holm procedure to the usual $p$-values gives the `Holm $z$-test'. 
%
In our simulation studies, we compare the different methods primarily by looking at the FWER. However, clearly another key consideration is the power of the different tests. To keep the comparisons simple, and as a similar measure to the FWER, we present results for the disjunctive power, which is the probability of rejecting at least one false null hypothesis.


\subsection{Fully sequential randomization}
\label{subsec:simul_seq}

We first consider a fully sequential response-adaptive trial, as presented in Section~\ref{sec:seq_rar}, with $m = 50$ patients allocated to the experimental treatments after the burn-in and $n_0 = 60/h$ patients allocated to the control. In the burn-in period, five patients are allocated to each of the experimental treatments. We set~$\alpha = 0.05$ and the true control mean $\mu = 0$ for simplicity. We compare the methods under two randomization schemes described below.

\textit{Type I error inflator}: For $h = 2$ treatments, this is the same randomization scheme as presented in Section~\ref{subsec:inflate_fwer_seq}.
For $h = 3$ treatments, if $\sum_{j = 1}^k (\mathds{1}_{\{a_j = 1\}} X_j / n_{1k}) > 0.5$, then we randomize patient~$(k+1)$ to treatments~2 and~3 with equal probability.

\textit{BAR}: The efficacy outcome for the $i$th experimental treatment follows a $N(\mu_i, 1)$ distribution. For simplicity, we assign independent normal priors to the~$\mu_i$, so that $\mu_i \sim N(\mu_{i,0} , \sigma_{i,0}^2)$, and let $n_{i,K} = \sum_{k=1}^K \mathds{1}_{\{a_k = i\}}$. After observing the efficacy outcomes $\bm{x} = (x_1, \ldots, x_K)$ for the first $K$ patients, the posterior for~$\mu_i$ is as follows: \[
\mu_i \mid \bm{X} = \bm{x} \sim N\left( \frac{\sigma_{i,0}^2}{1 + n_{i,K}\sigma_{i,0}^2} \sum_{k=1}^K \mathds{1}_{\{a_k = i\}} x_k + \frac{n_{i,K}}{1+n_{i,K} \sigma_{i,0}^2}\mu_{i,0} \, , \, \frac{\sigma_{i,0}^2}{1+n_{i,K} \sigma_{i,0}^2} \right) \]

We use a suggested BAR scheme of~\citet{Yin2012}. For $h = 2$ experimental treatments, the randomization probabilities $(\pi_1, 1 - \pi_1)$ after observing the~$K$th patient are: \[
\pi_1  = \frac{P(\mu_1 > \mu_2 \mid \bm{X} = \bm{x})^{\tau}}{P(\mu_1 > \mu_2 \mid \bm{X} = \bm{x})^{\tau} + \left\{1 - P(\mu_1 > \mu_2 \mid \bm{X} = \bm{x})\right\}^{\tau}}
\]


For $h>2$ experimental treatments, we first obtain the average of the posterior means $\bar{\mu} = \frac{1}{h} \sum_{i = 1}^h \mu_i$. The randomization probabilities~$\pi_i$ after  observing the~$K$th patient are:
\[ \pi_i = \frac{P(\mu_i > \bar{\mu} \mid \bm{X} = \bm{x})^{\tau}}{\sum_{j=1}^h P(\mu_j > \bar{\mu} \mid \bm{X} = \bm{x})^{\tau} }
\]

\noindent In our simulations, for simplicity we set the priors $\mu_{i,0} = 0$ and $\sigma_{i,0}^2 = 1$, while $\tau = 0.5$.

\textit{Simulation results}:
%
Table~\ref{tab:typeI_inflator_seq} gives the results for the type~I error inflator randomization scheme, while Table~\ref{tab:BAR_seq} gives the results for BAR. The auxiliary designs in all scenarios were simply $(m-1)$~random draws from a discrete uniform distribution on $\{1, \ldots, h\}$. 

\begin{sidewaystable}[htbp!]

\caption{\label{tab:typeI_inflator_seq}  \textit{Familywise error rate and disjunctive power for the type~I error inflator in the fully sequential setting. There were $10^5$ simulated trials for each set of parameter values.}}

\centering

		\begin{tabular}{p{0.5ex} l c c p{0cm} c c p{0cm} c c p{0cm} c c p{0cm} cc}
		\\ \hline \hline \Tstrut
		& & \multicolumn{2}{c}{Adaptive closed test} & & \multicolumn{2}{c}{Adaptive test (Holm)}
		& & \multicolumn{2}{c}{Closed $z$-test}  & &  \multicolumn{2}{c}{$z$-test (Holm)} 
		& & \multicolumn{2}{c}{$z$-test (Bonferroni)} \\
		\cline{3-4} \cline{6-7} \cline{9-10} \cline{12-13} \cline{15-16}\\ [-2ex]
		& {Parameter values} & Error & Power & & Error & Power
		& & Error & Power & & Error & Power & & Error & Power \\ \hline \Tstrut
		1.\ & $\delta_1 = \delta_2 = 0$
		  & 3.3 & - & & 4.7 & - & & 4.7 & - & & \textbf{7.0} & - & & \textbf{7.0} & -  \\ [1.5ex]
   		 
   		2.\ &  $\delta_1 = 0$, $\delta_2 = 1$ 
   		& 4.8 & 21.7 & & 3.7 & 27.5 & & \textbf{10.3} & 26.5 & & \textbf{9.9} & {63.6} & & 5.0 & 63.5 \\ [1.5ex]
		
		3.\ & $\delta_1 = \delta_2 = 0.5$
       & - & 62.4 & & - & 52.4 & & - &  {69.9} & & - & 61.6 & & - & 61.6 \\ [1.5ex]	
   		
		4.\ &  $\delta_1 = \delta_2 = \delta_3 = 0$ 
		& 2.8 & - & & 3.8 & - & & 4.1 & - & & \textbf{5.9} & - & & \textbf{5.9} \\ [1.5ex]
   		   		   		
   		5.\ &  $\delta_1 = \delta_2 = 0$, $\delta_3 = 1$
   		& 3.2 & 13.1 & & 4.2 & 24.2  & & \textbf{5.1} & 17.2 & & \textbf{6.4} &  {54.2} & & 4.5 & 54.1\\ [1.5ex]
   		
   		6.\ &  $\delta_1 = 0$, $\delta_2 = \delta_3 = 1$ 
   		& 4.6 & 22.2 & & 3.2 & 28.0  & & \textbf{9.7} & 27.0 & & \textbf{9.0} &  {75.4} & & 3.2 &  {75.4} \\ [1.5ex]

		7.\ &  $\delta_1 = 0$, $\delta_2 = 0.5$, $\delta_3 = 1$
		& 4.0 & 19.1 & & 2.6 & 24.5  & & \textbf{9.1} & 23.9 & & \textbf{7.4} & {58.5} & & 3.2 & 58.4 \\ [1.5ex]
		
   		8.\ &  $\delta_1 = \delta_2 = \delta_3 = 0.5$
   		& - & 51.3 & &  - & 41.7 & & - & {57.8} & & - & 49.7 & & - & 49.7 \\[1ex]  \hline

		\end{tabular}


\end{sidewaystable}


\begin{sidewaystable}[htbp!]

\caption{\label{tab:BAR_seq} \textit{Familywise error rate and disjunctive power for BAR in the fully sequential setting. There were $10^5$ simulated trials for each set of parameter values.}}	

\centering

\begin{tabular}{p{0.5ex} l c c p{0cm} c c p{0cm} c c p{0cm} c c p{0cm} cc}
		\\ \hline \hline \Tstrut
		& & \multicolumn{2}{c}{Adaptive closed test} & & \multicolumn{2}{c}{Adaptive test (Holm)}
		& & \multicolumn{2}{c}{Closed $z$-test}  & &  \multicolumn{2}{c}{$z$-test (Holm)} 
		& & \multicolumn{2}{c}{$z$-test (Bonferroni)} \\
		\cline{3-4} \cline{6-7} \cline{9-10} \cline{12-13} \cline{15-16}\\ [-2ex]
		& {Parameter values} & Error & Power & & Error & Power
		& & Error & Power & & Error & Power & & Error & Power \\ \hline \Tstrut
		1.\ & $\delta_1 = \delta_2 = 0$
		  & 4.7 & - & & 4.5 & - & & 4.8 & - & & 4.1 & - & & 4.1 & -  \\ [1.5ex]
   		 
   		2.\ &  $\delta_1 = 0$, $\delta_2 = 0.5$ 
   		& 4.6 & 46.4 & & 4.4 & 52.4 & & 3.9 & 46.7 & & 3.6 &  {53.6} & & 1.9 & 53.5 \\ [1.5ex]
		
		3.\ & $\delta_1 = \delta_2 = 0.5$
       & - & 70.8 & & - & 66.4 & & - &  {71.2} & & - & 65.9 & & - & 65.9 \\ [1.5ex]	
   		
		4.\ &  $\delta_1 = \delta_2 = \delta_3 = 0$
   		& 3.8 & - & & 4.1 & - & & 4.0 & - & & 3.8 & - & & 3.8 \\ [1.5ex]
   		   		   		
   		5.\ &  $\delta_1 = \delta_2 = 0$, $\delta_3 = 1$
   		& 4.4 & 59.9 & & 4.2 & 88.7 & & 4.3 & 60.1 & & 3.8 &  {90.6} & & 2.6 &  {90.6}\\ [1.5ex]
   		
   		6.\ &  $\delta_1 = 0$, $\delta_2 = \delta_3 = 1$
		& 4.8 & 89.8 & & 4.7 & 95.1 & & 4.0 & 90.1 & & 3.9 &  {96.0} & & 1.3 &  {96.0} \\ [1.5ex]
		 
		7.\ &  $\delta_1 = 0$, $\delta_2 = 0.5$, $\delta_3 = 1$ 
		& 4.3 & 74.8 & & 3.9 & 88.2 & & 3.9 & 75.7 & & 3.4 &  {90.0} &  & 1.4 &  {90.0}\\ [1.5ex]

   		8.\ &  $\delta_1 = \delta_2 = \delta_3 = 0.5$ 
   		& - & 56.5 & &  - & 51.8 & & - &  {57.9} & & - & 52.7 & & - & 52.7 \\ [1ex] \hline
		\end{tabular}
		
		
\end{sidewaystable}


Looking first at the results for the type~I error inflator in Table~\ref{tab:typeI_inflator_seq}, the closed $z$-test does not control the FWER in any of the scenarios where at least one null hypothesis is false, with an error rate as high as $10.3\%$ in scenario~2. Applying the Holm procedure to the $z$-test does not control the FWER, and actually increases the error rate in some scenarios (such as~1 and ~4). 
Applying the Bonferroni correction to the $z$-test also does not control the FWER, as can be seen in the scenarios where all null hypotheses are true. This may appear surprising at first, but the inflation occurs because the na\"ive $z$-test is not a valid level--$\alpha$ test for each elementary hypothesis.  In contrast, both the adaptive closed test and the Holm adaptive test strongly control the FWER. 

As for the power of the different methods, when at least one of the null hypotheses is true (as in scenarios~2, 5, 6 and~7), the Holm $z$-test has substantially higher power than the closed $z$-test. Indeed, the power more than doubles in all four scenarios, and even more than triples in scenario~5. This dramatic increase in power demonstrates that in these scenarios, the closed $z$-test is not very sensitive. This is because the test statistic for $H_I$ will be `diluted' by the contribution from responses belonging to the null hypotheses $(H_i)_{i \in I}$ that are true. It is only when all of the null hypotheses are false, as in scenarios~3 and~8, that the power of the closed $z$-test is reasonable, with a slightly higher power than the Holm $z$-test. 

As for the adaptive tests, the adaptive closed test has a slightly lower power than the closed $z$-test for all scenarios, with an absolute decrease of between $4.1\%$ in scenario 5 and $7.5\%$ in scenario 3. However, the Holm adaptive test has a substantially lower power than the Holm $z$-test, with the latter having more than double the power. This demonstrates the high cost in terms of power that controlling the FWER can incur for this randomization scheme. We return to this issue in Section~\ref{subsec:simul_summary}.

Turning to the BAR scheme in Table~\ref{tab:BAR_seq}, this time all of the methods strongly control the FWER. All methods are slightly conservative, with the adaptive closed test being generally the closest to the nominal $5\%$ level. The Bonferroni-corrected $z$-test is noticeably more conservative than all the other methods, particularly when there are three treatments.
%
%
In terms of disjunctive power, if at least one of the null hypotheses are true, we again see that the closed tests suffer from reduced power compared to the Holm versions. However, with BAR the loss of power is less dramatic, with a maximum of a $33\%$ relative decrease in power in scenario~5, but with much smaller decreases in scenarios~2 and~7 for example. This time, the adaptive closed test has almost the same power as the closed $z$-test, losing a maximum of only $1.4\%$ in scenario~8. In addition, the Holm adaptive test and Holm $z$-test now have comparable power, with a maximum loss of only $1.9\%$ in scenarios~6 and~7. This indicates that for BAR schemes, the adaptive tests do not lose out very much in terms of power. 




\subsection{Block randomization with a fixed control allocation}
\label{subsec:simul_block}

We now consider block randomized trials with a fixed control allocation, as presented in Section~\ref{subsec:block1}. We use the setup of a trial with $J = 3$ blocks, with sizes (40, 40, 40) for the experimental treatments and (20, 20, 20) for the control. In the burn-in period, five patients are allocated to each of the treatments including the control. We set the true control mean $\mu = 0$, and~$\alpha = 0.05$. We compare the methods under the randomization schemes below.


\textit{Type I error inflator}: The allocation probabilities for block $j  \in \{ 1, \ldots, J-1 \}$, patient $k =  D_j + 1, \ldots, D_{j+1}$ and treatment $l \in \{2, \ldots, h\}$ are: \begin{align*}
P(a_k = 1) & = \begin{cases}
0 & \text{if } \; \sum_{i = 1}^{D_j} \mathds{1}_{\{a_i = 1\}} \frac{X_i}{n_{1,j}} > 0.5 \\
1 & \text{otherwise}
\end{cases} \\
P(a_k = l) & = \begin{cases}
1/(h-1) & \text{if } \; \sum_{i = 1}^{D_j} \mathds{1}_{\{a_i = 1\}} \frac{X_i}{n_{1,j}} > 0.5 \\
0 & \text{otherwise}
\end{cases}
\end{align*}

\noindent where  $n_{1,j} = \sum_{i=1}^{D_j} \mathds{1}_{\{a_i = 1\}}$.

\textit{BAR}: 
%
The efficacy outcome for the $i$th treatment follows a $N(\mu_i, 1)$ distribution.  For notational convenience, let $\mu_0 = \mu$; that is, the mean of the control. We assign independent normal priors to the~$\mu_i$ ($i = 0, 1, \ldots, h$), such that $\mu_i \sim N(\mu_{i,0} , \sigma_{i,0}^2)$. At stage~$(j+1)$, when the efficacy outcomes $\bm{x} = (x_1, \ldots,  x_{\sss D_j})$ have been observed,  the posterior for~$\mu_i$ is as follows: \[
\mu_i \mid \bm{X} = \bm{x} \sim N\left( \frac{\sigma_{i,0}^2}{1 + n_{i,K}\sigma_{i,0}^2} \sum_{k=1}^{D_j}\mathds{1}_{\{a_k = i\}} x_k + \frac{n_{i,K}}{1+n_{i,K} \sigma_{i,0}^2}\mu_{i,0} \;  , \frac{\sigma_{i,0}^2}{1+n_{i,K} \sigma_{i,0}^2} \right)
\] where $n_{i,K} = \sum_{k=1}^{D_K} \mathds{1}_{\{a_k = i\}}$.

We use a similar BAR scheme to the one in~\citet{Wason2014a}. If there are $h$ experimental treatments, the randomization probabilities $(\pi_1, \ldots, \pi_h)$ for the experimental treatments at the~$(j+1)$th stage are: \[
\pi_i = \frac{P(\mu_i > \mu_0 \mid \bm{X} = \bm{x})^{\gamma}}{\sum_{l=1}^h P(\mu_l > \mu_0 \mid  \bm{X} = \bm{x})^{\gamma}}
\]

\noindent In our simulations, for simplicity we set the priors $\mu_{i,0} = 0$ and $\sigma_{i,0}^2 = 1$, while $\gamma = 0.5$.

\textit{Simulation results}:
%
Table~\ref{tab:typeI_inflator_seq} gives the results for the type~I error inflator randomization scheme, while Table~\ref{tab:BAR_seq} gives the results for BAR. The auxiliary designs in all scenarios were simply random draws from a discrete uniform distribution on $\{1, \ldots, h\}$. 

\begin{sidewaystable}[htbp!]

\caption{\label{tab:typeI_inflator_block}  \textit{Familywise error rate and disjunctive power for the type~I error inflator, for block randomization with a fixed control allocation. There were $10^5$ simulated trials for each set of parameter values.}}

\centering

\begin{tabular}{p{0.5ex} l c c p{0cm} c c p{0cm} c c p{0cm} c c p{0cm} cc}
		\\ \hline \hline \Tstrut
		& & \multicolumn{2}{c}{Adaptive closed test} & & \multicolumn{2}{c}{Adaptive test (Holm)}
		& & \multicolumn{2}{c}{Closed $z$-test}  & &  \multicolumn{2}{c}{$z$-test (Holm)} 
		& & \multicolumn{2}{c}{$z$-test (Bonferroni)} \\
		\cline{3-4} \cline{6-7} \cline{9-10} \cline{12-13} \cline{15-16}\\ [-2ex]
		& {Parameter values} & Error & Power & & Error & Power
		& & Error & Power & & Error & Power & & Error & Power \\ \hline \Tstrut
		1.\ & $\delta_1 = \delta_2 = 0$
		  & 3.8 & - & & 4.8 & - & & 4.6 & - & & \textbf{6.5} & - & & \textbf{6.5} & -  \\ [1.5ex]
   		 
   		2.\ &  $\delta_1 = 0$, $\delta_2 = 1$ 
   		& 4.8 & 22.0 & & 3.6 & 26.9 & & \textbf{8.3} & 25.6 & & \textbf{7.8} &  {61.1} & & 4.3 & 61.0 \\ [1.5ex]
		
		3.\ & $\delta_1 = \delta_2 = 0.5$
       & - & 92.7 & & - & 87.9 & & - & {94.6} & & - & 91.7 & & - & 91.7 \\ [1.5ex]	
   		
		4.\ &  $\delta_1 = \delta_2 = \delta_3 = 0$ 
		& 3.2 & - & & 4.1 & - & & 4.1 & - & & \textbf{6.1} & - & & \textbf{6.1} \\ [1.5ex]
   		   		   		
   		5.\ &  $\delta_1 = \delta_2 = 0$, $\delta_3 = 1$
   		& 3.7 & 14.2 & & 4.4 & 23.4  & & 4.7 & 18.1 & & \textbf{6.2} &  {61.2} & & 4.5 & 61.1\\ [1.5ex]
   		
   		6.\ &  $\delta_1 = 0$, $\delta_2 = \delta_3 = 1$ 
   		& 4.9 & 20.1 & & 3.2 & 26.1  & & \textbf{8.1} & 23.0 & & \textbf{7.3} &  {78.5} & & 3.2 & 78.4\\ [1.5ex]

		7.\ &  $\delta_1 = 0$, $\delta_2 = 0.5$, $\delta_3 = 1$ 
		& 4.7 & 17.7 & & 3.0 & 23.8  & & \textbf{8.0} & 21.1 & & \textbf{6.7} &  {66.2} & & 2.8 &  {66.2}\\ [1.5ex]
		
   		8.\ &  $\delta_1 = \delta_2 = \delta_3 = 0.5$
   		& - & 91.3 & &  - & 83.4 & & - &  {94.0} & & - & 89.7 & & - & 89.7 \\[1.5ex] \hline

		\end{tabular}


\end{sidewaystable}

\begin{sidewaystable}[htbp!]

	\caption{\label{tab:BAR_block}  \textit{Familywise error rate and disjunctive power for BAR, for block randomization with a fixed control allocation. There were $10^5$ simulated trials for each set of parameter values.}}	

\centering

\begin{tabular}{p{0.5ex} l c c p{0cm} c c p{0cm} c c p{0cm} c c p{0cm} cc}
		\\ \hline \hline \Tstrut
		& & \multicolumn{2}{c}{Adaptive closed test} & & \multicolumn{2}{c}{Adaptive test (Holm)}
		& & \multicolumn{2}{c}{Closed $z$-test}  & &  \multicolumn{2}{c}{$z$-test (Holm)} 
		& & \multicolumn{2}{c}{$z$-test (Bonferroni)} \\
		\cline{3-4} \cline{6-7} \cline{9-10} \cline{12-13} \cline{15-16}\\ [-2ex]
		& {Parameter values} & Error & Power & & Error & Power
		& & Error & Power & & Error & Power & & Error & Power \\ \hline \Tstrut
		1.\ & $\delta_1 = \delta_2 = 0$
		  & 4.8 & - & & 4.6 & - & & 4.8 & - & & 4.5 & - & & 4.5 & -  \\ [1.5ex]
   		 
   		2.\ &  $\delta_1 = 0$, $\delta_2 = 0.5$ 
   		& 5.0 & 61.2 & & 4.9 & 82.7 & & 4.9 & 61.2 & & 4.8 & {82.9} & & 2.5 & 82.8 \\ [1.5ex]
		
		3.\ & $\delta_1 = \delta_2 = 0.5$
       & - &  {94.5} & & - & 92.3 & & - &  {94.5} & & - & 92.2 & & - & 92.2 \\ [1.5ex]	
   		
		4.\ &  $\delta_1 = \delta_2 = \delta_3 = 0$
   		& 3.7 & - & & 4.5 & - & & 3.7 & - & & 4.2 & - & & 4.2 \\ [1.5ex]
   		   		   		
   		5.\ &  $\delta_1 = \delta_2 = 0$, $\delta_3 = 0.5$
   		& 4.4 & 36.1 & & 4.6 &  {71.8} & & 4.3 & 36.0 & & 4.4 &  {71.8} & & 3.0 & 71.7\\ [1.5ex]
   		
   		6.\ &  $\delta_1 = 0$, $\delta_2 = \delta_3 = 0.5$
		& 5.0 & 67.3 & & 4.6 &  {85.6} & & 4.8 & 66.8 & & 4.4 & 85.4 & & 1.6 & 85.4 \\ [1.5ex]
		 
		7.\ &  $\delta_1 = 0$, $\delta_2 = 0.25$, $\delta_3 = 0.5$ 
		& 4.6 & 51.1 & & 3.7 &  {73.0} & & 4.4 & 50.9 & & 3.5 & 72.6 &  & 1.6 & 72.6\\ [1.5ex]

   		8.\ &  $\delta_1 = \delta_2 = \delta_3 = 0.5$ 
   		& - &  {93.5} & &  - & 90.7 & & - & 93.4 & & - & 90.4 & & - & 90.4 \\ [1ex] \hline

		\end{tabular}


\end{sidewaystable}

The results are broadly similar to those for the fully sequential setting presented in Section~\ref{subsec:simul_seq}. For the type~I error inflator, we see that the closed $z$-test does not control the FWER in general (as seen in scenarios~2, 6 and~7), and neither does applying the Holm procedure to the $z$-test. The Bonferroni-corrected $z$-test has an inflated FWER when all null hypotheses are true, as in scenarios~1 and 4. In contrast, the adaptive tests strongly control the FWER in all scenarios. However, again this comes at the cost of reduced power. There is a slight reduction in power between the closed $z$-test and the closed adaptive test, of between $3-4\%$ in absolute terms. In scenarios where at least one null hypothesis is true, the Holm $z$-test has a much higher power than the Holm adaptive test, with the power more than doubling in these scenarios, and actually tripling in scenario~6.

As for the BAR scheme, all of the methods strongly control the FWER. This time, for some scenarios the adaptive closed test basically achieves the nominal $5\%$ level, as in scenarios~2 and~6. When there are three treatments, the Bonferroni-corrected $z$-test can again be overly conservative, as in scenarios~6 and~7. In contrast to the fully sequential setting, with block randomization we see that the adaptive tests actually have the highest power out of all the methods in all scenarios except scenario~2. When at least one null hypothesis is true, the Holm adaptive test has the highest power, while when all null hypotheses are false the adaptive closed test has the highest power. The power gains are small, but demonstrate that we do not always lose out in terms of power when using the proposed adaptive tests.

\textit{Block randomization with an adaptive control allocation}:
In Appendix~E.1, we present a simulation study considering  block randomization with an adaptive control allocation, as presented in Section~\ref{subsec:block_with_control}. The results are broadly similar to those presented above. 

\subsection{Summary}
\label{subsec:simul_summary}

In summary, the simulation results show that in the randomization settings considered, our proposed adaptive tests strongly control the FWER, as would be expected from theory. In contrast, the various $z$-tests can all fail to control the error rate, as seen in the results for the type~I error inflator. However, given a more realistic randomization scheme, such as the BAR schemes we considered, the $z$-tests achieve strong familywise error control. 
As for disjunctive power, we see that when at least one null hypothesis is true, the closed tests suffer a very large drop in power compared to the Holm versions. This is because of the `dilution' of the test statistic as mentioned in Section~\ref{subsec:simul_seq}. However, when all the null hypotheses are true, then the closed test has the higher power, although the gains are at most modest.

The adaptive tests can pay a large price in terms of power when compared with the $z$-tests, as seen in the results for the type~I error inflator. In Appendix~E.2, we give an additional simulation study with two treatments, where the randomization scheme used is simply a fixed allocation to the experimental treatments but with unequal randomization probabilities. We show that when the probability of assignment to treatment~2 is low (i.e. less than 0.2), there is a large drop in the power of the adaptive tests for testing~$H_1$.  This explains what is happening with the type~I error inflator when $\delta_1 = 0$, where in the majority of trial scenarios, apart from the unlikely event that treatment~1 stops early for `efficacy', the probability of assignment to treatment~2 is zero by design. Hence, the type~I inflator is in fact close to a worst-case scenario for the adaptive tests.

However, most adaptive randomization schemes are unlikely to have such extreme imbalances. Indeed, authors such as~\citet{Korn2011} recommend restricting the probability of arm assignment to between~0.2 and~0.8 in order to prevent extreme patient allocation. Hence, for `sensible' adaptive randomization schemes with such a restriction, we would not expect there to be a substantial loss of power when using the Holm adaptive test compared with the Holm $z$-test, particularly in the block randomized setting.



\section{Case study}
\label{sec:case_study}

Finally, we illustrate our proposed methodology using an example based on a phase~II placebo-controlled trial in primary hypercholesterolemia~\citep{Roth2012}. The purpose of the study was to compare the effects of using the SAR236553 antibody with high-dose or lose-dose atorvastatin, as compared with high-dose atorvastatin alone. The primary outcome was the least-squares mean percent reduction from baseline of low-density lipoprotein cholesterol (LDL-C). Patients were randomly assigned, in a 1:1:1 ratio, to receive 80 mg of atorvastatin plus placebo, 10 mg of atorvastatin plus SAR236553, or 80 mg of atorvastatin plus SAR236553. For convenience, we label these different interventions as the `control', `low dose' and `high dose' respectively.

In the trial, the observed least-squares mean $\pm$ SE percent reduction from baseline in LDL-C was $17.3 \pm 3.5$ for the control, $66.2 \pm 3.5$ for the low dose and $72.3 \pm 3.5$ for the high dose. There were $N = 31$ patients on the control, $N = 31$ patients on the low dose and $N = 30$ patients on the high dose, giving a total of $N = 61$ patients on the two experimental doses. For our illustrative case study, we use the observed values from the trial and assume that the distribution of the  least-squares standardized mean percent reduction from baseline of low-density LDL-C is $N(17.3/3.5, 1)$ for the control,  $N(66.2/3.5, 1)$ for the low dose, and $N(72.3/3.5, 1)$ for the high dose. 

Now suppose that the trial was carried out as an adaptive block randomized trial with a fixed control allocation, as described in Section~\ref{subsec:block1}. Let the trial have $J = 3$ blocks, with block sizes (15, 15, 15) for the experimental treatments and (8, 8, 8) for the placebo. In the burn-in period, 7~patients are allocated to the control and 8~patients are allocated to each of the experimental doses. Hence, a total of 31~patients are on the control and 61~on the experimental treatments, as in the original trial. We use the BAR scheme of Section~\ref{subsec:simul_block}, with priors $\mu_{i,0} = 5$ and $\sigma_{i,0}^2 = 1$ ($i = 0, 1, 2$), while $\gamma = 0.5$.

Table~\ref{tab:case_study} shows the results for a simulated trial with the above parameters, where the BAR scheme allocated 13~patients to the low dose and 32~patients to the high dose after the burn-in period. This yields the natural weights used in the na\"ive $z$-test of  $n_1' = 21$ for the low dose and $n_2' = 40$ for the high dose.  The natural weight for the control is $n_0 = 31$ by design. The auxiliary design randomly assigned 44~patients to the low or high dose in a 1:1 ratio, and allocated 21~patients to the low dose and 23~patients to the high dose. 

\begin{table}[H]
\centering
\caption{\label{tab:case_study} \textit{Test statistics, $p$-values and weights for a simulated block randomized trial using a BAR scheme.}}

\begin{tabular}{l c p{0ex} c }
\\ \hline \hline \Tstrut
& {Low dose} & & {High dose} \\ \hline
\Tstrut $z$-test statistic & 13.76 ($p < 0.001$) & & 15.50 ($p < 0.001$) \\[1.5ex]
Adaptive test statistic & 12.21 ($p < 0.001$) & & 16.22 ($p < 0.001$) \\[1.5ex]
Natural weights & $n_1' = 21$, $n_0 = 31$  & &  $n_2' = 40$, $n_0 = 31$  \\[1.5ex]
\multirow{2}{*}{Adaptive weights} & $w^{(1)} = (30, 28.05, 21.49, 16.43)$ & & $w^{(2)} = (32, 34.09, 42.68, 46.08)$ \\ 
&  $w^{(0)} = (31, 31.73, 35.91, 42.76)$ & &  $w^{(0)} = (31, 30.43, 28.86, 28.41)$ \\ [1ex] \hline

\end{tabular}

\end{table}

The adaptive test statistic is slightly smaller than the $z$-test statistic for the low dose, while the converse is true for the test statistics for the high dose. Looking at the adaptive weights for the burn-in period and the three blocks, we see that for the low dose, the weights for the low dose decrease for each block while the control weights increase. This pattern is reversed for the high dose. Given that all the $p$-values are less than~$0.001$, using either the $z$-test or the adaptive test we would conclude that adding the SAR236553 antibody to high-dose or low-dose atorvastatin leads to a statistically significant reduction in LDL-C levels.

\section{Discussion}
\label{sec:discuss}



A major regulatory concern over the use of response-adaptive trials in clinical practice has been ensuring control of the type~I error rate.
We have proposed procedures that guarantee strong familywise error control in the following multi-armed trial settings: \begin{enumerate}
\item Fully sequential response-adaptive trials with a fixed control allocation (where the randomization rule does not depend on the control information)
\item Block-randomized response-adaptive trials with a fixed control allocation
\item Block-randomized response-adaptive trials including an adaptive control allocation
\end{enumerate}
%
%
%

These procedures are applicable to a large class of response-adaptive randomization rules, particularly in settings~(2) and~(3) where there are no restrictions on the rule used. Hence both Bayesian and `optimal' response-adaptive randomization schemes proposed in the literature can be used without adjustment, with only the final test statistic having to be modified. 

In practice, to control the FWER we would recommend using the Holm adaptive test. Importantly, it has a much higher power than the adaptive closed test when at least one of the null hypotheses are true. As well, it only requires $h$~hypothesis tests as compared with $(2^h - 1)$~hypothesis tests for the adaptive closed test. 
%

Our adaptive tests lead to unequal weightings of patients, which may be controversial~\citep{Burman2006}. One solution is to use the so-called `dual test', and reject a hypothesis only if both the adaptive test and the na\"ive $z$-test rejects~\citep{Denne2001, Posch2003, Chen2004}, although this comes at the cost of reduced power.


We have assumed that the variances of the control and experimental treatments are known. Fully accounting for unknown variances would add considerable complexity to our approach. In Appendix~E.3, we show that estimating the common variance from the data does not inflate the FWER when using the Holm adaptive test, for any of the simulation scenarios considered in this paper. 


Our proposed procedures are designed for normally-distributed outcomes, and it would be useful to apply our approach to binary outcomes as well. As a starting point, it may be possible to use the asymptotically normal test statistic for contrasting each treatment arm with the control~\citep{Jennison2000, Wason2014a}, particularly in the block randomised setting. 

Finally, although we did not explicitly consider it in this paper, the adaptive randomization procedures used could also incorporate covariate information, so that the allocation probabilities vary across patients with different covariates. These covariate-adjusted response-adaptive randomization schemes are particularly useful when certain characteristics of the patients may be correlated with the primary outcome~\citep{Hu2006}. A related setting would be biomaker-guided response-adaptive trials, such as I-SPY 2.





{
\raggedright
\bibliography{FWER_control}

\begin{thebibliography}{}

\bibitem[\protect\citeauthoryear{Bartlett, Roloff, Cornell, Andrews, Dillon,
  and Zwischenberger}{Bartlett et~al.}{1985}]{Bartlett1985}
Bartlett, R.~H., Roloff, D.~W., Cornell, R.~G., Andrews, A.~F., Dillon, P.~W.,
  and Zwischenberger, J.~B. (1985).
\newblock Extracorporeal circulation in neonatal respiratory failure: a
  prospective randomized study.
\newblock {\em Pediatrics} {\bf 76,} 479--487.

\bibitem[\protect\citeauthoryear{Bello and Sabo}{Bello and
  Sabo}{2016}]{Bello2016}
Bello, G.~A. and Sabo, R.~T. (2016).
\newblock Outcome-adaptive allocation with natural lead-in for three-group
  trials with binary outcomes.
\newblock {\em Journal of Statistical Computation and Simulation} {\bf 86,}
  2441--2449.

\bibitem[\protect\citeauthoryear{Bergemann and Välimäki}{Bergemann and
  Välimäki}{2006}]{Bergemann2006}
Bergemann, D. and Välimäki, J. (2006).
\newblock Bandit problems.
\newblock Technical report, Cowles Foundation, http://ssrn.com/abstract=877173
  [accessed 1 Mar 2018].

\bibitem[\protect\citeauthoryear{Berry}{Berry}{2011}]{Berry2011}
Berry, D.~A. (2011).
\newblock Adaptive clinical trials: the promise and the caution.
\newblock {\em {Journal of Clinical Oncology}} {\bf 29,} 606--609.

\bibitem[\protect\citeauthoryear{Berry}{Berry}{2015}]{Berry2015}
Berry, D.~A. (2015).
\newblock {Commentary on Hey and Kimmelman}.
\newblock {\em {Clinical Trials}} {\bf 12,} 107--109.

\bibitem[\protect\citeauthoryear{Biswas, Bhattachary, and Zhang}{Biswas
  et~al.}{2007}]{Biswas2007}
Biswas, A., Bhattachary, R., and Zhang, L. (2007).
\newblock Optimal response-adaptive designs for continuous responses in phase
  {III} trials.
\newblock {\em Biometrical journal} {\bf 49,} 928--940.

\bibitem[\protect\citeauthoryear{Biswas and Bhattacharya}{Biswas and
  Bhattacharya}{2016}]{Biswas2016}
Biswas, A. and Bhattacharya, R. (2016).
\newblock Response-adaptive designs for continuous treatment responses in phase
  {III clinical trials: A review.}
\newblock {\em {Statistical Methods in Medical Research}} {\bf 25,} 81--100.

\bibitem[\protect\citeauthoryear{Brannath, Koenig, and Bauer}{Brannath
  et~al.}{2007}]{Brannath2007}
Brannath, W., Koenig, F., and Bauer, P. (2007).
\newblock Multiplicity and flexibility in clinical trials.
\newblock {\em Pharmaceutical statistics} {\bf 6,} 205--216.

\bibitem[\protect\citeauthoryear{Burman and Sonesson}{Burman and
  Sonesson}{2006}]{Burman2006}
Burman, C.-F. and Sonesson, C. (2006).
\newblock Are flexible designs sound?
\newblock {\em Biometrics} {\bf 62,} 664--9; discussion 670--83.

\bibitem[\protect\citeauthoryear{Chen, DeMets, and Lan}{Chen
  et~al.}{2004}]{Chen2004}
Chen, Y. H.~J., DeMets, D.~L., and Lan, K. K.~G. (2004).
\newblock Increasing the sample size when the unblinded interim result is
  promising.
\newblock {\em {Statistics in Medicine}} {\bf 23,} 1023--1038.

\bibitem[\protect\citeauthoryear{Denne}{Denne}{2001}]{Denne2001}
Denne, J.~S. (2001).
\newblock Sample size recalculation using conditional power.
\newblock {\em {Statistics in Medicine}} {\bf 20,} 2645--2660.

\bibitem[\protect\citeauthoryear{Du, Wang, and Lee}{Du et~al.}{2015}]{Du2015}
Du, Y., Wang, X., and Lee, J.~J. (2015).
\newblock Simulation study for evaluating the performance of response-adaptive
  randomization.
\newblock {\em {Contemporary Clinical Trials}} {\bf 40,} 15--25.

\bibitem[\protect\citeauthoryear{Eisele}{Eisele}{1994}]{Eisele1994}
Eisele, J.~R. (1994).
\newblock The doubly adaptive biased coin design for sequential clinical
  trials.
\newblock {\em {Journal of Statistical Planning and Inference}} {\bf 38,}
  249--261.

\bibitem[\protect\citeauthoryear{{European Medicines Agency}}{{European
  Medicines Agency}}{2002}]{European2002}
{European Medicines Agency} (2002).
\newblock {Points to Consider on Multiplicity Issues in Clinical Trials}.
\newblock {\em {London: CPMP}} .

\bibitem[\protect\citeauthoryear{{Food and Drug Administration}}{{Food and Drug
  Administration}}{2010}]{FDA2010}
{Food and Drug Administration} (2010).
\newblock {Guidance for Industry: Adaptive Design Clinical Trials for Drugs and
  Biologics; 2010}.
\newblock {\em Available from:
  https://www.fda.gov/downloads/drugs/guidances/ucm201790.pdf [accessed 1 Mar
  2018]} .

\bibitem[\protect\citeauthoryear{Giles, Kantarjian, Cortes, Garcia-Manero,
  Verstovsek, Faderl, et~al\mbox{.}}{Giles et~al.}{2003}]{Giles2003}
Giles, F.~J., Kantarjian, H.~M., Cortes, J.~E., Garcia-Manero, G., Verstovsek,
  S., Faderl, S., et~al. (2003).
\newblock Adaptive randomized study of idarubicin and cytarabine versus
  troxacitabine and cytarabine versus troxacitabine and idarubicin in untreated
  patients 50 years or older with adverse karyotype acute myeloid leukemia.
\newblock {\em {Journal of Clinical Oncology}} {\bf 21,} 1722--1727.

\bibitem[\protect\citeauthoryear{Gutjahr, Posch, and Brannath}{Gutjahr
  et~al.}{2011}]{Gutjahr2011}
Gutjahr, G., Posch, M., and Brannath, W. (2011).
\newblock Familywise error control in multi-armed response-adaptive two-stage
  designs.
\newblock {\em {Journal of Biopharmaceutical Statistics}} {\bf 21,} 818--830.

\bibitem[\protect\citeauthoryear{Hey and Kimmelman}{Hey and
  Kimmelman}{2015}]{Hey2015}
Hey, S.~P. and Kimmelman, J. (2015).
\newblock Are outcome-adaptive allocation trials ethical?
\newblock {\em {Clinical Trials}} {\bf 12,} 102--106.

\bibitem[\protect\citeauthoryear{Holm}{Holm}{1979}]{Holm1979}
Holm, S. (1979).
\newblock A simple sequentially rejective multiple test procedure.
\newblock {\em {Scandinavian Journal of Statistics}} pages 65--70.

\bibitem[\protect\citeauthoryear{Hu and Rosenberger}{Hu and
  Rosenberger}{2006}]{Hu2006}
Hu, F. and Rosenberger, W.~F. (2006).
\newblock {\em The theory of response-adaptive randomization in clinical
  trials}, volume 525.
\newblock John Wiley \& Sons.

\bibitem[\protect\citeauthoryear{Ivanova, Biswas, and Lurie}{Ivanova
  et~al.}{2006}]{Ivanova2006}
Ivanova, A., Biswas, A., and Lurie, A. (2006).
\newblock Response-adaptive designs for continuous outcomes.
\newblock {\em {Journal of Statistical Planning and Inference}} {\bf 136,}
  1845--1852.

\bibitem[\protect\citeauthoryear{Jennison and Turnbull}{Jennison and
  Turnbull}{2000}]{Jennison2000}
Jennison, C. and Turnbull, B. (2000).
\newblock Group sequential methods with applications to clinical trials.
\newblock {\em {Chapman-Hall/CRC, Boca Raton, FL}} .

\bibitem[\protect\citeauthoryear{Korn and Freidlin}{Korn and
  Freidlin}{2011}]{Korn2011}
Korn, E.~L. and Freidlin, B. (2011).
\newblock Outcome--adaptive randomization: is it useful?
\newblock {\em {Journal of Clinical Oncology}} {\bf 29,} 771--776.

\bibitem[\protect\citeauthoryear{Lee, Chen, and Yin}{Lee
  et~al.}{2012}]{Lee2012}
Lee, J.~J., Chen, N., and Yin, G. (2012).
\newblock {Worth adapting? Revisiting the usefulness of outcome-adaptive
  randomization}.
\newblock {\em {Clinical Cancer Research}} {\bf 18,} 4498--4507.

\bibitem[\protect\citeauthoryear{Marcus, Peritz, and Gabriel}{Marcus
  et~al.}{1976}]{Marcus1976}
Marcus, R., Peritz, E., and Gabriel, K.~R. (1976).
\newblock On closed testing procedures with special reference to ordered
  analysis of variance.
\newblock {\em Biometrika} pages 655--660.

\bibitem[\protect\citeauthoryear{Park, Liu, Yee, Yau, van~'t Veer, Symmans,
  et~al\mbox{.}}{Park et~al.}{2016}]{Park2016}
Park, J.~W., Liu, M.~C., Yee, D., Yau, C., van~'t Veer, L.~J., Symmans, W.~F.,
  et~al. (2016).
\newblock {Adaptive Randomization of Neratinib in Early Breast Cancer}.
\newblock {\em {The New England Journal of Medicine}} {\bf 375,} 11--22.

\bibitem[\protect\citeauthoryear{Posch, Bauer, and Brannath}{Posch
  et~al.}{2003}]{Posch2003}
Posch, M., Bauer, P., and Brannath, W. (2003).
\newblock Issues in designing flexible trials.
\newblock {\em {Statistics in Medicine}} {\bf 22,} 953--969.

\bibitem[\protect\citeauthoryear{Rosenberger and Hu}{Rosenberger and
  Hu}{2004}]{Rosenberger2004}
Rosenberger, W.~F. and Hu, F. (2004).
\newblock Maximizing power and minimizing treatment failures in clinical
  trials.
\newblock {\em Clinical Trials} {\bf 1,} 141--147.

\bibitem[\protect\citeauthoryear{Rosenberger, Stallard, Ivanova, Harper, and
  Ricks}{Rosenberger et~al.}{2001}]{Rosenberger2001}
Rosenberger, W.~F., Stallard, N., Ivanova, A., Harper, C.~N., and Ricks, M.~L.
  (2001).
\newblock Optimal adaptive designs for binary response trials.
\newblock {\em Biometrics} {\bf 57,} 909--913.

\bibitem[\protect\citeauthoryear{Roth, McKenney, Hanotin, Asset, and
  Stein}{Roth et~al.}{2012}]{Roth2012}
Roth, E.~M., McKenney, J.~M., Hanotin, C., Asset, G., and Stein, E.~A. (2012).
\newblock {Atorvastatin with or without an antibody to PCSK9 in primary
  hypercholesterolemia}.
\newblock {\em {The New England Journal of Medicine}} {\bf 367,} 1891--1900.

\bibitem[\protect\citeauthoryear{Rugo, Olopade, DeMichele, Yau, van~'t Veer,
  Buxton, et~al\mbox{.}}{Rugo et~al.}{2016}]{Rugo2016}
Rugo, H.~S., Olopade, O.~I., DeMichele, A., Yau, C., van~'t Veer, L.~J.,
  Buxton, M.~B., et~al. (2016).
\newblock {Adaptive Randomization of Veliparib-Carboplatin Treatment in Breast
  Cancer}.
\newblock {\em {The New England Journal of Medicine}} {\bf 375,} 23--34.

\bibitem[\protect\citeauthoryear{Scott}{Scott}{2010}]{Scott2010}
Scott, S.~L. (2010).
\newblock {A modern Bayesian look at the multi-armed bandit}.
\newblock {\em {Applied Stochastic Models in Business and Industry}} {\bf 26,}
  639--658.

\bibitem[\protect\citeauthoryear{Smith and Villar}{Smith and
  Villar}{2017}]{Smith2017}
Smith, A. and Villar, S.~S. (2017).
\newblock Bayesian adaptive bandit-based designs using the gittins index for
  multi-armed trials with normally distributed endpoints.
\newblock {\em Journal of Applied Statistics} doi:
  10.1080/02664763.2017.1342780.

\bibitem[\protect\citeauthoryear{Thall, Fox, and Wathen}{Thall
  et~al.}{2015}]{Thall2015}
Thall, P., Fox, P., and Wathen, J. (2015).
\newblock Statistical controversies in clinical research: scientific and
  ethical problems with adaptive randomization in comparative clinical trials.
\newblock {\em {Annals of Oncology}} {\bf 26,} 1621--1628.

\bibitem[\protect\citeauthoryear{Thall and Wathen}{Thall and
  Wathen}{2007}]{Thall2007}
Thall, P.~F. and Wathen, J.~K. (2007).
\newblock Practical {Bayesian} adaptive randomisation in clinical trials.
\newblock {\em {European Journal of Cancer}} {\bf 43,} 859--866.

\bibitem[\protect\citeauthoryear{Trippa, Lee, Wen, Batchelor, Cloughesy,
  Parmigiani, and Alexander}{Trippa et~al.}{2012}]{Trippa2012}
Trippa, L., Lee, E.~Q., Wen, P.~Y., Batchelor, T.~T., Cloughesy, T.,
  Parmigiani, G., and Alexander, B.~M. (2012).
\newblock Bayesian adaptive randomized trial design for patients with recurrent
  glioblastoma.
\newblock {\em {Journal of Clinical Oncology}} {\bf 30,} 3258--3263.

\bibitem[\protect\citeauthoryear{Tymofyeyev, Rosenberger, and Hu}{Tymofyeyev
  et~al.}{2007}]{Tymofyeyev2007}
Tymofyeyev, Y., Rosenberger, W.~F., and Hu, F. (2007).
\newblock Implementing optimal allocation in sequential binary response
  experiments.
\newblock {\em Journal of the American Statistical Association} {\bf 102,}
  224--234.

\bibitem[\protect\citeauthoryear{Wason, Stecher, and Mander}{Wason
  et~al.}{2014}]{Wason2014}
Wason, J.~M., Stecher, L., and Mander, A.~P. (2014).
\newblock Correcting for multiple-testing in multi-arm trials: is it necessary
  and is it done?
\newblock {\em Trials} {\bf 15,} 364.

\bibitem[\protect\citeauthoryear{Wason and Trippa}{Wason and
  Trippa}{2014}]{Wason2014a}
Wason, J. M.~S. and Trippa, L. (2014).
\newblock {A comparison of Bayesian adaptive randomization and multi-stage
  designs for multi-arm clinical trials}.
\newblock {\em {Statistics in Medicine}} {\bf 33,} 2206--2221.

\bibitem[\protect\citeauthoryear{Wathen and Thall}{Wathen and
  Thall}{2017}]{Wathen2017}
Wathen, J.~K. and Thall, P.~F. (2017).
\newblock A simulation study of outcome adaptive randomization in multi-arm
  clinical trials.
\newblock {\em Clinical Trials} {\bf 14,} 432--440.

\bibitem[\protect\citeauthoryear{Wei and Durham}{Wei and
  Durham}{1978}]{Wei1978}
Wei, L. and Durham, S. (1978).
\newblock The randomized play-the-winner rule in medical trials.
\newblock {\em {Journal of the American Statistical Association}} {\bf 73,}
  840--843.

\bibitem[\protect\citeauthoryear{Yin, Chen, and Jack~Lee}{Yin
  et~al.}{2012}]{Yin2012}
Yin, G., Chen, N., and Jack~Lee, J. (2012).
\newblock {Phase II trial design with Bayesian adaptive randomization and
  predictive probability}.
\newblock {\em Journal of the Royal Statistical Society: Series C (Applied
  Statistics)} {\bf 61,} 219--235.

\bibitem[\protect\citeauthoryear{Zhang and Rosenberger}{Zhang and
  Rosenberger}{2006}]{Zhang2006}
Zhang, L. and Rosenberger, W.~F. (2006).
\newblock Response-adaptive randomization for clinical trials with continuous
  outcomes.
\newblock {\em Biometrics} {\bf 62,} 562--569.

\bibitem[\protect\citeauthoryear{Zhu and Hu}{Zhu and Hu}{2010}]{Zhu2010}
Zhu, H. and Hu, F. (2010).
\newblock Sequential monitoring of response-adaptive randomized clinical
  trials.
\newblock {\em {The Annals of Statistics}} {\bf 38,} 2218--2241.

\end{thebibliography}
}



\section*{Appendix~A: Derivation of the weights for familywise error control in fully sequential response-adaptive trials}
\label{Asec:fully_seq}
\doublespace

Below is a diagrammatic representation of the assignments and observations for the auxiliary design compared to the actual design for the patients on the experimental treatments: \vspace{12pt}

\textbf{Actual design} \vspace{0pt}
\begin{equation*}
\small
\begin{matrix}
\tikzmark{left1} a_1 & \cdots & a_r & a_{r+1} & a_{r+2} & \cdots & a_n\\
X_1 & \cdots & X_r \tikzmark{right1} & X_{r+1} & X_{r+2} & \cdots & X_n \tikzmark{right2} \\[12pt]
\end{matrix}
\DrawBox[thick, black, dashed]{left1}{right1}{\textcolor{black}{\footnotesize$B$}} 
\end{equation*}

\textbf{Auxiliary design} \vspace{0pt}
\begin{equation*}
\small
\begin{matrix}
\tikzmark{left1} b_1 & \cdots & b_r & b_{r+1} & b_{r+2} & \cdots & b_{n-1} & b_n \\
Y_1 & \cdots & Y_r \tikzmark{right1} & Y_{r+1} & Y_{r+2} & \cdots & Y_{n-1} & Y_n \\[12pt]
\end{matrix}
\DrawBox[thick, black, dashed]{left1}{right1}{\textcolor{black}{\footnotesize$B$}}
\vspace{6pt}
\end{equation*}

\noindent where $b_k = a_k$, $Y_k = X_k$ ($k = 1, \ldots,  r$) and $b_n \in I$ by design.

\subsection*{Step 1}

In step 1 we only consider the first response-adaptive allocation $a_{r+1}$. We view the auxiliary and actual trials as coming from a two-stage design, where the first stage for both is the burn-in period~$B$, as shown below. \vspace{12pt}


\textbf{Auxiliary design (step 1)}\vspace{0ex}
\begin{equation*}
\begin{array}{c c c | c c c c}
\bovermat{Stage 1}{\tikzmark{left1} a_1 & \cdots & a_r &} \bovermat{Stage 2}{b_{r+1} & b_{r+2} & \cdots & b_n} \\
X_1 & \cdots & X_r \tikzmark{right1} & Y_{r+1} & Y_{r+2} & \cdots & Y_n
\end{array}
\DrawBox[thick, black, dashed]{left1}{right1}{\textcolor{black}{\footnotesize$B$}}
\end{equation*}

\newpage

\textbf{Actual design (step 1)}\vspace{0ex}
\begin{equation*}
\begin{array}{c c c | c c c c}
\bovermat{Stage 1}{\tikzmark{left1} a_1 & \cdots & a_r &} \bovermat{Modified stage 2}{a_{r+1} & b_{r+2} & \cdots & b_n}\\
X_1 & \cdots & X_r \tikzmark{right1} & X_{r+1} & Y_{r+2} & \cdots & Y_n \tikzmark{right2} \\
\end{array}
\DrawBox[thick, black, dashed]{left1}{right1}{\textcolor{black}{\footnotesize$B$}} 
\end{equation*}
\vspace{-12pt}

\noindent Given the interim data from~$B$, we can determine the actual allocation~$a_{r+1}$. Indeed, $a_{r+1}$ needs to be known in order for the trial to continue. Hence the second stage for the actual design in step~1 is a data-dependent modification of the auxiliary design, where the allocation $b_{r+1}$ is set to $a_{r+1}$. At this step, all  other allocations for the actual design remain the same as the auxiliary design. The modification to the second stage in the actual design can only depend on data available at the end of the interim stage; that is, the burn-in period. Hence when considering a fully sequential response-adaptive scheme, we cannot adapt $b_k$ to $ a_k$ for $k > r+1$ at this step, and can only consider the modification $b_{r+1}$ to $a_{r+1}$.

Under the auxiliary two-stage design, the test statistic~$T_I = T_I^{(1)} + T_I^{(2)}$ for the experimenal treatments is decomposed into two parts, where $T_I^{(1)}$ is calculated from the first stage data and $T_I^{(2)}$ is calculated from the second stage data. More explicitly,  \vspace{-6pt} \begin{align*}
T_I^{(1)} & =  \sum_{k = 1}^{r} \left( \mathbbm{1}_{\{a_k \in I\}} \frac{X_k}{n_I'} \right)\\
T_I^{(2)} & =  \sum_{k = r+1}^{n} \left( \mathbbm{1}_{\{b_k \in I\}} \frac{Y_k}{n_I'} \right) - \sum_{j = 1}^{n_0} \frac{X_{0j}}{n_0}.
\end{align*}
\noindent Since the control data are independent of the design adaptations, we can consider these data as coming from the second stage of the actual and auxiliary designs.

Using the conditional invariance principle, we seek a statistic $\tilde{T}_I^{(2)}$ from the second stage  data of the actual design so that under~$H_I$ and conditional on the interim data, the statistics $T_I^{(2)}$ and $\tilde{T}_I^{(2)}$ have identical conditional distributions. Hence the modified statistic~$\tilde{T}_I = T_I^{(1)} + \tilde{T}_I^{(2)}$ can be used as the test statistic in the actual design, since $\tilde{T}_I$ has the same unconditional distribution under~$H_I$. 

We now select weights $w_{r+1}^{(I)}$ and $w_{r+1}^{(0)}$ so that under~$H_I$, the statistic \vspace{-6pt} \[
\tilde{T}_I^{(2)} = \mathbbm{1}_{\{a_{r+1} \in I\}} \frac{X_{r+1}}{w_{r+1}^{(I)}} + \sum_{k = r+2}^{n} \left( \mathbbm{1}_{\{b_k \in I\}} \frac{Y_k}{w_{r+1}^{(I)}} \right) -  \sum_{j = 1}^{n_0} \frac{X_{0j}}{w_{r+1}^{(0)}}\]
has the same distribution as~$T_I^{(2)}$ conditional on the interim data $\mathcal{D}^{(1)}$. Under~$H_I$, we have \vspace{-6pt}
\begin{align*}
& T_I^{(2)} \mid \mathcal{D}^{(1)} \sim N\!\left(\mu \frac{m_{I,r+1}}{n_I'} - \mu \, , \frac{m_{I,r+1}}{(n_I')^2} + \frac{1}{n_0} \right) \\
& \tilde{T}_I^{(2)} \mid \mathcal{D}^{(1)} \sim N\!\left(\mu \frac{\tilde{m}_{I,r+1}}{w_{r+1}^{(I)}} - \mu  \frac{n_0}{w_{r+1}^{(0)}} \, , \frac{\tilde{m}_{I,r+1}}{\left(w_{r+1}^{(I)}\right)^2} + \frac{n_0}{\left(w_{r+1}^{(0)}\right)^2} \right)  \\[-48pt]
\end{align*}
where $\tilde{m}_{I,r+1} = m_{I,r+1} + \mathbbm{1}_{\{ a_{r+1} \in I,  b_{r+1} \notin I\}}  - \mathbbm{1}_{\{ a_{r+1} \notin I,  b_{r+1} \in I\}}$.

To match the conditional distributions we equate the conditional means and variances to give \vspace{-6pt}\begin{align*}
w_{r+1}^{(I)} & = \frac{\lambda_{r+1} \tilde{m}_{I,r+1} - [ n_0 \tilde{m}_{I,r+1} \{\eta_{r+1}(n_0 +  \tilde{m}_{I,r+1}) - \lambda_{r+1}^2\}]^{1/2} }{\lambda_{r+1}^2 - n_0 \eta_{r+1}} \\
w_{r+1}^{(0)} & = \frac{n_0 w_{r+1}^{(I)}}{\tilde{m}_{I,r+1} - \lambda_{r+1} w_{r+1}^{(I)} }
\end{align*} \vspace{-6pt}
\noindent where $\lambda_{r+1} = m_{I,r+1}/n_I' - 1$ and $\eta_{r+1} = m_{I,r+1}/(n_I')^2 + 1/{n_0}$.  


Hence the full modified statistic for the actual design in step~1 is \[
\tilde{T}_{I,r+1} = \tilde{T}_I = \sum_{k = 1}^{r+1} \left( \mathbbm{1}_{\{a_k \in I\}} \frac{X_k}{w_k^{(I)}} \right) + 
\sum_{k = r+2}^{n} \left( \mathbbm{1}_{\{b_k \in I\}} \frac{Y_k}{w_{r+1}^{(I)}} \right) - \sum_{j = 1}^{n_0} \frac{X_{0j}}{w_{r+1}^{(0)}}
\] where we define $w_k^{(I)} = n_I'$ ($k = 1, \ldots, r$), and the $(r+1)$ subscript on $\tilde{T}_{I,r+1}$ indicates that this is the modified test statistic for the actual design after the first $(r+1)$ patients. By the conditional invariance principle, $\tilde{T}_{I,r+1}$ is a valid test statistic for the actual design.

\subsection*{Step 2}

In step 2, we take the actual design from step 1 as the new auxiliary design. This means that the modified test statistic $\tilde{T}_{I,r+1}$, as defined above, is also taken forward from step~1 and is the valid test statistic for the new auxiliary design. We again view the auxiliary and actual trials as two-stage designs, where this time the first stage is the data from the first $(r+1)$ patients, as shown below. \\[-12pt]

\textbf{Auxiliary design (step 2)}\vspace{0ex}
\begin{equation*}
\begin{array}{c c c c | c c c c}
\bovermat{Stage 1}{\tikzmark{left1} a_1 & \cdots & a_r & a_{r+1} &} \bovermat{Stage 2}{b_{r+2} & b_{r+3} & \cdots & b_n} \\
X_1 & \cdots & X_r \tikzmark{right1} & X_{r+1} & Y_{r+2} & Y_{r+3} & \cdots & Y_n  \\
\end{array}
\DrawBox[thick, black, dashed]{left1}{right1}{\textcolor{black}{\footnotesize$B$}} 
\end{equation*}

\textbf{Actual design (step 2)}\vspace{0ex}
\begin{equation*}
\begin{array}{c c c c | c c c c}
\bovermat{Stage 1}{\tikzmark{left1} a_1 & \cdots & a_r & a_{r+1} &} \bovermat{Modified stage 2}{a_{r+2} & b_{r+3} & \cdots & b_n}\\
X_1 & \cdots & X_r \tikzmark{right1} & X_{r+1} & X_{r+2} & Y_{r+3} & \cdots & Y_n \tikzmark{right2} \\
\end{array}
\DrawBox[thick, black, dashed]{left1}{right1}{\textcolor{black}{\footnotesize$B$}} 
\end{equation*}
\vspace{-12pt}

\noindent  Here the second stage for the new actual design is a modification of the new auxiliary design where the allocation $b_{r+2}$ is set to $a_{r+2}$. 

Under the auxiliary two-stage design, the test statistic~$\tilde{T}_{I,r+1}$ for the experimental treatments is decomposed into the statistics calculated from the first and second stage data, $\tilde{T}_{I,r+1} = \tilde{T}_{I,r+1}^{(1)} + \tilde{T}_{I,r+1}^{(2)}$, where now \vspace{-12pt} \begin{align*}
\tilde{T}_{I,r+1}^{(1)} & = \sum_{k = 1}^{r+1} \left( \mathbbm{1}_{\{a_k \in I\}} \frac{X_k}{w_k^{(I)}} \right) \\[6pt]
\tilde{T}_{I,r+1}^{(2)} & = \sum_{k = r+2}^{n} \left( \mathbbm{1}_{\{b_k \in I\}} \frac{Y_k}{w_{r+1}^{(I)}} \right) - \sum_{j = 1}^{n_0} \frac{X_{0j}}{w_{r+1}^{(0)}}.
\end{align*}

\noindent Following the conditional invariance principle like in step 1, we seek weights $w_{r+2}^{(I)}$ and $w_{r+2}^{(0)}$ so that under~$H_I$, the statistic \vspace{-6pt}
\[ \tilde{T}_{I,r+2}^{(2)} = \mathbbm{1}_{\{a_{r+2} \in I\}} \frac{X_{r+2}}{w_{r+2}^{(I)}} + \sum_{k = r+3}^{n} \left( \mathbbm{1}_{\{b_k \in I\}} \frac{Y_k}{w_{r+2}^{(I)}} \right) - \sum_{j = 1}^{n_0} \frac{X_{0j}}{w_{r+2}^{(0)}}\]
has the same distribution as~$\tilde{T}_{I,r+1}^{(2)}$ conditional on the interim data $\mathcal{D}^{(1)}$.
%
%
%
%
Matching the conditional distributions by equating the conditional means and variances gives \begin{align*}
w_{r+2}^{(I)} & = \frac{\lambda_{r+2} \tilde{m}_{I,r+2} - [n_0 \tilde{m}_{I,r+2} \{  \eta_{r+2}(n_0 + \tilde{m}_{I,r+2}) - \lambda_{r+2}^2\}]^{1/2} }{\lambda_{r+2}^2 - n_0 \eta_{r+2}} \\
w_{r+2}^{(0)} & = \frac{n_0 w_{r+2}^{(I)}}{\tilde{m}_{I,r+2} - \lambda_{r+2} w_{r+2}^{(I)} }
\end{align*}
\noindent where \vspace{-16pt} \begin{align*}
\lambda_{r+2} & = \frac{m_{I,r+2}}{w_{1}^{(I)}} - \frac{n_0}{w_{1}^{(0)}} \, , \quad  \eta_{r+2} = \frac{m_{I,r+2}}{\left(w_{1}^{(I)}\right)^2} + \frac{n_0}{\left(w_{1}^{(0)}\right)^2}\\[6pt]
\tilde{m}_{I,r+2} & = m_{I,r+2} + \mathbbm{1}_{\{ a_{r+2} \in I,  b_{r+2} \notin I\}}  - \mathbbm{1}_{\{ a_{r+2} \notin I,  b_{r+2} \in I\}}.
\end{align*}

\noindent Hence the full modified statistic for the actual design in step~2 is \[
\tilde{T}_{I,r+2} = \sum_{k = 1}^{r+2} \left( \mathbbm{1}_{\{a_k \in I\}} \frac{X_k}{w_k^{(I)}} \right) + 
\sum_{k = r+3}^{n} \left( \mathbbm{1}_{\{b_k \in I\}} \frac{Y_k}{w_{r+2}^{(I)}} \right) - \sum_{j = 1}^{n_0} \frac{X_{0j}}{w_{r+2}^{(0)}}
\]

\noindent This statistic is taken forward to the next step of the process as the valid test statistic for the new auxiliary design.

\subsection*{Inductive step}

We now repeat the process above, at each step taking forward the actual design as the new auxiliary design. The actual design at step~$l$ of the process ($l = 2, \ldots, n-r-1$) is a modification of the  new auxiliary design where $b_{r+l}$ is set to $a_{r+l}$. The valid test statistic for the new auxiliary design is $\tilde{T}_{I, r+l-1}$, taken forward from the previous step of the process, where we provide an explicit expression for the test statistics shortly. The diagrammatic representation of step~$l$ of the process is given below. \\[-12pt]

\textbf{Auxiliary design (step $l$)}\vspace{1ex}
\begin{equation*}
\begin{array}{c c c c c c c | c c c c}
\bovermat{Stage 1}{\tikzmark{left1} a_1 & \cdots & a_r & a_{r+1} & a_{r+2} & \cdots & a_{r+l-1} &} \bovermat{Stage 2}{ b_{r+l} & b_{r+l+1} & \cdots & b_n}\\
X_1 & \cdots & X_r \tikzmark{right1} & X_{r+1} & X_{r+2} & \cdots & X_{r+l-1} & Y_{r+l} & Y_{r+l+1} & \cdots & Y_n \\ \tikzmark{right2}
\end{array}
\DrawBox[thick, black, dashed]{left1}{right1}{\textcolor{black}{\footnotesize$B$}}
\end{equation*}

\textbf{Actual design (step $l$)}\vspace{1ex}
\begin{equation*}
\begin{array}{c c c c c c c | c c c c}
\bovermat{Stage 1}{\tikzmark{left1} a_1 & \cdots & a_r & a_{r+1} & a_{r+2} & \cdots & a_{r+l-1} &} \bovermat{Modified stage 2}{ a_{r+l} & b_{r+l+1} & \cdots & b_n}\\
X_1 & \cdots & X_r \tikzmark{right1} & X_{r+1} & X_{r+2} & \cdots & X_{r+l-1} & X_{r+l} & Y_{r+l+1} & \cdots & Y_n \\ \tikzmark{right2}
\end{array}
\DrawBox[thick, black, dashed]{left1}{right1}{\textcolor{black}{\footnotesize$B$}}
\end{equation*}
\vspace{-30pt}

Using these auxiliary and actual designs, we select new weights $w_{r+l}^{(I)}$ and $w_{r+l}^{(0)}$ so that under~$H_I$ the conditional distributions of the second stage statistics are the same. This yields the new test statistic $\tilde{T}_{I, r+l}$ for the actual design in step~$l$. For notational convenience, we introduce the following functions: \vspace{-12pt} \begin{align*}
f(\lambda, \eta, m) & = \frac{\lambda m - \{m n_0 (n_0 \eta - \lambda^2)\}^{1/2} }{\lambda - n_0 \eta} \\
g(\lambda, w, m) & = \frac{n_0 w}{m - \lambda w }
\end{align*}

 \noindent Given the weights $w_{r+l-1}^{(I)}$ and $w_{r+l-1}^{(0)}$ found in the previous step, let\begin{align*}
\lambda_{r+l} & = \frac{m_{I,r+l}}{w_{r+l-1}^{(I)}} - \frac{n_0}{w_{r+l-1}^{(0)}} \, , \quad  \eta_{r+l} = \frac{m_{I,r+l}}{\left(w_{r+l-1}^{(I)}\right)^2} + \frac{n_0}{\left(w_{r+l-1}^{(0)}\right)^2}\\[6pt]
\tilde{m}_{I,r+l} & = m_{I,r+l} + \mathbbm{1}_{\{ a_{r+l} \in I,  b_{r+l} \notin I\}}  - \mathbbm{1}_{\{ a_{r+l} \notin I,  b_{r+l} \in I\}}.
\end{align*}

\noindent The weights $w_{r+l}^{(I)}$ and $w_{r+l}^{(0)}$ are given by
\begin{align*}
w_{r+l}^{(I)} & = f(\lambda_{r+l}, \eta_{r+l}, \tilde{m}_{I,r+l})\\
w_{r+l}^{(0)} &=  g(\lambda_{r+l},  w_{r+l}^{(I)},  \tilde{m}_{I,r+l})
\end{align*}

\noindent The corresponding test statistics $\tilde{T}_{I, r+k}$ for $k = 1, \ldots, l$ are:
\[\tilde{T}_{I,r+k} = \sum_{j= 1}^{r+k} \left( \mathbbm{1}_{\{a_j \in I\}} \frac{X_j}{w_j^{(I)}} \right) + 
\sum_{j = r+k+1}^{n} \left( \mathbbm{1}_{\{b_j \in I\}} \frac{Y_j}{w_{r+k}^{(I)}} \right) - \sum_{j = 1}^{n_0} \frac{X_{0j}}{w_{r+k}^{(0)}}
\]

\subsection*{Final step}

In the final step of the process, the second stage data for the auxiliary and actual designs is a single allocation: \\[-12pt]

\textbf{Auxiliary design (final step)}\vspace{-2ex}
\begin{equation*}
\begin{array}{c c c c c c c | c}
\tikzmark{left1} a_1 & \cdots & a_r & a_{r+1} & a_{r+2} & \cdots & a_{n-1} & b_n \\
X_1 & \cdots & X_r \tikzmark{right1} & X_{r+1} & X_{r+2} & \cdots & X_{n-1} & Y_n \\
\end{array}
\DrawBox[thick, black, dashed]{left1}{right1}{\textcolor{black}{\footnotesize$B$}} 
\end{equation*}

\textbf{Actual design (final step)}\vspace{-2ex}
\begin{equation*}
\begin{array}{c c c c c c c | c}
\tikzmark{left1} a_1 & \cdots & a_r & a_{r+1} & a_{r+2} & \cdots & a_{n-1} & a_n\\
X_1 & \cdots & X_r \tikzmark{right1} & X_{r+1} & X_{r+2} & \cdots & X_{n-1} & X_n \tikzmark{right2} \\
\end{array}
\DrawBox[thick, black, dashed]{left1}{right1}{\textcolor{black}{\footnotesize$B$}}
\end{equation*}
\vspace{-12pt}

Under the auxiliary design, the test statistic~$\tilde{T}_{I,n-1}$ is decomposed into the following first and second stage statistics, since $b_n \in I$ by design:
\[
\tilde{T}_{I,n-1}^{(1)} = \sum_{k = 1}^{n-1} \left( \mathbbm{1}_{\{a_k \in I\}} \frac{X_k}{w_k^{(I)}} \right), \qquad
\tilde{T}_{I,n-1}^{(2)} = \frac{Y_n}{w_{n-1}^{(I)}} - \sum_{j = 1}^{n_0} \frac{X_{0j}}{w_{n-1}^{(0)}}
\]
where $w_{n-1}^{(0)}$ and $w_k^{(I)}$ have been defined in the previous steps for $k = r+1, \ldots, n-1$. As before, we want to select a weight $w_n^{(I)}$ so that \[
\tilde{T}_{I,n}^{(2)} = \mathbbm{1}_{\{a_n \in I\}} \frac{X_n}{w_n^{(I)}} - \sum_{j = 1}^{n_0} \frac{X_{0j}}{w_{n}^{(0)}} \] has the same conditional distribution as~$T_{I,n-1}^{(2)}$ under~$H_I$. If $a_n \in I$ then the auxiliary and actual designs are identical, and so $w_n^{(I)} = w_{n-1}^{(I)}$ and $w_n^{(0)} = w_{n-1}^{(0)}$.

However, if $a_n \notin I$, then we do not have enough degrees of freedom with a single weight $w_n^{(0)}$ to match both the conditional means and variances. Hence, we split the $n_0$ control observations into two groups of size $m_{0,1}$ and $m_{0,2}$, where $m_{0,1} \geq 1$, $m_{0,2} \geq 1$ and $m_{0,1} + m_{0,2} = n_0$. In practice, to keep the weights as close to the natural weight $n_0$ for as many of the control observations as possible, we recommend setting $m_{0,1} = n_0 - 1$ and $m_{0,2} = 1$, which is what we use for the simulation studies in Section~4.2 of the paper.

We select weights $w_{n,1}^{(0)}$ and $w_{n,2}^{(0)}$ so that under~$H_I$, the statistic \[
\tilde{T}_{I,n}^{(2)} = - \sum_{j = 1}^{m_{0,1}} \frac{X_{0j}}{w_{n,1}^{(0)}}  - \sum_{j = m_{0,1}+1}^{n_0} \frac{X_{0j}}{w_{n,2}^{(0)}}\] has the same distribution as~$T_{I,n}^{(2)}$ conditional on the interim data $\mathcal{D}^{(1)}$.
Under~$H_I$, we have \vspace{-12pt}\begin{align*}
& T_{I,n}^{(2)} | \mathcal{D}^{(1)} \sim N\!\left(\mu \frac{1}{w_{n-1}^{(I)}} - \mu \frac{n_0}{w_{n-1}^{(0)}} \, , \frac{1}{\left(w_{n-1}^{(I)}\right)^2} + \frac{n_0}{\left(w_{n-1}^{(0)}\right)^2} \right) \\
& \tilde{T}_{I,n}^{(2)} | \mathcal{D}^{(1)} \sim N\!\left( - \mu \frac{m_{0,1}}{w_{n,1}^{(0)}} - \mu  \frac{m_{0,2}}{w_{n,2}^{(0)}} \, , \frac{m_{0,1}}{\left(w_{1,n}^{(0)}\right)^2} + \frac{m_{0,2}}{\left(w_{n,2}^{(0)}\right)^2} \right)
\end{align*}

\noindent Equating the conditional means and variances gives \begin{align*}
w_{n,1}^{(0)} & = \frac{ -2m_{0,1} \lambda_{n} - [m_{0,1} m_{0,2} \{ \eta_{n}(m_{0,1} + m_{0,2}) - \lambda_{n}^2\}]^{1/2}}{\lambda_{n}^2 - m_{0,2} \eta_{n}} \\
w_{n,2}^{(0)} & = - \frac{m_{0,2} w_{n,1}^{(0)}}{m_{0,1} + \lambda_{n} w_{n,1}^{(0)} }
\end{align*}
\noindent where $\lambda_{n} = 1/w_{n-1}^{(I)} - n_0/w_{n-1}^{(0)}$ and $\eta_{n} = 1/(w_{n-1}^{(I)})^2 + n_0/(w_{n-1}^{(0)})^2$.

For notational convenience, define the following functions \begin{align*} 
F_1(w, m_1, m_2) & = \begin{cases} w & \text{if } a_n \in I \\[6pt]
\displaystyle \frac{ -2m_{1} \lambda_{n} - \sqrt{m_1 m_2(\eta_{n}(m_1 + m_2) - \lambda_{n}^2)}}{\lambda_{n}^2 - m_{2} \eta_{n}} & \text{if } a_n \notin I
\end{cases} \\[6pt]
F_2(w, m_1, m_2) & = \begin{cases} w & \text{if } a_n \in I \\[6pt]
\displaystyle  - \frac{m_{2} F_1(w, m_1, m_2)}{m_{1} + \lambda_{n} F_1(w, m_1, m_2)} & \text{if } a_n \notin I
\end{cases}
\end{align*} We can express the weights $w_{n,1}^{(0)}$ and $w_{n,2}^{(0)}$ for the controls, for either $a_n \in I$ or $a_n \notin I$, as $w_{n,1}^{(0)} = F_1(w_{n-1}^{(0)}, m_{0,1}, m_{0,2})$ and $w_{n,2}^{(0)} = F_2(w_{n-1}^{(0)}, m_{0,1}, m_{0,2})$.


The final test statistic for testing hypothesis $H_I$ is as follows:\[
\tilde{T}_I = \sum_{k = 1}^{n} \left( \mathbbm{1}_{\{a_k \in I\}} \frac{X_k}{w_k^{(I)}} \right) - \sum_{j=1}^{n_0} \frac{X_{0j}}{w_{n,j}^{(0)}}
\]
where \vspace{-24pt} \begin{alignat*}{2}
w_k^{(I)} & = n_I' , \quad w_k^{(0)} = n_0 && \qquad (k = 1, \ldots, r)\\
w_{r+l}^{(I)} & = f(\lambda_{r+l}, \eta_{r+l}, \tilde{m}_{I,r+l}) && \qquad (l = 1, \ldots, n-r) \\
w_{r+l}^{(0)} & = g(\lambda_{r+l}, w_{r+l}^{(I)}, \tilde{m}_{I,r+l}) && \qquad (l = 1, \ldots, n-r) \\
w_{n,j}^{(0)} & = F_1(w_{n-1}^{(0)}, m_{0,1}, m_{0,2}) && \qquad (j = 1, \ldots, m_{0,1}) \\
w_{n,j}^{(0)} & = F_2(w_{n-1}^{(0)}, m_{0,1}, m_{0,2}) && \qquad (j = m_{0,1}+1, \ldots, n_0) 
\end{alignat*}

\noindent We reject $H_I$ if $\tilde{T}_I$ is greater than $z_{\alpha} \left( 1/n_I' + 1/n_0\right)^{1/2} $.


\section*{Appendix B: Numerical example for fully sequential response-adaptive trials}
\label{Asec:num_ex}
\doublespace

As a simple illustration of how the weights change over the course of a trial, consider testing $h = 2$ experimental treatments. We set $\alpha = 0.05$, $n_0 = 10$, $n = 11$ and $r_1 = r_2 = 1$. Suppose we have no a priori reason to favour one treatment over the other, and so we simply choose the auxiliary design to be an equal randomization of the two treatments:
\[ b = \begin{array}{c c | c c c c c c c c c}
1 & 2 & 2 & 1 & 2 & 2 & 1 & 1 & 2 & 1 & *
\end{array}\]

Here the vertical line indicates where the burn-in period ends, and the * represents the allocation for $b_n$, which by design must satisfy $b_n \in I$. We set $m_{0,1} = 9$ and $m_{0,2} = 1$, so that $w_1^{(0)} = \cdots = w_9^{(0)}$. Below are the weights $w^{(I)}$, $w_{1}^{(0)}$ and $w_{10}^{(0)}$ for a variety of actual allocations.

\singlespace

\begin{table}[H]

\caption{\textit{An actual allocation $a$ that is almost the same as the auxiliary design $b$. The weights that would be used in the na\"ive $z$-test are $n_0 = 10$, $n_1 = 4$ and $n_2 = 7$.}}

\resizebox{\textwidth}{!}{
\begin{tabular}{l c c | c c c c c c c c c l l}
\\
 $a$ = & 1 &   2 &   2 &   2 &   1 &   2 &   2 &   1 &   2 &   1 &   2  \\
 $b$ =  & 1 & 2 & 2 & 1 & 2 & 2 & 1 & 1 & 2 & 1 & * \\
 $w^{(1)}$ = & 6 & 6 & 6 & 5.16 & 6 & 6 & 4.94 & 4.94 & 4.94 & 4.94 & -  & & $w_{1}^{(0)} = 9.74,  w_{10}^{(0)} = -5.38$\\
 $w^{(2)}$ = &6 & 6 & 6 & 7.01 & 5.74 & 5.74 & 7.63 & 7.63 & 7.63 & 7.63 & 7.63 & &  $w_{1}^{(0)} = w_{10}^{(0)} = 9.58$
\end{tabular}}

\end{table}

\begin{table}[H]

\caption{\textit{An actual allocation $a$ that is the opposite of the auxiliary design $b$. The weights that would be used in the na\"ive $z$-test are $n_0 = 10$, $n_1 = 6$ and $n_2 = 5$.}}

\resizebox{\textwidth}{!}{
\begin{tabular}{l c c | c c c c c c c c c l l}
\\
 $a$ = & 1 &   2 &   1 &   2 &   1 &   1 &   2 &   2 &   1 &   2 &   1  \\
 $b$ =  & 1 & 2 & 2 & 1 & 2 & 2 & 1 & 1 & 2 & 1 & * \\
 $w^{(1)}$ = &  6 & 6 & 6.81 & 5.83 & 6.81 & 8 & 6.51 & 4.94 & 6.51 & 4.09 & 4.09  & & $w_{1}^{(0)} = w_{10}^{(0)} = 10.17$\\
 $w^{(2)}$ = & 6 & 6 & 5.16 & 6 & 4.94 & 3.81 & 4.94 & 6.51 & 4.10 & 6.51 & - & &  $w_{1}^{(0)} = 9.23,  w_{10}^{(0)} = -7.59$
\end{tabular}}

\end{table}

\begin{table}[H]

\caption{ \textit{An extreme actual allocation $a$ that is equal to 1 after the burn-in period. The weights that would be used in the na\"ive $z$-test are $n_0 = 10$, $n_1 = 10$ and $n_2 = 2$.}}

\resizebox{\textwidth}{!}{
\begin{tabular}{l c c | c c c c c c c c c l l}
\\
 $a$ = &  1 &   2 &   1 &   1 &   1 &   1 &   1 &   1 &   1 &   1 &   1  \\
 $b$ =  & 1 & 2 & 2 & 1 & 2 & 2 & 1 & 1 & 2 & 1 & * \\
 $w^{(1)}$ = &  6 & 6 & 6.81 & 6.81 & 8.00 & 9.45 & 9.45 & 9.45 & 12.95 & 12.95 & 12.95 & &  $w_{1}^{(0)} = w_{10}^{(0)} = 8.82$\\
 $w^{(2)}$ = & 6 & 6 & 5.16 & 5.16 & 4.28 & 3.33 & 3.33 & 3.33 & 2.23 & 2.23 & -& &  $w_{1}^{(0)} = 14.73,  w_{10}^{(0)} =-2.25$
\end{tabular}}

\end{table}

\begin{table}[H]

\caption{\textit{An extreme actual allocation $a$ that is equal to 2 after the burn-in period. The weights that would be used in the na\"ive $z$-test are $n_0 = 10$, $n_1 = 1$ and $n_2 = 10$.}}

\resizebox{\textwidth}{!}{
\begin{tabular}{l c c | c c c c c c c c c l l}
\\
 $a$ = &  1 &   2 &   2 &   2 &   2 &   2 &   2 &   2 &   2 &   2 &   2   \\
 $b$ =  & 1 & 2 & 2 & 1 & 2 & 2 & 1 & 1 & 2 & 1 & * \\
 $w^{(1)}$ = & 6 & 6 & 6 & 5.16 & 5.16 & 5.16 & 4.28 & 3.33 & 3.33 & 2.23 & -  & &  $w_{1}^{(0)} = 14.73,  w_{10}^{(0)} = -2.25$\\
 $w^{(2)}$ = &  6 & 6 & 6 & 7.01 & 7.01 & 7.01 & 9.44 & 12.91 & 12.91 & 22.89 & 22.89& &  $w_{1}^{(0)} = w_{10}^{(0)} = 9.01$
\end{tabular}}

\end{table}

\doublespace



\section*{Appendix C: Derivation of the weights for familywise error control in block randomized response-adaptive trials with a fixed control allocation}
\label{Asec:block_without_control}
\doublespace

Below is a diagrammatic representation of the assignments and observations for the auxiliary design compared to the actual design (for the patients on the experimental treatments): \\[-12pt]

\textbf{Actual design} \vspace{-12pt}
\begin{equation*}
\begin{matrix}
\bm{a}_{\sss B} & \bm{a}_1 & \bm{a}_2 & \cdots & \bm{a}_{\sss J}\\
\bm{X}_{\sss B} & \bm{X}_1 & \bm{X}_2 & \cdots & \bm{X}_{\sss J}
\end{matrix}
\end{equation*}

\textbf{Auxiliary design}  \vspace{-12pt}
\begin{equation*}
\small
\begin{matrix}
\bm{b}_{\sss B} & \bm{b}_1 & \bm{b}_2 & \cdots & \bm{b}_{\sss J}\\
\bm{Y}_{\sss B} & \bm{Y}_1 & \bm{Y}_2 & \cdots & \bm{Y}_{\sss J} \\[6pt]
\end{matrix}
\end{equation*}

\noindent Here $\bm{a}_{\sss B} = (a_1, \ldots, a_r)$ and $\bm{X}_{\sss B} = (X_1, \ldots, X_r)$ refer to the burn-in period~$B$, while \hfill \\ $\bm{a}_j = (a_{\sss D_{j-1} + 1}, \ldots, a_{\sss D_j})$ and $\bm{X}_j = (X_{\sss D_{j-1} + 1}, \ldots, X_{\sss D_j})$ represent the response-adaptive allocations and observations in block~$j$ ($j = 1, \ldots, J$). By design, $\bm{b}_{\sss B} = \bm{a}_{\sss B}$, $\bm{Y}_{\sss B} = \bm{X}_{\sss B}$, while $\bm{b}_j = (b_{\sss D_{j-1} + 1}, \ldots, b_{\sss D_j})$ and $\bm{Y}_j = (Y_{\sss D_{j-1} + 1}, \ldots, Y_{\sss D_j})$ represent the auxiliary allocations and observations in block~$j$ ($j = 1, \ldots, J$). As before, we require $b_n \in I$.

\subsection*{Step 1}

In step 1 we only consider the response-adaptive allocations for the first block $\bm{a}_1$. We view the auxiliary and actual trials as coming from a two-stage design, where the first stage for both is the burn-in period~$B$, as shown below.\\[-16pt]

\textbf{Auxiliary design (step 1)}\vspace{-1ex}
\begin{equation*}
\begin{array}{c | c c c c c}
\bm{a}_{\sss B} & \bovermat{Stage 2}{\bm{b}_1 & \bm{b}_2 & \cdots & \bm{b}_{\sss J} & }\\
\bm{X}_{\sss B} & \bm{X}_1 & \bm{X}_2 & \cdots & \bm{X}_{\sss J}
\end{array}
\end{equation*}


\textbf{Actual design (step 1)}\vspace{3ex}
\begin{equation*}
\begin{array}{c | c c c c c}
\bm{a}_{\sss B} & \bovermat{Modified stage 2}{\bm{a}_1 & \bm{b}_2 & \cdots & \bm{b}_{\sss J} & }\\
\bm{X}_{\sss B} & \bm{X}_1 & \bm{Y}_2 & \cdots & \bm{Y}_{\sss J}
\end{array}
\end{equation*}

\vspace{6pt}

\noindent Given the interim data from~$B$, we can determine the actual allocations~$\bm{a}_1$ for the first block. Hence the second stage for the actual design in step~1 is a data-dependent modification of the auxiliary design, where the allocations $\bm{b}_1$ are set to~$\bm{a}_1$. All the other allocations for the actual design remain the same as the auxiliary design.


Under the auxiliary two-stage design, the test statistic~$T_I$  is decomposed into two parts, with $T_I = T_I^{(1)} + T_I^{(2)}$, where $T_I^{(1)}$ is calculated from the first stage data and $T_I^{(2)}$ is calculated from the second stage data. More explicitly,  \begin{align*}
T_I^{(1)} & =  \sum_{k = 1}^{r} \left( \mathbbm{1}_{\{a_k \in I\}} \frac{Y_k}{n_I'} \right) -  \sum_{k = 1}^{r_0} \frac{X_{0k}}{n_0} \\
T_I^{(2)} & =  \sum_{k = r+1}^{n} \left( \mathbbm{1}_{\{b_k \in I\}} \frac{Y_k}{n_I'} \right) -  \sum_{k = r_0+1}^{n_0}  \frac{X_{0k}}{n_0} 
\end{align*}

Following the conditional invariance principle, we select weights $w_{1}^{(I)}$ and $w_{1}^{(0)}$ so that under~$H_I$, the statistic \[
\tilde{T}_I^{(2)} = \sum_{k = r+1}^{D_1} \mathbbm{1}_{\{a_k \in I\}} \frac{X_k}{w_1^{(I)}} + \sum_{k = D_1 + 1}^{n} \mathbbm{1}_{\{b_k \in I\}} \frac{Y_k}{w_1^{(I)}} -  \sum_{k = r_0+1}^{n_0} \frac{X_{0k}}{w_1^{(0)}} \]
has the same distribution as~$T_I^{(2)}$ conditional on the interim data $\mathcal{D}^{(1)}$. Under~$H_I$, we have \vspace{-6pt}
\begin{align*}
& T_I^{(2)} | \mathcal{D}^{(1)} \sim N\!\left(\mu \frac{m_{I,1}}{n_I'} - \mu \frac{m_{0,1}}{n_0'} \, , \frac{m_{I,1}}{(n_I')^2} + \frac{m_{0,1}}{(n_0')^2} \right) \\
& \tilde{T}_I^{(2)} | \mathcal{D}^{(1)} \sim N\!\left(\mu \frac{\tilde{m}_{I,1}}{w_1^{(I)}} - \mu  \frac{m_{0,1}}{w_1^{(0)}} \, , \frac{\tilde{m}_{I,1}}{\left(w_{1}^{(I)}\right)^2} + \frac{m_{0,1}}{\left(w_{1}^{(0)}\right)^2} \right)
\end{align*}
where $\tilde{m}_{I,1} = m_{I,2} + \sum_{k = r + 1}^{D_1} \mathbbm{1}_{\{a_k \in I\}}$.

To match the conditional distributions we equate the conditional means and variances to give \vspace{-6pt} \begin{align*}
w_{1}^{(I)} & = \frac{\lambda_{1} \tilde{m}_{I,1} - [\tilde{m}_{I,1} m_{0,1} \{ \eta_{1} (m_{0,1} + \tilde{m}_{I,1}) - \lambda_{1}^2\}]^{1/2} }{\lambda_{1}^2 - m_{0,1} \eta_{1}} \\
w_{1}^{(0)} & = \frac{m_{0,1} w_{1}^{(I)}}{\tilde{m}_{I,1} - \lambda_{1} w_{1}^{(I)} }
\end{align*}
\noindent where $\lambda_{1} = m_{I,1}/n_I' - m_{0,1}/n_0$ and $\eta_{1} = m_{I,1}/(n_I')^2 + m_{0,1}/(n_0)^2$.

Hence the full modified statistic for the actual design in step~1 is \begin{equation*}
\tilde{T}_{I,1} = \tilde{T}_I = \sum_{j = 0}^{1} \sum_{k = D_{j-1} + 1}^{D_{j}} \mathbbm{1}_{\{a_k \in I\}} \frac{X_k}{w_j^{(I)}} + \sum_{k = D_1 + 1}^{n} \mathbbm{1}_{\{b_k \in I\}} \frac{Y_k}{w_1^{(I)}} - \sum_{k = 1}^{r_0} \frac{X_{0k}}{w_0^{(0)}} - \sum_{k = r_0 + 1}^{n_0} \frac{X_{0k}}{w_1^{(0)}}
\end{equation*}
where $w_0^{(0)} = n_0$ and $w_0^{(I)} = n_I'$. This statistic is taken forward to the next step of the process as the valid test statistic for the new auxiliary design.
\subsection*{Inductive step}

We continue the process above, at each step taking forward the actual design as the new auxiliary design. The actual design at step~$l$ of the process ($l \in \{1, \ldots, J-1\}$) is a modification of the  new auxiliary design where the allocations $\bm{b}_l$ are set to $\bm{a}_l$. The valid test statistic for the new auxiliary design is $\tilde{T}_{I, l}$, taken forward from the previous step of the process. The diagrammatic representation of step~$l$ of the process is given below: \\[-12pt]

\textbf{Auxiliary design (step $l$)}\vspace{1ex}
\begin{equation*}
\begin{array}{c c c c| c c c c c}
\bovermat{Stage 1}{\bm{a}_{\sss B} & \bm{a}_1 & \cdots & \bm{a}_{l-1} &} \bovermat{Stage 2}{\bm{b}_l & \bm{b}_{l+1} & \cdots & \bm{b}_{\sss J} & }\\
\bm{X}_{\sss B} & \bm{X}_1 & \cdots & \bm{X}_{l-1} & \bm{Y}_l & \bm{Y}_{l+1} & \cdots & \bm{Y}_{\sss J}
\end{array}
\end{equation*}

\textbf{Actual design (step $l$)}\vspace{1ex}
\begin{equation*}
\begin{array}{c c c c| c c c c c}
\bovermat{Stage 1}{\bm{a}_{\sss B} & \bm{a}_1 & \cdots & \bm{a}_{l-1} &} \bovermat{Modified stage 2}{\bm{a}_l & \bm{b}_{l+1} & \cdots & \bm{b}_{\sss J} & }\\
\bm{X}_{\sss B} & \bm{X}_1 & \cdots & \bm{X}_{l-1} & \bm{X}_l & \bm{Y}_{l+1} & \cdots & \bm{Y}_{\sss J}
\end{array}
\end{equation*}

Using these auxiliary and actual designs, we select new weights $w_{l}^{(I)}$ and $w_{l}^{(0)}$ so that under~$H_I$ the conditional distributions of the second stage statistics are the same. This yields the new test statistic~$\tilde{T}_{I, l}$ for the actual design in step~$l$.  For notational convenience, we introduce the following functions: \vspace{-12pt} \begin{align*}
f(\lambda, \eta, m_I, m_0) & = \frac{\lambda m_I - [m_I m_0 \{ \eta (m_0 + m_I) - \lambda^2 \}]^{1/2} }{\lambda - m_0 \eta} \\
g(\lambda, w, m_I, m_0) & = \frac{m_0 w}{m_I - \lambda w }
\end{align*}

\noindent Given the weights $w_{l-1}^{(I)}$ and $w_{l-1}^{(0)}$ found in the previous step, let\begin{align*}
\lambda_l & = \frac{m_{I,l}}{w_{l-1}^{(I)}} - \frac{m_{0,l}}{w_{l-1}^{(0)}} \, , \quad  \eta_l = \frac{m_{I,l}}{\left(w_{l-1}^{(I)}\right)^2} + \frac{m_{0,l}}{\left(w_{l-1}^{(0)}\right)^2}\\[6pt]
\tilde{m}_{I,l} & = m_{I,l+1} + \sum_{k=D_{l-1}+1}^{D_l} \mathbbm{1}_{\{a_k \in I\}}
\end{align*}

\noindent The weights $w_{l}^{(I)}$ and $w_{l}^{(0)}$ are given by \vspace{-12pt}
\begin{align*}
w_{l}^{(I)} & = f(\lambda_l, \eta_l, \tilde{m}_{I,l},  m_{0,l})\\
w_{l}^{(0)} &=  g(\lambda_l, w_{l}^{(I)}, \tilde{m}_{I,l}, m_{0,l})
\end{align*}

\noindent The corresponding test statistic $\tilde{T}_{I,l}$ is: \vspace{-12pt}
\begin{equation*}
\begin{split}
\tilde{T}_{I,l} = \sum_{j = 0}^{l} \sum_{k = D_{j-1} + 1}^{D_{j}} \mathbbm{1}_{\{a_k \in I\}} \frac{X_k}{w_j^{(I)}} + \sum_{k = D_l + 1}^{n} \mathbbm{1}_{\{b_k \in I\}} \frac{Y_k}{w_l^{(I)}} \\
- \sum_{j = 0}^{l} \sum_{k = D_{0,j-1} + 1}^{D_{0,j}} \frac{X_{0k}}{w_j^{(0)}} - \sum_{k = D_{0,l} + 1}^{n_0} \frac{X_{0k}}{w_l^{(0)}}
\end{split}
\end{equation*}
where we define $D_{0, -1} = 0$.

\subsection*{Final step}

In the final step of the process, the second stage data for the auxiliary and actual designs is the final block: \\[-12pt]


\textbf{Auxiliary design (final step)}\vspace{0ex}
\begin{equation*}
\begin{array}{c c c c | c c}
\bovermat{Stage 1}{\bm{a}_{\sss B} & \bm{a}_1 & \cdots & \bm{a}_{\sss J-1} &}  \bm{b}_{\sss J} & \\
\bm{X}_{\sss B} & \bm{X}_1 & \cdots & \bm{X}_{\sss J-1} & \bm{Y}_{\sss J}
\end{array}
\end{equation*}

\textbf{Actual design (step 2)}\vspace{2ex}
\begin{equation*}
\begin{array}{c c c c | c c}
\bovermat{Stage 1}{\bm{a}_{\sss B} & \bm{a}_1 & \cdots & \bm{a}_{\sss J-1} &}  \bm{a}_{\sss J} & \\
\bm{X}_{\sss B} & \bm{X}_1 & \cdots & \bm{X}_{\sss J-1} & \bm{X}_{\sss J}
\end{array}
\end{equation*}

\vspace{0pt}
Under the auxiliary design, the test statistic~$\tilde{T}_{I,J-1}$ is decomposed into the following first and second stage statistics, where $b_n \in I$ by design:
\begin{align*}
\tilde{T}_{I,J-1}^{(1)} & = \sum_{j = 0}^{J-1} \sum_{k = D_{j-1} + 1}^{D_{j}} \mathbbm{1}_{\{a_k \in I\}} \frac{X_k}{w_j^{(I)}} - \sum_{j = 0}^{J-1} \sum_{k = D_{0,j-1} + 1}^{D_{0,j}} \frac{X_{0k}}{w_j^{(0)}} \\[6pt]
\tilde{T}_{I,J-1}^{(2)} & = \sum_{k = D_{J-1} + 1}^{n} \mathbbm{1}_{\{b_k \in I \}} \frac{Y_k}{w_{J-1}^{(0)}} - \sum_{k = D_{0,J-1} + 1}^{n_0} \frac{X_{0k}}{w_{J-1}^{(0)}}
\end{align*}
where $w_j^{(0)}$ and $w_j^{(I)}$ have been defined in the previous steps for $j = 1, \ldots, J-1$. We want to select weights $w_J^{(I)}$ and $w_J^{(0)}$ so that \[ \tilde{T}_{I,J}^{(2)} = \sum_{k = D_{J-1}+1}^{n} \mathbbm{1}_{\{a_k \in I\}} \frac{X_k}{w_J^{(I)}}  -  \sum_{k = D_{0,J-1}+1}^{n_0} \frac{X_{0k}}{w_J^{(0)}} \] has the same conditional distribution as~$T_{I,J-1}^{(2)}$ under~$H_I$.  Let \[ \lambda_J  = \frac{m_{I,J}}{w_{J-1}^{(I)}} - \frac{m_{0,J}}{w_{J-1}^{(0)}} \, , \quad  \eta_J = \frac{m_{I,J}}{\left(w_{J-1}^{(I)}\right)^2} + \frac{m_{0,J}}{\left(w_{J-1}^{(0)}\right)^2} \]

If $m_{I,J} > 0$, then let $\tilde{m}_{I,J} = \sum_{k=D_{J-1}+1}^{n} \mathbbm{1}_{\{a_k \in I\}}$. As before, the weights $w_{J}^{(I)}$ and $w_{J}^{(0)}$ are given by \vspace{-16pt}
\begin{align*}
w_{J}^{(I)} & = f(\lambda_J, \eta_J, \tilde{m}_{I,J}, m_{0,J})\\
w_{J}^{(0)} &=  g(\lambda_l, w_{J}^{(I)}, \tilde{m}_{I,J}, m_{0,J})
\end{align*}

\noindent In this case, the final test statistic for testing hypothesis $H_I$ is as follows:\[
\tilde{T}_{I} = \sum_{j = 0}^{J} \sum_{k = D_{j-1} + 1}^{D_{j}} \mathbbm{1}_{\{a_k \in I\}} \frac{X_k}{w_j^{(I)}}  - 
\sum_{j = 0}^{J} \sum_{k = D_{0,j-1} + 1}^{D_{0,j}}\frac{X_{0k}}{w_j^{(0)}} \]
where \vspace{-24pt} \begin{alignat*}{2}
w_0^{(I)} & = n_I' \, , \quad w_0^{(0)} = n_0 \\
w_{j}^{(I)} & = f(\lambda_j, \eta_j, \tilde{m}_{I,j},  m_{0,j}) && \qquad (j = 1, \ldots, J)  \\
w_{j}^{(0)} & = g(\lambda_j, w_{j}^{(I)}, \tilde{m}_{I,j}, m_{0,j}) && \qquad (j = 1, \ldots, J)
\end{alignat*}
\noindent We reject $H_I$ if $\tilde{T}_I$ is greater than $z_{\alpha} ( 1/n_I' + 1/n_0')^{1/2} $. 

However, if $m_{I,J} = 0$ then we do not have enough degrees of freedom with a single weight $w_J^{(0)}$ to match both the conditional means and variances.  In this case, since by design $m_{0,J} = d_{0J} > 1$, we consider separately the first $n_{0,J,1}$ control observations and the next $n_{0,J,2}$ control observations, where $n_{0,J,1} > 0$, $n_{0,J,2}> 0$ and $n_{0,J,1} + n_{0,J,2} = d_{0J}$.  In order to keep the weights as close to the natural weight $n_0$ for as many of the control observations as possible, we recommend setting $n_{0,J,1} = d_{0J} - 1$ and $n_{0,J,2} = 1$, which is what we use for the simulation studies in Section~4.3 of the paper.

Letting $D_{0,J,1} = D_{0,J-1} + n_{0,J,1}$, we select weights $w_{J,1}^{(0)}$ and $w_{J,2}^{(0)}$ so that under~$H_I$, the statistic \[
\tilde{T}_{I,J}^{(2)} = - \sum_{k = D_{0,J-1}+1}^{D_{0,J,1}} \frac{X_{0k}}{w_{J,1}^{(0)}} - \sum_{k =D_{0,J,1}+1}^{n_0} \frac{X_{0k}}{w_{J,2}^{(0)}}\] has the same distribution as~$T_{I,J-1}^{(2)}$ conditional on the interim data $\mathcal{D}^{(1)}$. Under~$H_I$, we have \vspace{-6pt} \begin{align*}
& \tilde{T}_{I,J}^{(2)} | \mathcal{D}^{(1)} \sim N\!\left(- \mu \frac{n_{0,J,1}}{w_{J,1}^{(0)}} - \mu  \frac{n_{0,J,2}}{w_{J,2}^{(0)}} \, , \frac{n_{0,J,1}}{\left(w_{J,1}^{(0)}\right)^2} + \frac{n_{0,J,2}}{\left(w_{J,2}^{(0)}\right)^2} \right)
\end{align*}

\noindent Equating the conditional means and variances gives \vspace{-6pt} \begin{align*}
w_{J,1}^{(I)} & = \frac{ -n_{0,J,1} \lambda_J - [n_{0,J,1} n_{0,J,2} \{ \eta_J (n_{0,J,1} + n_{0,J,2})- \lambda_{J}^2 \}]^{1/2}}{\lambda_{J}^2 - n_{0,J,2} \eta_{J}} \\
w_{J,2}^{(I)} & = - \frac{n_{0,J,2} w_{J,1}^{(I)}}{n_{0,J,1} + \lambda_{J} w_{J,1}^{(I)} }
\end{align*}

\noindent In this case, the final test statistic for testing hypothesis $H_I$ is as follows:\[
\begin{split}
\tilde{T}_{I} = \sum_{j = 0}^{J-1} \sum_{k = D_{j-1} + 1}^{D_{j}} \mathbbm{1}_{\{a_k \in I\}} \frac{X_k}{w_j^{(I)}}  - \sum_{j = 0}^{J-1} \sum_{k = D_{0,j-1} + 1}^{D_{0,j}} \frac{X_{0k}}{w_j^{(0)}} \\ - \sum_{k = D_{0,J-1}+1}^{D_{0,J,1}} \frac{X_{0k}}{w_{J,1}^{(0)}} - \sum_{k = D_{0,J,1}+1}^{n_0} \frac{X_{0k}}{w_{J,2}^{(0)}}
\end{split} \]
\noindent We reject $H_I$ if $\tilde{T}_I$ is greater than $z_{\alpha} ( 1/n_I' + 1/n_0)^{1/2} $.


\section*{Appendix D: Derivation of the weights for familywise error control in block randomized response-adaptive trials with an  adaptive control allocation}
\label{Asec:block_with_control}
\doublespace

Let~$a_k = 0$ if the $k$th patient is allocated to the control and $n_0 = \sum_{k=1}^n  \mathbbm{1}_{\{a_k = 0\}}$ denote the total number of allocations to the control. The na\"ive $z$-test for~$H_I$ rejects $H_I$ if the test statistic \[ T_I = 
\sum_{k = 1}^{n} \left( \mathbbm{1}_{\{a_k \in I\}} \frac{X_k}{n_I} \right) - \sum_{k=1}^{n} \left( \mathbbm{1}_{\{a_k = 0\}} \frac{X_k}{n_0} \right) \] is greater than $z_{\alpha}\left( 1/n_I + 1/n_0\right)^{1/2}$.

The trial starts with a a burn-in period $B$, which allocates $r_0 > 0$ patients to the control and $r_i > 0$ patients to the $i$th treatment $(i = 1, \ldots, h)$, where $r_0$ and the $r_i$ are again fixed in advance. Hence a total of~$r = \sum_{i=0}^h r_i$ patients are allocated to the experimental treatments during the burn-in period.


The auxiliary design for hypothesis~$H_I$ starts with a burn-in period~$B$ with $r$ patients that is identical to the actual design. The subsequent~$n-r-2$ allocations are given by a fixed sequence $(b_{r+1},\ldots, b_{n-2})$. The allocation $b_{n-1}$ is to the control, while the allocation~$b_n$ must be in~$I$. For the auxiliary design, let $n_0'$ and $n_i'$ denote the total number of allocations to the control and the $i$th treatment respectively ($i = 1, \ldots, h$), including the burn-in period.

\subsection*{Step 1}

In step 1 we only consider the response-adaptive allocations for the first block $\bm{a}_1$. We view the auxiliary and actual trials as coming from a two-stage design, where the first stage for both is the burn-in period~$B$, as shown below.\\[-12pt]

\textbf{Auxiliary design (step 1)}\vspace{-1ex}
\begin{equation*}
\begin{array}{c | c c c c c}
\bm{a}_{\sss B} & \bovermat{Stage 2}{\bm{a}_1 & \bm{a}_2 & \cdots & \bm{a}_{\sss J} & }\\
\bm{X}_{\sss B} & \bm{X}_1 & \bm{X}_2 & \cdots & \bm{X}_{\sss J}
\end{array}
\end{equation*}

\textbf{Actual design (step 1)}\vspace{3ex}
\begin{equation*}
\begin{array}{c | c c c c c}
\bm{a}_{\sss B} & \bovermat{Modified stage 2}{\bm{a}_1 & \bm{b}_2 & \cdots & \bm{b}_{\sss J} & }\\
\bm{X}_{\sss B} & \bm{X}_1 & \bm{Y}_2 & \cdots & \bm{Y}_{\sss J}
\end{array}
\end{equation*}

\vspace{6pt}



\noindent Under the auxiliary two-stage design, the test statistic~$T_I = T_I^{(1)} + T_I^{(2)}$ for the experimenal treatments is decomposed into two parts, where $T_I^{(1)}$ is calculated from the first stage data and $T_I^{(2)}$ is calculated from the second stage data. More explicitly,  \begin{align*}
T_I^{(1)} & =  \sum_{k = 1}^{r} \left( \mathbbm{1}_{\{a_k \in I\}} \frac{Y_k}{n_I'} \right) -  \sum_{k = 1}^{r} \left( \mathbbm{1}_{\{a_k = 0\}} \frac{Y_k}{n_0'} \right) \\
T_I^{(2)} & =  \sum_{k = r+1}^{n} \left( \mathbbm{1}_{\{b_k \in I\}} \frac{Y_k}{n_I'} \right) -  \sum_{k = r+1}^{n} \left( \mathbbm{1}_{\{b_k = 0\}} \frac{Y_k}{n_0'} \right)
\end{align*}


We now select weights $w_{1}^{(I)}$ and $w_{1}^{(0)}$ so that under~$H_I$, the statistic \[
\tilde{T}_I^{(2)} = \sum_{k = r+1}^{D_1} \mathbbm{1}_{\{a_k \in I\}} \frac{X_k}{w_1^{(I)}} + \sum_{k = D_1 + 1}^{n} \mathbbm{1}_{\{b_k \in I\}} \frac{Y_k}{w_1^{(I)}} -  \sum_{k = r+1}^{D_1} \mathbbm{1}_{\{a_k = 0 \}} \frac{X_k}{w_1^{(0)}} - \sum_{k = D_1 + 1}^{n} \mathbbm{1}_{\{b_k = 0 \}} \frac{Y_k}{w_1^{(0)}} \]
has the same distribution as~$T_I^{(2)}$ conditional on the interim data $\mathcal{D}^{(1)}$. Under~$H_I$, we have \vspace{-16pt}
\begin{align*}
& T_I^{(2)} | \mathcal{D}^{(1)} \sim N\!\left(\mu \frac{m_{I,1}}{n_I'} - \mu \frac{m_{0,1}}{n_0'} \, , \frac{m_{I,1}}{(n_I')^2} + \frac{m_{0,1}}{(n_0')^2} \right) \\
& \tilde{T}_I^{(2)} | \mathcal{D}^{(1)} \sim N\!\left(\mu \frac{\tilde{m}_{I,1}}{w_1^{(I)}} - \mu  \frac{\tilde{m}_{0,1}}{w_1^{(0)}} \, , \frac{\tilde{m}_{I,1}}{\left(w_{1}^{(I)}\right)^2} + \frac{\tilde{m}_{0,1}}{\left(w_{1}^{(0)}\right)^2} \right)
\end{align*}
where \[
\tilde{m}_{I,1} = m_{I,2} + \sum_{k = r + 1}^{D_1} \mathbbm{1}_{\{a_k \in I\}}, \quad
\tilde{m}_{0,1} = m_{0,2} + \sum_{k = r + 1}^{D_1} \mathbbm{1}_{\{a_k = 0\}} .
\]

\noindent To match the conditional distributions we equate the conditional means and variances to give \vspace{-6pt}\begin{align*}
w_{1}^{(I)} & = \frac{\lambda_{1} \tilde{m}_{I,1} - [\tilde{m}_{I,1}\tilde{m}_{0,1} \{ \eta_{1} (\tilde{m}_{0,1} + \tilde{m}_{I,1}) - \lambda_{1}^2 \}]^{1/2} }{\lambda_{1}^2 - \tilde{m}_{0,1} \eta_{1}} \\
w_{1}^{(0)} & = \frac{\tilde{m}_{0,1} w_{1}^{(I)}}{\tilde{m}_{I,1} - \lambda_{1} w_{1}^{(I)} }
\end{align*}
\noindent where $\lambda_{1} = m_{I,1}/n_I' - m_{0,1}/n_0'$ and $\eta_{1} = m_{I,1}/(n_I')^2 + m_{0,1}/(n_0')^2$.

Hence the full modified statistic for the actual design in step~1 is \begin{equation*}
\begin{split}
\tilde{T}_{I,1} = \tilde{T}_I = \sum_{j = 0}^{1} \sum_{k = D_{j-1} + 1}^{D_{j}} \mathbbm{1}_{\{a_k \in I\}} \frac{X_k}{w_j^{(I)}} + \sum_{k = D_1 + 1}^{n} \mathbbm{1}_{\{b_k \in I\}} \frac{Y_k}{w_1^{(I)}} \\
- \sum_{j = 0}^{1} \sum_{k = D_{j-1} + 1}^{D_{j}} \mathbbm{1}_{\{a_k = 0 \}} \frac{X_k}{w_j^{(0)}} - \sum_{k = D_1 + 1}^{n} \mathbbm{1}_{\{b_k = 0 \}} \frac{Y_k}{w_1^{(0)}}
\end{split}
\end{equation*}
where we define $D_{-1} = 0$, $w_0^{(I)} = n_I'$ and $w_0^{(0)} = n_0'$. By the conditional invariance principle, $\tilde{T}_{I,1}$ is a valid test statistic for the actual design.

\subsection*{Inductive step}

We now repeat the process above, at each step taking forward the actual design as the new auxiliary design. The actual design at step~$l$ of the process ($l \in \{1, \ldots, J-1\}$) is a modification of the  new auxiliary design where the allocations $\bm{b}_l$ are set to $\bm{a}_l$. The valid test statistic for the new auxiliary design is $\tilde{T}_{I, l}$, taken forward from the previous step of the process. The diagrammatic representation of step~$l$ of the process is given below. \\[-12pt]

\textbf{Auxiliary design (step $l$)}\vspace{1ex}
\begin{equation*}
\begin{array}{c c c c| c c c c c}
\bovermat{Stage 1}{\bm{a}_{\sss B} & \bm{a}_1 & \cdots & \bm{a}_{l-1} &} \bovermat{Stage 2}{\bm{b}_l & \bm{b}_{l+1} & \cdots & \bm{b}_{\sss J} & }\\
\bm{X}_{\sss B} & \bm{X}_1 & \cdots & \bm{X}_{l-1} & \bm{Y}_l & \bm{Y}_{l+1} & \cdots & \bm{Y}_{\sss J}
\end{array}
\end{equation*}

\newpage

\textbf{Actual design (step $l$)}\vspace{1ex}
\begin{equation*}
\begin{array}{c c c c| c c c c c}
\bovermat{Stage 1}{\bm{a}_{\sss B} & \bm{a}_1 & \cdots & \bm{a}_{l-1} &} \bovermat{Modified stage 2}{\bm{a}_l & \bm{b}_{l+1} & \cdots & \bm{b}_{\sss J} & }\\
\bm{X}_{\sss B} & \bm{X}_1 & \cdots & \bm{X}_{l-1} & \bm{X}_l & \bm{Y}_{l+1} & \cdots & \bm{Y}_{\sss J}
\end{array}
\end{equation*}

Using these auxiliary and actual designs, we select new weights $w_{l}^{(I)}$ and $w_{l}^{(0)}$ so that under~$H_I$ the conditional distributions of the second stage statistic are the same. This yields the new test statistic $\tilde{T}_{I, l}$ for the actual design in step~$l$.  

For notational convenience, we introduce the following functions: \begin{align*}
f(\lambda, \eta, m_I, m_0) & = \frac{\lambda m_I - [m_I m_0 \{ \eta (m_0  + m_I) - \lambda^2 \}]^{1/2} }{\lambda - m_0 \eta} \\
g(\lambda, w, m_I, m_0) & = \frac{m_0 w}{m_I - \lambda w }
\end{align*}

\noindent Given the weights $w_{l-1}^{(I)}$ and $w_{l-1}^{(0)}$ found in the previous step, let\begin{align*}
\lambda_l & = \frac{m_{I,l}}{w_{l-1}^{(I)}} - \frac{m_{0,l}}{w_{l-1}^{(0)}} \, , \quad  \eta_l = \frac{m_{I,l}}{\left(w_{l-1}^{(I)}\right)^2} + \frac{m_{0,l}}{\left(w_{l-1}^{(0)}\right)^2}\\[6pt]
\tilde{m}_{I,l} & = m_{I,l+1} + \sum_{k=D_{l-1}+1}^{D_l} \mathbbm{1}_{\{a_k \in I\}} \\[12pt]
\tilde{m}_{0,l} & = m_{0,l+1} + \sum_{k=D_{l-1}+1}^{D_l} \mathbbm{1}_{\{a_k = 0 \}} 
\end{align*}

\noindent The weights $w_{l}^{(I)}$ and $w_{l}^{(0)}$ are given by \vspace{-12pt}
\begin{align*}
w_{l}^{(I)} & = f(\lambda_l, \eta_l, \tilde{m}_{I,l}, \tilde{m}_{0,l})\\
w_{l}^{(0)} &=  g(\lambda_l, w_{l}^{(I)}, \tilde{m}_{I,l}, \tilde{m}_{0,l})
\end{align*}

\noindent The corresponding test statistic $\tilde{T}_{I,l}$ is:
\begin{equation*}
\begin{split}
\tilde{T}_{I,l} = \sum_{j = 0}^{l} \sum_{k = D_{j-1} + 1}^{D_{j}} \mathbbm{1}_{\{a_k \in I\}} \frac{X_k}{w_j^{(I)}} + \sum_{k = D_l + 1}^{n} \mathbbm{1}_{\{b_k \in I\}} \frac{Y_k}{w_l^{(I)}} \\
- \sum_{j = 0}^{l} \sum_{k = D_{j-1} + 1}^{D_{j}} \mathbbm{1}_{\{a_k = 0 \}} \frac{X_k}{w_j^{(0)}} - \sum_{k = D_l + 1}^{n} \mathbbm{1}_{\{b_k = 0 \}} \frac{Y_k}{w_l^{(0)}} \\[12pt]
\end{split} 
\end{equation*}

\subsection*{Final step}

%
%
%

Under the auxiliary design, the test statistic~$\tilde{T}_{I,J-1}$ is decomposed into the following first and second stage statistics, where $b_{n-1} = 0$ and $b_n \in I$ by design:
\begin{align*}
\tilde{T}_{I,J-1}^{(1)} & = \sum_{j = 0}^{J-1} \sum_{k = D_{j-1} + 1}^{D_{j}} \mathbbm{1}_{\{a_k \in I\}} \frac{X_k}{w_j^{(I)}} - \sum_{j = 0}^{J-1} \sum_{k = D_{j-1} + 1}^{D_{j}} \mathbbm{1}_{\{a_k = 0 \}} \frac{X_k}{w_j^{(0)}} \\[6pt]
\tilde{T}_{I,J-1}^{(2)} & = \sum_{k = D_{J-1} + 1}^{n} \mathbbm{1}_{\{b_k \in I \}} \frac{Y_k}{w_{J-1}^{(0)}} - \sum_{k = D_{J-1} + 1}^{n} \mathbbm{1}_{\{b_k = 0 \}} \frac{Y_k}{w_{J-1}^{(0)}}
\end{align*}
where $w_{j}^{(0)}$ and $w_j^{(I)}$ have been defined in the previous steps for $j = 1, \ldots, J-1$. We want to select weights $w_J^{(I)}$ and $w_J^{(0)}$ so that \[ \tilde{T}_{I,J}^{(2)} = \sum_{k = D_{J-1}+1}^{n} \mathbbm{1}_{\{a_k \in I\}} \frac{X_k}{w_J^{(I)}}  -  \sum_{k = D_{J-1}+1}^{n} \mathbbm{1}_{\{a_k = 0 \}} \frac{X_k}{w_J^{(0)}} \] has the same conditional distribution as~$T_{I,J-1}^{(2)}$ under~$H_I$.  For notational convenience, let \[
\lambda_J = \frac{m_{I,J}}{w_{J-1}^{(I)}} - \frac{m_{0,J}}{w_{J-1}^{(0)}} \, , \quad  \eta_J = \frac{m_{I,J}}{\left(w_{J-1}^{(I)}\right)^2} + \frac{m_{0,J}}{\left(w_{J-1}^{(0)}\right)^2}\]

If $m_{I,J} > 0$ and $m_{0,J} > 0$, then let \begin{align*}
\tilde{m}_{I,J} & = \sum_{k=D_{J-1}+1}^{n} \mathbbm{1}_{\{a_k \in I\}} \, , \quad \tilde{m}_{0,J}  = \sum_{k=D_{J-1}+1}^{n} \mathbbm{1}_{\{a_k = 0 \}}.
\end{align*}

\noindent The weights $w_{J}^{(I)}$ and $w_{J}^{(0)}$ are given by \vspace{-12pt}
\begin{align*}
w_{J}^{(I)} & = f(\lambda_J, \eta_J, \tilde{m}_{I,J}, \tilde{m}_{0,J})\\
w_{J}^{(0)} &=  g(\lambda_l, w_{J}^{(I)}, \tilde{m}_{I,J}, \tilde{m}_{0,J})
\end{align*}

\noindent In this case, the final test statistic for testing hypothesis $H_I$ is as follows:\[
\tilde{T}_{I} = \sum_{j = 0}^{J} \sum_{k = D_{j-1} + 1}^{D_{j}} \mathbbm{1}_{\{a_k \in I\}} \frac{X_k}{w_j^{(I)}}  - 
\sum_{j = 0}^{J} \sum_{k = D_{j-1} + 1}^{D_{j}} \mathbbm{1}_{\{a_k = 0 \}} \frac{X_k}{w_j^{(0)}} \]
where \vspace{-24pt} \begin{alignat*}{2}
w_0^{(I)} & = n_I' \, , \quad w_0^{(0)} = n_0' \\
w_{j}^{(I)} & = f(\lambda_j, \eta_j, \tilde{m}_{I,j}, \tilde{m}_{0,j}) && \qquad (j = 1, \ldots, J)  \\
w_{j}^{(0)} & = g(\lambda_j, w_{j}^{(I)}, \tilde{m}_{I,j}, \tilde{m}_{0,j}) && \qquad (j = 1, \ldots, J)
\end{alignat*}
\noindent We reject $H_I$ if $\tilde{T}_I$ is greater than $z_{\alpha} \left( 1/n_I' + 1/n_0'\right)^{1/2} $. 

However, if $m_{I,J} = 0$ and $m_{0,J} > 1$, then we do not have enough degrees of freedom with a single weight $w_J^{(0)}$ to match both the conditional means and variances.  In this case, we consider separately the first $n_{0,J,1}$ control observations and the next $n_{0,J,2}$ control observations, where $n_{0,J,1} > 0$, $n_{0,J,2}> 0$ and $n_{0,J,1} + n_{0,J,2} = m_{0,J}$. As before, we recommend setting $n_{0,J,1} = m_{0,J}-1$ and $n_{0,J,2} = 1$, which is what we use for the simulation studies in Section~\ref{Asubsec:block_rand_error}.

Suppose the $(D_{0,J,1})$th patient receives the $(n_{0,J,1})$th allocation to the control in block~$J$. We select weights $w_{J,1}^{(0)}$ and $w_{J,2}^{(0)}$ so that under~$H_I$, the statistic \[
\tilde{T}_{I,J}^{(2)} = - \sum_{k = D_{J-1}+1}^{D_{0,J,1}} \mathbbm{1}_{\{a_k = 0\}} \frac{X_k}{w_{J,1}^{(0)}} - \sum_{k = D_{0,J,1}+1}^{n} \mathbbm{1}_{\{a_k= 0\}} \frac{X_k}{w_{J,2}^{(0)}}\] has the same distribution as~$T_{I,J-1}^{(2)}$ conditional on the interim data $\mathcal{D}^{(1)}$. Under~$H_I$, we have \vspace{-6pt}\begin{align*}
& \tilde{T}_{I,J}^{(2)} | \mathcal{D}^{(1)} \sim N\!\left(- \mu \frac{n_{0,J,1}}{w_{J,1}^{(0)}} - \mu  \frac{n_{0,J,2}}{w_{J,2}^{(0)}} \, , \frac{n_{0,J,1}}{\left(w_{J,1}^{(0)}\right)^2} + \frac{n_{0,J,2}}{\left(w_{J,2}^{(0)}\right)^2} \right)
\end{align*}

\noindent Equating the conditional means and variances gives \vspace{-6pt} \begin{align*}
w_{J,1}^{(I)} & = \frac{ -n_{0,J,1} \lambda_J - [n_{0,J,1} n_{0,J,2} \{ \eta_J (n_{0,J,1} + n_{0,J,2}) - \lambda_{J}^2 \}]^{1/2}}{\lambda_{J}^2 - n_{0,J,2} \eta_{J}} \\
w_{J,2}^{(I)} & = - \frac{n_{0,J,2} w_{J,1}^{(I)}}{n_{0,J,1} + \lambda_{J} w_{J,1}^{(I)} }
\end{align*}

\noindent In this case, the final test statistic for testing hypothesis $H_I$ is as follows:\[
\begin{split}
\tilde{T}_{I} = \sum_{j = 0}^{J-1} \sum_{k = D_{j-1} + 1}^{D_{j}} \mathbbm{1}_{\{a_k \in I\}} \frac{X_k}{w_j^{(I)}}  - \sum_{j = 0}^{J-1} \sum_{k = D_{j-1} + 1}^{D_{j}} \mathbbm{1}_{\{a_k = 0 \}} \frac{X_k}{w_j^{(0)}} \\ - \sum_{k = D_{J-1}+1}^{D_{0,J,1}} \mathbbm{1}_{\{a_k = 0\}} \frac{X_k}{w_{J,1}^{(0)}} - \sum_{k = D_{0,J,1}+1}^{n} \mathbbm{1}_{\{a_k = 0\}} \frac{X_k}{w_{J,2}^{(0)}}
\end{split} \]
\noindent We reject $H_I$ if $\tilde{T}_I$ is greater than $z_{\alpha} \left( 1/n_I' + 1/n_0'\right)^{1/2} $.

If $m_{0,J} = 0$ and $m_{I,J} > 1$, then we do not have enough degrees of freedom with a single weight $w_J^{(I)}$ to match both the conditional means and variances.  In this case, we consider separately the first $n_{I,J,1}$ and the next $n_{I,J,2}$ observations from treatments in~$I$, where $n_{I,J,1} > 0$, $n_{I,J,2} > 0$ and $n_{I,J,1} + n_{I,J,2} = m_{I,J}$. We recommend setting $n_{I,J,1} = m_{I,J}-1$ and $n_{I,J,2} = 1$, which is used for the simulation studies in Section~\ref{Asubsec:block_rand_error}.

Suppose the $(D_{I,J,1})$th patient receives the $(n_{I,J,1})$th allocation to a treatment in~$I$ in block~$J$. We select weights $w_{J,1}^{(I)}$ and $w_{J,2}^{(I)}$ so that under~$H_I$, the statistic \[
\tilde{T}_{I,J}^{(2)} = \sum_{k = D_{J-1}+1}^{D_{I,J,1}} \mathbbm{1}_{\{a_k \in I\}} \frac{X_k}{w_{J,1}^{(I)}} + \sum_{k = D_{I,J,1}+1}^{n} \mathbbm{1}_{\{a_k \in I\}} \frac{X_k}{w_{J,2}^{(I)}}\] has the same distribution as~$T_{I,J-1}^{(2)}$ conditional on the interim data $\mathcal{D}^{(1)}$. Under~$H_I$, we have \vspace{-6pt} \begin{align*}
& T_{I,J-1}^{(2)} | \mathcal{D}^{(1)} \sim N\!\left(\mu \frac{m_{I,J}}{w_{J-1}^{(I)}} - \mu \frac{m_{0,J}}{w_{J-1}^{(0)}} \, , \frac{m_{I,J}}{\left(w_{J-1}^{(I)}\right)^2} + \frac{m_{0,J}}{\left(w_{J-1}^{(0)}\right)^2} \right) \\
& \tilde{T}_{I,J}^{(2)} | \mathcal{D}^{(1)} \sim N\!\left(\mu \frac{n_{I,J,1}}{w_{J,1}^{(I)}} + \mu  \frac{n_{I,J,2}}{w_{J,2}^{(I)}} \, , \frac{n_{I,J,1}}{\left(w_{J,1}^{(I)}\right)^2} + \frac{n_{I,J,2}}{\left(w_{J,2}^{(I)}\right)^2} \right)
\end{align*}

\noindent Equating the conditional means and variances gives \begin{align*}
w_{J,1}^{(I)} & = \frac{ n_{I,J,1} \lambda_J - [ n_{I,J,1}  n_{I,J,2} \{ \eta_J (n_{I,J,1}  + n_{I,J,2}) - \lambda_{J}^2 \}]^{1/2}}{\lambda_{J}^2 - n_{I,J,2} \eta_{J}} \\
w_{J,2}^{(I)} & = \frac{n_{I,J,2} w_{J,1}^{(I)}}{\lambda_{J} w_{J,1}^{(I)} - n_{I,J,1} }
\end{align*}

\noindent In this case, the final test statistic for testing hypothesis $H_I$ is as follows:\[
\begin{split}
\tilde{T}_{I} = \sum_{j = 0}^{J-1} \sum_{k = D_{j-1} + 1}^{D_{j}} \mathbbm{1}_{\{a_k \in I\}} \frac{X_k}{w_j^{(I)}}  - \sum_{j = 0}^{J-1} \sum_{k = D_{j-1} + 1}^{D_{j}} \mathbbm{1}_{\{a_k = 0 \}} \frac{X_k}{w_j^{(0)}} \\ + \sum_{k = D_{J-1}+1}^{D_{I,J,1}} \mathbbm{1}_{\{a_k \in I\}} \frac{X_k}{w_{J,1}^{(I)}} + \sum_{k = D_{I,J,1}+1}^{n} \mathbbm{1}_{\{a_k \in I\}} \frac{X_k}{w_{J,2}^{(I)}}
\end{split} \]
\noindent We reject $H_I$ if $\tilde{T}_I$ is greater than $z_{\alpha} \left( 1/n_I' + 1/n_0'\right)^{1/2} $. 

Finally, if $\max(m_{0,J}, m_{I,J}) \leq 1$ and $\min(m_{0,J}, m_{I,J}) = 0$ then we cannot match the conditional means and variances and hence out adaptive procedure fails to give a valid test statistic. However, such a scenario is unlikely given reasonably large block sizes and a minimum allocation probability to the control, for example. In our simulation study in Section~\ref{Asubsec:block_rand_error}, this scenario was never observed.

The weights for the final block may not be real-valued and hence the procedure can fail to give a valid test statistic. It can also happen that the weights for the experimental treatment are negative. In this case, our procedure no longer necessarily controls the FWER for the composite null hypotheses~$H_i: \delta_i \leq 0$, but only the point null hypotheses~$H_i: \delta_i = 0$. Hence the adaptive test that allows for response-adaptive allocation to the control does so at the cost of being less robust and flexible. Appendix~E.1 gives some simulation results to illustrate these issues.

\section*{Appendix E: Additional simulation results}
\label{Asec:additional_simul}

\renewcommand\thesubsection{E.\arabic{subsection}}

\subsection{Block randomization with an adaptive allocation to the control}
\label{Asubsec:block_rand_error}
\doublespace

We consider block randomization with an adaptive control allocation, as presented in Section~3.4 in the paper. We use the setup of a trial with $J = 3$ blocks and sizes (50, 50, 50). In the burn-in period, 5~patients are allocated to each of the treatments including the control. We again set the true control mean $\mu = 0$, and~$\alpha = 0.05$.

\textit{Type I error inflator}:
The allocation probabilities for block $j \in \{1, \ldots, J-1\}$, patient $k =  D_j + 1, \ldots, D_{j+1}$ and treatment $l \in \{0, 2, \ldots, h\}$ are:
\begin{align*}
P(a_k = 1) & = \begin{cases}
0 & \text{if } \; \sum_{i = 1}^{D_j} \mathbbm{1}_{\{a_i = 1\}} \frac{X_i}{n_{1,j}} > 0.5 \\
1 & \text{otherwise}
\end{cases} \\
P(a_k = l) & = \begin{cases}
1/h & \text{if } \; \sum_{i = 1}^{D_j} \mathbbm{1}_{\{a_i = 1\}} \frac{X_i}{n_{1,j}} > 0.5 \\
0 & \text{otherwise}
\end{cases}
\end{align*}

\noindent where  $n_{1,j} = \sum_{i=1}^{D_j} \mathbbm{1}_{\{a_i = 1\}}$.

\textit{Bayesian adaptive randomization}:
%
%
The priors and posteriors are the same as in Section~4.3 in the paper, and we use a similar Bayesian adaptive randomization scheme. If there are $h$ experimental treatments, then the randomization probabilities $(\pi_0, \pi_1, \ldots, \pi_h)$ at the~$(j+1)$th stage are: \[
\pi_i \propto \begin{cases}
\frac{P(\mu_i > \mu_0 \mid X_1 = x_1, \ldots, X_{\sss D_j}= x_{\sss D_j})^{\gamma}}{ \sum_{l=1}^h P(\mu_l > \mu_0 \mid X_1 = x_1, \ldots, X_{\sss D_j}= x_{\sss D_j})^{\gamma}} \qquad (i = 1, \ldots, h) \vspace{12pt}\\ 
\frac{1}{h} \exp \left( \max(\hat{m}_{1j},\hat{m}_{2j}, \ldots, \hat{m}_{hj}) - \hat{m}_{0j}\right)^{\nu} \qquad \quad (i = 0)
\end{cases}
\] where $\hat{m}_{ij}$ is the current arm-specific sample size for the $i$th treatment at the end of the $j$th stage. In our simulations, for simplicity we set the priors $\mu_{i,0} = 0$ and $\sigma_{i,0}^2 = 1$, while $\gamma = 0.5$ and $\nu = 0.1$.

\textit{Simulation results}:
%
Table~\ref{Atab:typeI_inflator_block2} gives the results for the type~I error inflator randomization scheme, while Table~\ref{Atab:BAR_block2} gives the result for BAR. For each scenario, we ran $10^5$ simulated trials. The auxiliary designs in all scenarios were random draws from a discrete uniform distribution on $\{0, 1, \ldots, h\}$. 

\begin{table}[ht!]

\caption{\label{Atab:typeI_inflator_block2}   \textit{Familywise error rate and disjunctive power for the type~I error inflator, for block randomization with an adaptive control allocation. There were $10^5$ simulated trials for each set of parameter values.}}

\resizebox{\linewidth}{!} {
\begin{tabular}{p{0.5ex} l c c p{0cm} c c p{0cm} c c p{0cm} c c p{0cm} cc}
		\\ \hline \hline \Tstrut
		& & \multicolumn{2}{c}{Adaptive closed test} & & \multicolumn{2}{c}{Adaptive test (Holm)}
		& & \multicolumn{2}{c}{Closed $z$-test}  & &  \multicolumn{2}{c}{$z$-test (Holm)} 
		& & \multicolumn{2}{c}{$z$-test (Bonferroni)} \\
		\cline{3-4} \cline{6-7} \cline{9-10} \cline{12-13} \cline{15-16}\\ [-2ex]
		& {Parameter values} & Error & Power & & Error & Power
		& & Error & Power & & Error & Power & & Error & Power \\ \hline \Tstrut
		1.\ & $\delta_1 = \delta_2 = 0$
		  & 3.8 & - & & 4.9 & - & & 4.4 & - & & \textbf{6.7} & - & & \textbf{6.7} & -  \\ [1.5ex]
   		 
   		2.\ &  $\delta_1 = 0$, $\delta_2 = 1$ 
   		& 4.8 & 19.1 & & 3.7 & 25.1 & & \textbf{8.2} & 25.4 & & \textbf{7.8} & {67.2} & & 4.3 & 67.1 \\ [1.5ex]
		
		3.\ & $\delta_1 = \delta_2 = 0.5$
       & - & 90.1 & & - & 84.4 & & - & {93.3} & & - & 89.9 & & - & 89.9 \\ [1.5ex]	
   		
		4.\ &  $\delta_1 = \delta_2 = \delta_3 = 0$ 
		& 3.2 & - & & 4.1 & - & & 3.9 & - & & \textbf{6.2} & - & & \textbf{6.2} & -\\ [1.5ex]
   		   		   		
   		5.\ &  $\delta_1 = \delta_2 = 0$, $\delta_3 = 1$
   		& 3.8 & 14.0 & & 4.4 & 21.9 & & 4.8 & 20.0 & & \textbf{6.5} & {61.7} & & 4.8 & 61.6 \\ [1.5ex]
   		
   		6.\ &  $\delta_1 = 0$, $\delta_2 = \delta_3 = 1$ 
   		& 4.8 & 19.1 & & 3.4 & 24.6 & & \textbf{8.4} & 26.4 & & \textbf{7.5} &{80.6} & & 3.2 & {80.6} \\ [1.5ex]
		 
		7.\ &  $\delta_1 = 0$, $\delta_2 = 0.5$, $\delta_3 = 1$ 
		& 4.5 & 17.0 & & 3.0 & 22.6 &  & \textbf{8.0} & 23.7 & & \textbf{6.6} & {66.9} & & 2.9 & 66.8 \\ [1.5ex]
		
   		8.\ &  $\delta_1 = \delta_2 = \delta_3 = 0.5$
   		& - & 87.5 & & - & 78.4 & & - & {91.8} & & - & 86.9 & & - &  86.9 \\[1ex] \hline \\
		\end{tabular}
		}
		

\end{table}

\begin{table}[ht!]

\caption{\label{Atab:BAR_block2} \textit{Familywise error rate and disjunctive power for BAR, for block randomization with an adaptive control allocation. There were $10^5$ simulated trials for each set of parameter values.}}	

\resizebox{\linewidth}{!} {
\begin{tabular}{p{0.5ex} l c c p{0cm} c c p{0cm} c c p{0cm} c c p{0cm} cc}
		\\ \hline \hline \Tstrut
		& & \multicolumn{2}{c}{Adaptive closed test} & & \multicolumn{2}{c}{Adaptive test (Holm)}
		& & \multicolumn{2}{c}{Closed $z$-test}  & &  \multicolumn{2}{c}{$z$-test (Holm)} 
		& & \multicolumn{2}{c}{$z$-test (Bonferroni)} \\
		\cline{3-4} \cline{6-7} \cline{9-10} \cline{12-13} \cline{15-16}\\ [-2ex]
		& {Parameter values} & Error & Power & & Error & Power
		& & Error & Power & & Error & Power & & Error & Power \\ \hline \Tstrut
		1.\ & $\delta_1 = \delta_2 = 0$
		  & 4.6 & - & & 4.5 & - & & 4.6 & - & & 4.4 & - & & 4.4 & -  \\ [1.5ex]
   		 
   		2.\ &  $\delta_1 = 0$, $\delta_2 = 0.5$ 
   		& 5.0 & 54.4 & & 4.9 & 76.4 & & 4.8 & 56.1 & & 4.7 & {78.0} & & 2.4 & {78.0} \\ [1.5ex]
		
		3.\ & $\delta_1 = \delta_2 = 0.5$
       & - & 90.6 & & - & 87.4 & & - & {91.5} & & - & 88.3 & & - & 88.3 \\ [1.5ex]	
   		
		4.\ &  $\delta_1 = \delta_2 = \delta_3 = 0$ 
		& 4.0 & - & & 4.3 & - & & 3.9 & - & & 4.2 & - & & 4.2 & - \\ [1.5ex]
   		   		   		
   		5.\ &  $\delta_1 = \delta_2 = 0$, $\delta_3 = 0.5$
   		& 4.6 & 29.0 & & 4.5 & 61.4 & & 4.6 & 29.9 & & 4.4 & {62.9} & & 3.1 & {62.9} \\ [1.5ex]
   		
   		6.\ &  $\delta_1 = 0$, $\delta_2 = \delta_3 = 0.5$ 
   		& 4.9 & 56.4 & & 4.5 & 76.1 & & 4.7 & 57.4 & & 4.4 & {77.3} & & 1.6 & {77.3}\\ [1.5ex]
		 
		7.\ &  $\delta_1 = 0$, $\delta_2 = 0.25$, $\delta_3 = 0.5$
		& 4.5 & 41.7 & & 3.6 & 62.9 & & 4.3 & 42.7 & & 3.5 & {63.7} & & 1.7 & {63.7} \\ [1.5ex]

   		8.\ &  $\delta_1 = \delta_2 = \delta_3 = 0.5$ 
   		& - & 85.9 & & - & 82.0 & & - &  {86.9} & & - & 83.0 & & - & 83.0 \\[1ex] \hline
		\end{tabular}
		}
		
		
\end{table}

The results here are again broadly similar to those for the fully sequential setting, and the block randomization setting with a fixed allocation to the control. For the type~I error inflator, the various $z$-tests do not strongly control the FWER. The adaptive tests do achieve strong error control, but this comes at the cost of a very large decrease in power when compared with the Holm $z$-test.

For the BAR scheme, again all methods strongly control the FWER. This time, the $z$-tests have the highest power. When at least one null hypothesis is true the Holm $z$-test has the highest power, although there is only a small gain compared to the Holm adaptive test. When all null hypotheses are false, the closed $z$-test has a slightly higher power than the closed adaptive test.

With an adaptive control allocation, the weights of the adaptive test can become imaginary, or negative for the experimental treatments. In the former case, we set the modified test statistics $\tilde{T}_I = -\infty$ and do not reject the null hypothesis~$H_I$, which will preserve the FWER at the cost of lower power. In the latter case, we cannot use the adaptive test for the composite null hypotheses $H_i : \delta_i \leq 0$, although it will still be a valid test for the point null hypotheses $H_i : \delta_i = 0$.

We also considered how often at least one imaginary or negative weight occurs over the $10^5$ simulations for the two randomization schemes. Tables~\ref{Atab:typeI_inflator_block_error} and~\ref{Atab:BAR_block_error} give the percentage of simulations where the weights for the experimental treatments are imaginary or negative, for the type I error inflator and BAR scheme respectively.

\begin{table}[ht!]
\centering
\caption{\label{Atab:typeI_inflator_block_error} \textit{Percentage of simulations where at least one imaginary or negative weight occurs for the type~I error inflator, with $10^5$ simulated trials for each set of parameter values.}}

	\begin{tabular}{p{0.5ex} l c c p{0cm} c c }
	\\ \hline \hline \Tstrut
		& & \multicolumn{2}{c}{{Adaptive closed test}} & & \multicolumn{2}{c}{{Adaptive test (Holm)}} \\
		\cline{3-4}  \cline{6-7} \Tstrut
		& {Parameter values} & Imaginary & Negative & & Imaginary & Negative \\ \hline \Tstrut
		1.\ & $\delta_1 = \delta_2 = 0$
		  & 0.09 & 0.00 & & 0.09 & 0.00 \\ [1.5ex]
   		 
   		2.\ &  $\delta_1 = 0$, $\delta_2 = 1$ 
   		& 0.09 & 0.00 & & 0.09 & 0.00 \\ [1.5ex]
		
		3.\ & $\delta_1 = \delta_2 = 0.5$
       & 0.07 & 0.00 & & 0.07 & 0.00  \\ [1.5ex]	
   		
		4.\ &  $\delta_1 = \delta_2 = \delta_3 = 0$ 
		& 6.81 & 0.00 & & 0.18 & 0.00 \\ [1.5ex]
   		   		   		
   		5.\ &  $\delta_1 = \delta_2 = 0$, $\delta_3 = 1$
   		& 6.86 & 0.00 & & 0.17 & 0.00  \\ [1.5ex]
   		
   		6.\ &  $\delta_1 = 0$, $\delta_2 = \delta_3 = 1$ 
   		& 6.84 & 0.00 & & 0.17 & 0.00 \\ [1.5ex]
		 
		7.\ &  $\delta_1 = 0$, $\delta_2 = 0.5$, $\delta_3 = 1$ 
		& 6.89 & 0.00 & & 0.17 & 0.00 \\ [1.5ex]
		
   		8.\ &  $\delta_1 = \delta_2 = \delta_3 = 0.5$
   		& 2.67 & 0.00 & & 0.53 & 0.00  \\[1ex] \hline \\[12pt]
		\end{tabular}
\end{table}


\begin{table}[ht!]

\centering
	\caption{\label{Atab:BAR_block_error} \textit{Percentage of simulations where at least one imaginary or negative weight occurs for BAR, with $10^5$ simulated trials for each set of parameter values.}}
	
	\begin{tabular}{p{0.5ex} l c c p{0cm} c c }
	\\ \hline \hline \Tstrut
		& & \multicolumn{2}{c}{{Adaptive closed test}} & & \multicolumn{2}{c}{{Adaptive test (Holm)}} \\
		\cline{3-4}  \cline{6-7} \Tstrut
		& {Parameter values} & Imaginary & Negative & & Imaginary & Negative \\ \hline \Tstrut
		1.\ & $\delta_1 = \delta_2 = 0$
		& 0.10 & 0.09 & & 0.10 & 0.09 \\ [1.5ex]
   		 
   		2.\ &  $\delta_1 = 0$, $\delta_2 = 1$ 
   		& 0.14 & 0.10 & & 0.14 & 0.10  \\ [1.5ex]
		
		3.\ & $\delta_1 = \delta_2 = 0.5$
       & 0.00 & 0.08 & & 0.00 & 0.08  \\ [1.5ex]	
   		
		4.\ &  $\delta_1 = \delta_2 = \delta_3 = 0$ 
		& 0.29 & 0.12 & & 0.29 & 0.12 \\ [1.5ex]
   		   		   		
   		5.\ &  $\delta_1 = \delta_2 = 0$, $\delta_3 = 1$
   		& 0.35 & 0.07 & & 0.34 & 0.07  \\ [1.5ex]
   		
   		6.\ &  $\delta_1 = 0$, $\delta_2 = \delta_3 = 1$ 
   		& 0.25 & 0.08 & & 0.25 & 0.08 \\ [1.5ex]
		 
		7.\ &  $\delta_1 = 0$, $\delta_2 = 0.5$, $\delta_3 = 1$ 
		& 0.20 & 0.11 & & 0.19 & 0.11 \\ [1.5ex]
		
   		8.\ &  $\delta_1 = \delta_2 = \delta_3 = 0.5$
   		& 0.01 & 0.08 & & 0.01 & 0.08  \\[1ex] \hline
		\end{tabular}

\end{table}

For the type~I error inflator, there are essentially no negative weights for either adaptive test. The Holm adaptive tests has imaginary weights less than than $0.6\%$ of the time in all scenarios. In contrast, when there are three treatments, the adaptive closed tests can have up to $7\%$ of the simulations having imaginary weights. This shows that for more extreme randomization schemes, the adaptive closed test is not very robust, and is much less robust than the Holm adaptive test. For BAR, the percentage of imaginary weights is less than $0.4\%$ for either adaptive test. This time there are negative weights in some simulations, but the percentage is very low, at less than $0.2\%$.

\clearpage

\subsection{Power of the adaptive test}
\label{Asubsec:power_adaptive}


The adaptive tests can pay a large price in terms of power when compared with the $z$-tests, as seen in the results for the type~I error inflator. In order to understand what is happening in this setting, we conducted an additional simulation study. Suppose we are testing $h = 2$ treatments, and that the randomization scheme used is simply a fixed allocation to the experimental treatments, but with unequal randomization probabilities. Let $p_2$~denote the probability of assignment to treatment~2.

\begin{figure}[ht!]
\centering

\includegraphics[scale=0.75]{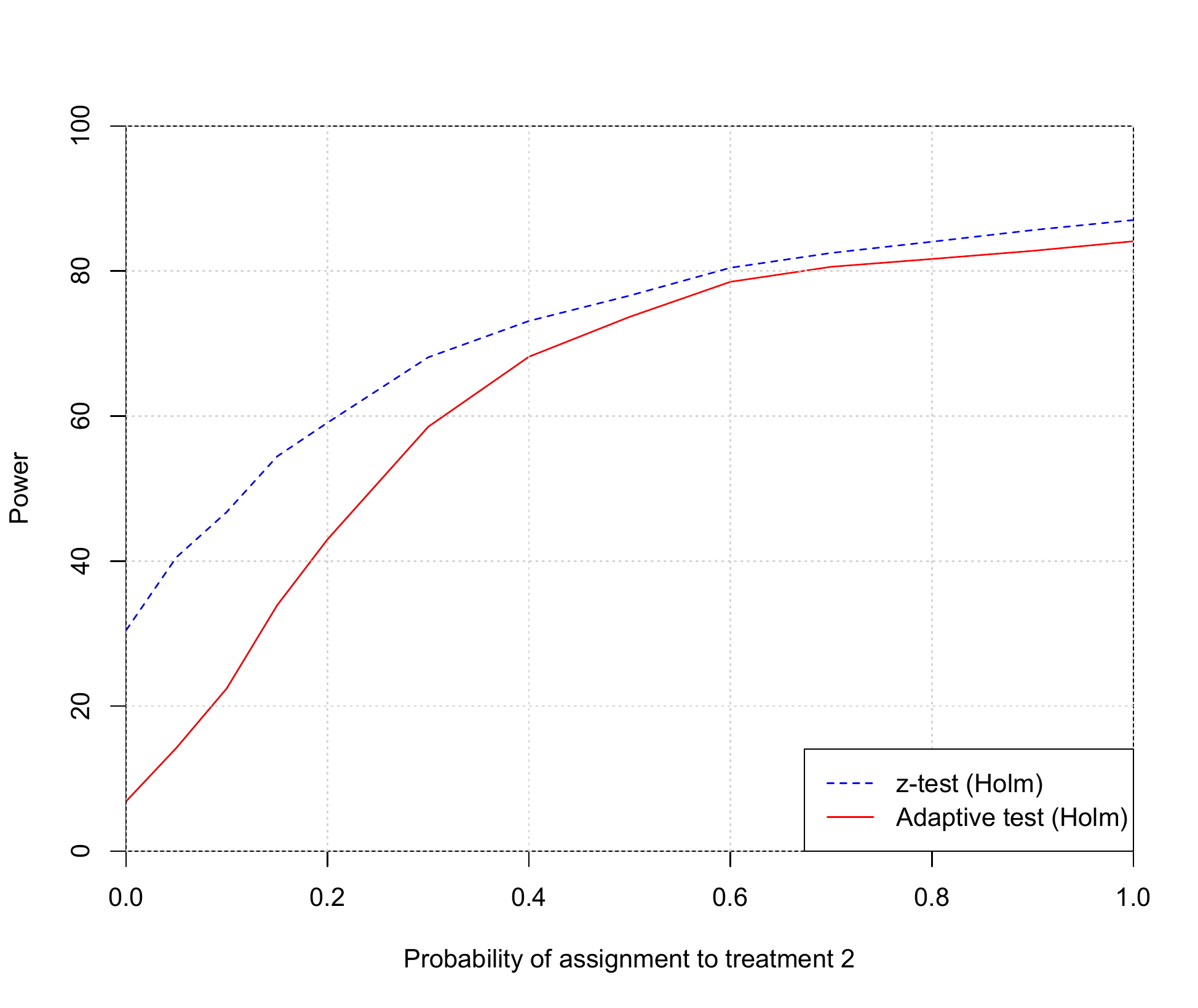}

\caption{\label{fig:unequal_rand_seq} \textit{Power of the Holm adaptive test and Holm $z$-test as a function of the probability of assignment to treatment~2. We use the fully sequential trial setup of Section~4.2 in the paper, with $h = 2$ treatments and $\delta_1 = 0$, $\delta_2 = 0.7$. }}

\end{figure}

Firstly consider the fully sequential trial setup of Section~4.2 in the paper, with $\delta_1 = 0$, $\delta_2 = 0.7$. Figure~\ref{fig:unequal_rand_seq} shows how the power of the Holm adaptive test and $z$-test compares as  $p_2$ varies. We see that when $p_2 >0.5$, the adaptive test only suffers a small loss of power compared to the $z$-test. However, when $p_2 < 0.5$, the adaptive test loses an increasing amount of power.

\begin{figure}[ht!]
\centering
\includegraphics[scale=0.75]{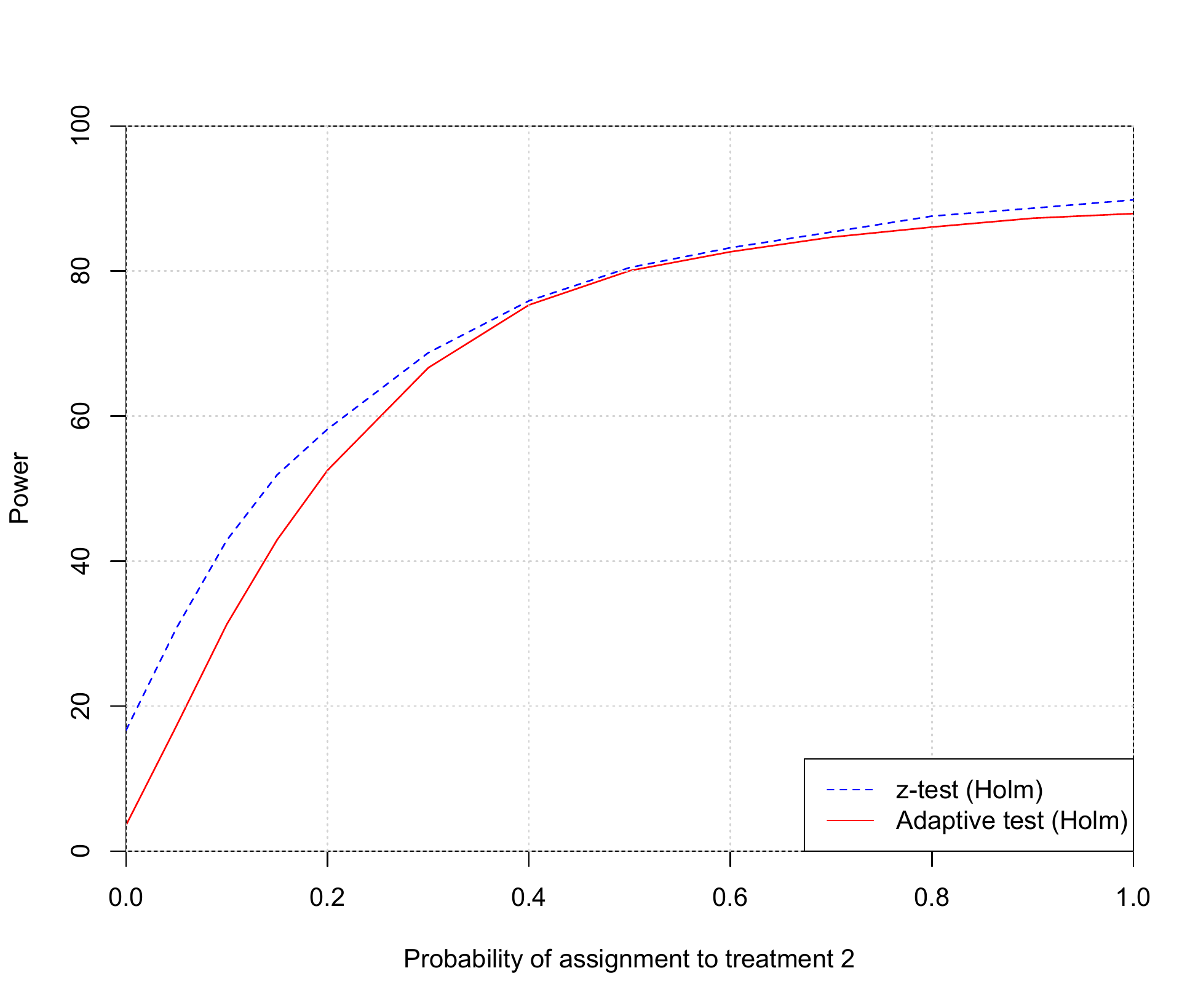}

\caption{\label{fig:unequal_rand_block} \textit{Power of the Holm adaptive test and Holm $z$-test as a function of the probability of assignment to treatment~2. We use the block randomized trial setup of Section~4.3 in the paper, with $h = 2$ treatments and $\delta_1 = 0$, $\delta_2 = 0.5$.}}
\end{figure}

Now consider the block randomization setup of Section~4.3 in the paper, with $\delta_1 = 0$, $\delta_2 = 0.5$. Figure~\ref{fig:unequal_rand_block} shows that this time, the power of the adaptive test is very close, or even equal, to the $z$-test when $p_2 > 1/3$. This shows how the adaptive test is more robust in terms of power in the block randomization setting compared to the fully sequential version.

\begin{figure}[ht!]
\centering
\includegraphics[scale=0.75]{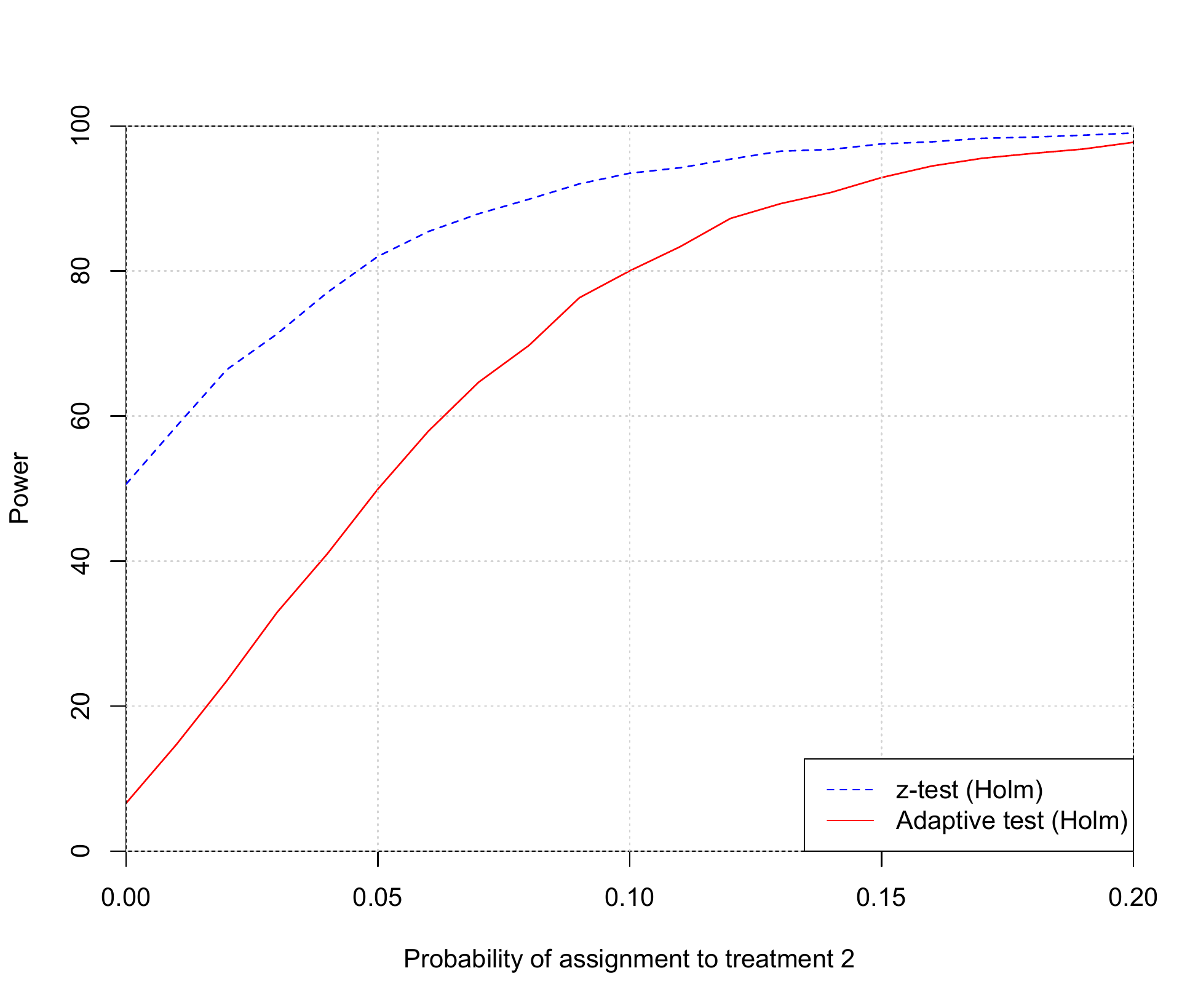}

\caption{\label{fig:unequal_rand_block2} \textit{Power of the Holm adaptive  test and Holm $z$-test as a function of the probability of assignment to treatment~2. We use the block randomized trial setup of Section~4.3 in the paper, with $h = 2$ treatments and $\delta_1 = 0$, $\delta_2 = 1$.}}
\end{figure}

Figure~\ref{fig:unequal_rand_block2} shows how the powers differ for $\delta_1 = 0$, $\delta_2 = 1$ and $p_2 < 0.2$. We can see that when $p_2<0.15$, there is a noticeable and increasing divergence between the powers of the two tests. Indeed, when $p_2 = 0$ the power of the Holm $z$-test is three times that of the Holm adaptive test. This shows what is happening with the type~I error inflator when $\delta_1 = 0$, where in the majority of trial scenarios, apart from the unlikely event that treatment~1 stops early for `efficacy', $p_2 = 0$ by design. Hence, the type~I inflator is in fact close to a worst-case scenario for the adaptive tests.

\clearpage

\subsection{Using the pooled sample variance}

In this section, we assume that $\sigma^2$ is unknown and will be estimated at the end of the trial using the pooled sample variance~$\hat{\sigma}^2$ (as defined below) of the experimental treatments and control. More precisely, given the adaptive test statistic~$\tilde{T}_I$, to test hypothesis~$H_I$ we compare $\tilde{T}_I/\hat{\sigma}$ with the critical value $z_{\alpha}(1/n_I' + 1/n_0')$. We rerun all the simulation studies in Section~4 of the paper and Appendix~E.1, with exactly the same setup except for this change in the test procedure. Due to the extra variability induced by estimating the pooled sample variance, we simulate $10^6$ trials for each set of parameter values.

Given that there are~$h$ experimental treatments, the formula for the pooled sample variance is  \begin{equation}
\hat{\sigma}^2 = \frac{\sum_{i = 0}^h \, (n_i - 1)s_i^2}{\sum_{i = 0}^h \, (n_i - 1)}
\end{equation} where $s_i^2$ is the sample variance for treatment~$i$ ($i = 0, 1, \ldots, h$).

For an adaptive control allocation,\[
s_i^2 = \frac{1}{n_i - 1} \sum_{k = 1}^{n} \left( \mathbbm{1}_{\{a_k = i\}} \left( X_k - \bar{X}_i \right)^2 \right) \qquad \text{ for } i = 0, 1, \ldots, h \]
where \[
\bar{X}_i = \frac{1}{n_i} \sum_{k =1}^{n_i} \mathbbm{1}_{\{a_k = i\}} X_k \qquad \text{ for } i = 0, 1, \ldots, h
\]
\noindent
For a fixed control allocation, the formulae for $s_i^2$ and $\bar{X}_i$ are the same for $i = 1, \ldots, h$, but now \begin{align*}
s_0^2 & = \frac{1}{n_0 - 1} \sum_{j = 1}^{n_0} \left( X_{0j} - \bar{X}_0 \right)^2\\
\bar{X}_0 &=  \frac{1}{n_0} \sum_{j =1}^{n_0} X_{0j}
\end{align*}

\newpage
\subsubsection{Fully sequential randomization (with a fixed control allocation)}

Table~\ref{Atab:typeI_inflator_seq} gives the results for the type~I error inflator, and Table~\ref{Atab:BAR_seq} the results for BAR. Compared to assuming a known variance of $\sigma^2 = 1$, for all sets of parameter values both the FWER and disjunctive power increase slightly. For the type~I error inflator, the Bonferroni-corrected $z$-test now has a FWER above 5\% in scenarios~2 and~5, while the adaptive closed test now has a FWER of 5.1\% in scenario~6. However, the Holm adaptive test still achieves strong FWER control.  For BAR, as before all the testing strategies control the FWER. 

\singlespace
\begin{table}[ht!]

\caption{\label{Atab:typeI_inflator_seq}  \textit{Familywise error rate and disjunctive power for the type~I error inflator in the fully sequential setting. There were $10^6$ simulated trials for each set of parameter values.}}

\resizebox{\linewidth}{!} {
		\begin{tabular}{p{0.5ex} l c c p{0cm} c c p{0cm} c c p{0cm} c c p{0cm} cc}
		\\ \hline \hline \Tstrut
		& & \multicolumn{2}{c}{Adaptive closed test} & & \multicolumn{2}{c}{Adaptive test (Holm)}
		& & \multicolumn{2}{c}{Closed $z$-test}  & &  \multicolumn{2}{c}{$z$-test (Holm)} 
		& & \multicolumn{2}{c}{$z$-test (Bonferroni)} \\
		\cline{3-4} \cline{6-7} \cline{9-10} \cline{12-13} \cline{15-16}\\ [-2ex]
		& {Parameter values} & Error & Power & & Error & Power
		& & Error & Power & & Error & Power & & Error & Power \\ \hline \Tstrut
		1.\ & $\delta_1 = \delta_2 = 0$
		  & 3.4 & - & & 4.9 & - & & 4.9 & - & & \textbf{7.5} & - & & \textbf{7.5} & -  \\ [1.5ex]
   		 
   		2.\ &  $\delta_1 = 0$, $\delta_2 = 1$ 
   		& 5.0 & 21.8 & & 3.8 & 27.6 & & \textbf{10.5} & 26.8 & & \textbf{10.1} & {63.9} & & \textbf{5.3} & 63.8 \\ [1.5ex]
		
		3.\ & $\delta_1 = \delta_2 = 0.5$
       & - & 62.4 & & - & 52.7 & & - &  {69.9} & & - & 61.6 & & - & 61.6 \\ [1.5ex]	
   		
		4.\ &  $\delta_1 = \delta_2 = \delta_3 = 0$ 
		& 3.0 & - & & 4.3 & - & & 4.3 & - & & \textbf{6.4} & - & & \textbf{6.4} \\ [1.5ex]
   		   		   		
   		5.\ &  $\delta_1 = \delta_2 = 0$, $\delta_3 = 1$
   		& 3.6 & 13.2 & & 4.6 & 24.9  & & \textbf{5.4} & 17.3 & & \textbf{7.0} &  {54.6} & & \textbf{5.1} & 54.5\\ [1.5ex]
   		
   		6.\ &  $\delta_1 = 0$, $\delta_2 = \delta_3 = 1$ 
   		& \textbf{5.1} & 22.6 & & 3.6 & 28.8  & & \textbf{10.3} & 27.5 & & \textbf{9.6} &  {72.8} & & 3.6 &  {72.8} \\ [1.5ex]

		7.\ &  $\delta_1 = 0$, $\delta_2 = 0.5$, $\delta_3 = 1$
		& 4.5 & 19.5 & & 3.0 & 25.5  & & \textbf{9.6} & 24.4 & & \textbf{7.9} & {59.0} & & 3.7 & 58.9 \\ [1.5ex]
		
   		8.\ &  $\delta_1 = \delta_2 = \delta_3 = 0.5$
   		& - & 51.6 & &  - & 43.0 & & - & {57.8} & & - & 50.0 & & - & 50.0 \\[1ex] \hline

		\end{tabular}
}

\end{table}

\begin{table}[ht!]

\caption{\label{Atab:BAR_seq} \textit{Familywise error rate and disjunctive power for BAR in the fully sequential setting. There were $10^6$ simulated trials for each set of parameter values.}}	

\resizebox{\linewidth}{!} {
\begin{tabular}{p{0.5ex} l c c p{0cm} c c p{0cm} c c p{0cm} c c p{0cm} cc}
		\\ \hline \hline \Tstrut
		& & \multicolumn{2}{c}{Adaptive closed test} & & \multicolumn{2}{c}{Adaptive test (Holm)}
		& & \multicolumn{2}{c}{Closed $z$-test}  & &  \multicolumn{2}{c}{$z$-test (Holm)} 
		& & \multicolumn{2}{c}{$z$-test (Bonferroni)} \\
		\cline{3-4} \cline{6-7} \cline{9-10} \cline{12-13} \cline{15-16}\\ [-2ex]
		& {Parameter values} & Error & Power & & Error & Power
		& & Error & Power & & Error & Power & & Error & Power \\ \hline \Tstrut
		1.\ & $\delta_1 = \delta_2 = 0$
		  & 4.8 & - & & 4.7 & - & & 4.9 & - & & 4.2 & - & & 4.2 & -  \\ [1.5ex]
   		 
   		2.\ &  $\delta_1 = 0$, $\delta_2 = 0.5$ 
   		& 4.7 & 46.7 & & 4.4 & 52.6 & & 4.0 & 46.9 & & 3.7 &  {53.6} & & 2.0 & 53.5 \\ [1.5ex]
		
		3.\ & $\delta_1 = \delta_2 = 0.5$
       & - & 70.9 & & - & 66.7 & & - &  {71.1} & & - & 65.8 & & - & 65.8 \\ [1.5ex]	
   		
		4.\ &  $\delta_1 = \delta_2 = \delta_3 = 0$
   		& 4.2 & - & & 4.6 & - & & 4.3 & - & & 4.2 & - & & 4.2 \\ [1.5ex]
   		   		   		
   		5.\ &  $\delta_1 = \delta_2 = 0$, $\delta_3 = 1$
   		& 4.7 & 59.9 & & 4.8 & 88.8 & & 4.6 & 60.1 & & 4.2 &  {90.2} & & 2.9 &  {90.2}\\ [1.5ex]
   		
   		6.\ &  $\delta_1 = 0$, $\delta_2 = \delta_3 = 1$
		& 5.0 & 89.9 & & 5.0 & 95.3 & & 4.2 & 89.9 & & 4.2 &  {95.8} & & 1.4 &  {95.8} \\ [1.5ex]
		 
		7.\ &  $\delta_1 = 0$, $\delta_2 = 0.5$, $\delta_3 = 1$ 
		& 4.7 & 75.1 & & 4.2 & 88.6 & & 4.1 & 75.6 & & 3.6 &  {89.7} &  & 1.5 &  {89.7}\\ [1.5ex]

   		8.\ &  $\delta_1 = \delta_2 = \delta_3 = 0.5$ 
   		& - & 57.2 & &  - & 53.2 & & - &  {58.0} & & - & 52.8 & & - & 52.8 \\[1ex] \hline
		\end{tabular}
		}

\end{table}

\doublespace

\clearpage
\subsubsection{Block randomization with a fixed control allocation}

Table~\ref{Atab:typeI_inflator_block} gives the results for the type~I error inflator, and Table~\ref{Atab:BAR_block} the results for BAR. Compared to assuming a known variance of $\sigma^2 = 1$, for all sets of parameter values both the FWER and disjunctive power increase slightly. For the type~I error inflator, the same scenarios as before lead to an inflation of the FWER. In particular, both the adaptive closed test and the Holm adaptive test achieve FWER control. For BAR, as before all the testing strategies control the FWER. \\[-6pt]

\begin{table}[ht!]

\caption{\label{Atab:typeI_inflator_block}  \textit{Familywise error rate and disjunctive power for the type~I error inflator, for block randomization with a fixed control allocation. There were $10^6$ simulated trials for each set of parameter values.}}

\resizebox{\linewidth}{!} {
\begin{tabular}{p{0.5ex} l c c p{0cm} c c p{0cm} c c p{0cm} c c p{0cm} cc}
		\\ \hline \hline \Tstrut
		& & \multicolumn{2}{c}{Adaptive closed test} & & \multicolumn{2}{c}{Adaptive test (Holm)}
		& & \multicolumn{2}{c}{Closed $z$-test}  & &  \multicolumn{2}{c}{$z$-test (Holm)} 
		& & \multicolumn{2}{c}{$z$-test (Bonferroni)} \\
		\cline{3-4} \cline{6-7} \cline{9-10} \cline{12-13} \cline{15-16}\\ [-2ex]
		& {Parameter values} & Error & Power & & Error & Power
		& & Error & Power & & Error & Power & & Error & Power \\ \hline \Tstrut
		1.\ & $\delta_1 = \delta_2 = 0$
		  & 3.9 & - & & 4.9 & - & & 4.7 & - & & \textbf{6.9} & - & & \textbf{6.9} & -  \\ [1.5ex]
   		 
   		2.\ &  $\delta_1 = 0$, $\delta_2 = 1$ 
   		& 4.9 & 22.0 & & 3.7 & 27.0 & & \textbf{8.4} & 25.6 & & \textbf{7.9} &  {61.1} & & 4.4 & 61.0 \\ [1.5ex]
		
		3.\ & $\delta_1 = \delta_2 = 0.5$
       & - & 92.6 & & - & 87.7 & & - & {94.6} & & - & 91.6 & & - & 91.6 \\ [1.5ex]	
   		
		4.\ &  $\delta_1 = \delta_2 = \delta_3 = 0$ 
		& 3.2 & - & & 4.1 & - & & 4.1 & - & & \textbf{6.3} & - & & \textbf{6.3} \\ [1.5ex]
   		   		   		
   		5.\ &  $\delta_1 = \delta_2 = 0$, $\delta_3 = 1$
   		& 3.8 & 14.2 & & 4.4 & 23.4  & & 4.8 & 18.3 & & \textbf{6.4} &  {61.4} & & 4.7 & 61.2\\ [1.5ex]
   		
   		6.\ &  $\delta_1 = 0$, $\delta_2 = \delta_3 = 1$ 
   		& 5.0 & 20.1 & & 3.3 & 26.1  & & \textbf{8.3} & 23.1 & & \textbf{7.5} &  {79.8} & & 3.0 & 79.8\\ [1.5ex]

		7.\ &  $\delta_1 = 0$, $\delta_2 = 0.5$, $\delta_3 = 1$ 
		& 4.8 & 17.7 & & 3.0 & 23.9  & & \textbf{8.1} & 21.3 & & \textbf{6.8} &  {66.3} & & 3.0 &  {66.3}\\ [1.5ex]
		
   		8.\ &  $\delta_1 = \delta_2 = \delta_3 = 0.5$
   		& - & 91.2 & &  - & 83.1 & & - &  {93.9} & & - & 89.5 & & - & 89.5 \\[1ex] \hline

		\end{tabular}
		}

\end{table}

\begin{table}[ht!]

	\caption{\label{Atab:BAR_block}  \textit{Familywise error rate and disjunctive power for BAR, for block randomization with a fixed control allocation. There were $10^6$ simulated trials for each set of parameter values. There were $10^6$ simulated trials for each set of parameter values.}}	

\resizebox{\linewidth}{!} {
\begin{tabular}{p{0.5ex} l c c p{0cm} c c p{0cm} c c p{0cm} c c p{0cm} cc}
		\\ \hline \hline \Tstrut
		& & \multicolumn{2}{c}{Adaptive closed test} & & \multicolumn{2}{c}{Adaptive test (Holm)}
		& & \multicolumn{2}{c}{Closed $z$-test}  & &  \multicolumn{2}{c}{$z$-test (Holm)} 
		& & \multicolumn{2}{c}{$z$-test (Bonferroni)} \\
		\cline{3-4} \cline{6-7} \cline{9-10} \cline{12-13} \cline{15-16}\\ [-2ex]
		& {Parameter values} & Error & Power & & Error & Power
		& & Error & Power & & Error & Power & & Error & Power \\ \hline \Tstrut
		1.\ & $\delta_1 = \delta_2 = 0$
		  & 4.7 & - & & 4.6 & - & & 4.7 & - & & 4.5 & - & & 4.5 & -  \\ [1.5ex]
   		 
   		2.\ &  $\delta_1 = 0$, $\delta_2 = 0.5$ 
   		& 5.0 & 61.3 & & 5.0 & 82.6 & & 4.9 & 61.3 & & 4.9 & {82.9} & & 2.5 & 82.8 \\ [1.5ex]
		
		3.\ & $\delta_1 = \delta_2 = 0.5$
       & - &  {94.5} & & - & 92.2 & & - &  {94.5} & & - & 92.2 & & - & 92.2 \\ [1.5ex]	
   		
		4.\ &  $\delta_1 = \delta_2 = \delta_3 = 0$
   		& 3.9 & - & & 4.6 & - & & 3.8 & - & & 4.4 & - & & 4.4 \\ [1.5ex]
   		   		   		
   		5.\ &  $\delta_1 = \delta_2 = 0$, $\delta_3 = 0.5$
   		& 4.5 & 36.0 & & 4.6 &  {71.7} & & 4.5 & 36.0 & & 4.4 &  {71.7} & & 3.1 & 71.7\\ [1.5ex]
   		
   		6.\ &  $\delta_1 = 0$, $\delta_2 = \delta_3 = 0.5$
		& 5.0 & 67.3 & & 4.7 &  {85.4} & & 4.8 & 66.8 & & 4.5 & 85.2 & & 1.6 & 85.2 \\ [1.5ex]
		 
		7.\ &  $\delta_1 = 0$, $\delta_2 = 0.25$, $\delta_3 = 0.5$ 
		& 4.6 & 51.1 & & 3.8 &  {72.9} & & 4.4 & 50.9 & & 3.6 & 72.6 &  & 1.6 & 72.6\\ [1.5ex]

   		8.\ &  $\delta_1 = \delta_2 = \delta_3 = 0.5$ 
   		& - &  {93.3} & &  - & 90.4 & & - & 93.2 & & - & 90.2 & & - & 90.2 \\[1ex] \hline

		\end{tabular}
		}

\end{table}

\clearpage
\subsubsection{Block randomisation with an adaptive control allocation}

Table~\ref{Atab:typeI_inflator_block2b} gives the results for the type~I error inflator, and Table~\ref{Atab:BAR_block2b} the results for BAR. Compared to assuming a known variance of $\sigma^2 = 1$, for all sets of parameter values both the FWER and disjunctive power increase slightly. For the type~I error inflator, the same scenarios as before lead to an inflation of the FWER. In particular, both the adaptive closed test and the Holm adaptive test achieve FWER control. However, for BAR, the adaptive closed test now has a FWER of 5.1\% in scenario~2 (the FWER is controlled for all other scenarios and testing procedures). \\[-6pt]

\begin{table}[ht!]

\caption{\label{Atab:typeI_inflator_block2b}   \textit{Familywise error rate and disjunctive power for the type~I error inflator, for block randomization with an adaptive control allocation. There were $10^6$ simulated trials for each set of parameter values.}}

\resizebox{\linewidth}{!} {
\begin{tabular}{p{0.5ex} l c c p{0cm} c c p{0cm} c c p{0cm} c c p{0cm} cc}
		\\ \hline \hline \Tstrut
		& & \multicolumn{2}{c}{Adaptive closed test} & & \multicolumn{2}{c}{Adaptive test (Holm)}
		& & \multicolumn{2}{c}{Closed $z$-test}  & &  \multicolumn{2}{c}{$z$-test (Holm)} 
		& & \multicolumn{2}{c}{$z$-test (Bonferroni)} \\
		\cline{3-4} \cline{6-7} \cline{9-10} \cline{12-13} \cline{15-16}\\ [-2ex]
		& {Parameter values} & Error & Power & & Error & Power
		& & Error & Power & & Error & Power & & Error & Power \\ \hline \Tstrut
		1.\ & $\delta_1 = \delta_2 = 0$
		  & 3.9 & - & & 5.0 & - & & 4.6 & - & & \textbf{7.0} & - & & \textbf{7.0} & -  \\ [1.5ex]
   		 
   		2.\ &  $\delta_1 = 0$, $\delta_2 = 1$ 
   		& 5.0 & 19.1 & & 3.8 & 25.1 & & \textbf{8.4} & 25.7 & & \textbf{7.9} & {67.4} & & 4.5 & 67.3 \\ [1.5ex]
		
		3.\ & $\delta_1 = \delta_2 = 0.5$
       & - & 90.0 & & - & 84.4 & & - & {93.2} & & - & 89.8 & & - & 89.8 \\ [1.5ex]	
   		
		4.\ &  $\delta_1 = \delta_2 = \delta_3 = 0$ 
		& 3.3 & - & & 4.1 & - & & 4.0 & - & & \textbf{6.6} & - & & \textbf{6.6} & -\\ [1.5ex]
   		   		   		
   		5.\ &  $\delta_1 = \delta_2 = 0$, $\delta_3 = 1$
   		& 3.8 & 14.0 & & 4.5 & 22.1 & & 4.9 & 20.0 & & \textbf{6.6} & {62.0} & & 4.9 & 61.9 \\ [1.5ex]
   		
   		6.\ &  $\delta_1 = 0$, $\delta_2 = \delta_3 = 1$ 
   		& 5.0 & 19.2 & & 3.5 & 24.7 & & \textbf{8.5} & 26.5 & & \textbf{7.7} &{80.6} & & 3.2 & {80.6} \\ [1.5ex]
		 
		7.\ &  $\delta_1 = 0$, $\delta_2 = 0.5$, $\delta_3 = 1$ 
		& 4.7 & 17.1 & & 3.1 & 22.6 &  & \textbf{8.2} & 24.0 & & \textbf{6.8} & {67.1} & & 3.2 & 67.0 \\ [1.5ex]
		
   		8.\ &  $\delta_1 = \delta_2 = \delta_3 = 0.5$
   		& - & 87.4 & & - & 78.3 & & - & {91.7} & & - & 86.8 & & - &  86.8 \\[1ex] \hline
		\end{tabular}
		}

\end{table}

\begin{table}[ht!]

\caption{\label{Atab:BAR_block2b} \textit{Familywise error rate and disjunctive power for BAR, for block randomization with an adaptive control allocation. There were $10^6$ simulated trials for each set of parameter values.}}	

\resizebox{\linewidth}{!} {
\begin{tabular}{p{0.5ex} l c c p{0cm} c c p{0cm} c c p{0cm} c c p{0cm} cc}
		\\ \hline \hline \Tstrut
		& & \multicolumn{2}{c}{Adaptive closed test} & & \multicolumn{2}{c}{Adaptive test (Holm)}
		& & \multicolumn{2}{c}{Closed $z$-test}  & &  \multicolumn{2}{c}{$z$-test (Holm)} 
		& & \multicolumn{2}{c}{$z$-test (Bonferroni)} \\
		\cline{3-4} \cline{6-7} \cline{9-10} \cline{12-13} \cline{15-16}\\ [-2ex]
		& {Parameter values} & Error & Power & & Error & Power
		& & Error & Power & & Error & Power & & Error & Power \\ \hline \Tstrut
		1.\ & $\delta_1 = \delta_2 = 0$
		  & 4.7 & - & & 4.7 & - & & 4.7 & - & & 4.6 & - & & 4.6 & -  \\ [1.5ex]
   		 
   		2.\ &  $\delta_1 = 0$, $\delta_2 = 0.5$ 
   		& \textbf{5.1} & 54.6 & & 5.0 & 76.5 & & 5.0 & 56.2 & & 4.9 & {78.1} & & 2.5 & {78.1} \\ [1.5ex]
		
		3.\ & $\delta_1 = \delta_2 = 0.5$
       & - & 90.5 & & - & 87.3 & & - & {91.4} & & - & 88.3 & & - & 88.3 \\ [1.5ex]	
   		
		4.\ &  $\delta_1 = \delta_2 = \delta_3 = 0$ 
		& 4.0 & - & & 4.5 & - & & 4.0 & - & & 4.4 & - & & 4.4 & - \\ [1.5ex]
   		   		   		
   		5.\ &  $\delta_1 = \delta_2 = 0$, $\delta_3 = 0.5$
   		& 4.7 & 29.2 & & 4.6 & 61.3 & & 4.6 & 30.1 & & 4.4 & {62.8} & & 3.1 & {62.8} \\ [1.5ex]
   		
   		6.\ &  $\delta_1 = 0$, $\delta_2 = \delta_3 = 0.5$ 
   		& 5.0 & 56.5 & & 4.7 & 75.9 & & 4.9 & 57.5 & & 4.5 & {77.1} & & 1.7 & {77.1}\\ [1.5ex]
		 
		7.\ &  $\delta_1 = 0$, $\delta_2 = 0.25$, $\delta_3 = 0.5$
		& 4.5 & 41.8 & & 3.7 & 62.4 & & 4.4 & 42.8 & & 3.5 & {63.7} & & 1.7 & {63.6} \\ [1.5ex]

   		8.\ &  $\delta_1 = \delta_2 = \delta_3 = 0.5$ 
   		& - & 85.9 & & - & 81.9 & & - &  {86.9} & & - & 82.9 & & - & 82.9 \\[1ex] \hline
		\end{tabular}
		}

\end{table}

%

\end{document}